\documentclass[iop,revtex4]{emulateapj}
 \usepackage{color}
\newcommand{\Kkms}{\mbox{K km s$^{-1}$}}
\newcommand{\Lsun}{\mbox{$\mathrm{L}_{\sun}$}}
\newcommand{\Msun}{\mbox{$\mathrm{M}_{\sun}$}}
\newcommand{\um}{\mbox{$\mu$m}} 
\newcommand{\skipthis}[1]{}
\newcommand{\hii}{\mbox{\ion{H}{2}}}
\newcommand{\gtres}{\mbox{G331.5$-$0.1}}
\newcommand{\kms}{\mbox{km s$^{-1}$}}
\newcommand{\outflow}{G331.512-0.103}

\newcommand{\csduj}{\mbox{CS$(J = 2\rightarrow$1)}}
\newcommand{\csccj}{\mbox{CS$(J = 5\rightarrow$4)}}
\newcommand{\csssj}{\mbox{CS$(J = 7\rightarrow$6)}}
\newcommand{\tcotdj}{\mbox{$^{13}$CO$(J = 3\rightarrow$2)}}

\newcommand{\coucj}{\mbox{CO$(J = 1\rightarrow$0)}}
\newcommand{\cDOoj}{\mbox{C$^{18}$O$(J = 1\rightarrow$0)}}
\newcommand{\siossj}{\mbox{SiO$(J = 7\rightarrow$6)}}
\newcommand{\sioosj}{\mbox{SiO$(J = 8\rightarrow$7)}}

\newcommand{\csdu}{\mbox{CS$(2\rightarrow$1)}}
\newcommand{\cscc}{\mbox{CS$(5\rightarrow$4)}}
\newcommand{\csss}{\mbox{CS$(7\rightarrow$6)}}
\newcommand{\tcotd}{\mbox{$^{13}$CO$(3\rightarrow$2)}}

\newcommand{\coss}{\mbox{CO$(7\rightarrow$6)}}

\newcommand{\couc}{\mbox{CO$(1\rightarrow$0)}}
\newcommand{\cDOo}{\mbox{C$^{18}$O$(1\rightarrow$0)}}
\newcommand{\sioss}{\mbox{SiO$(7\rightarrow$6)}}
\newcommand{\sioos}{\mbox{SiO$(8\rightarrow$7)}}
\newcommand{\sooo}{\mbox{SO$(8_8\rightarrow7_7$)}}
\newcommand{\soos}{\mbox{SO$(8_7\rightarrow7_6$)}}

\shorttitle{Physical characteristics of G331.5-0.1}
\shortauthors{Merello et al.}

\begin{document}


\title{Physical characteristics of G331.5-0.1: \\ The luminous central region of a Giant Molecular Cloud}


\author{Manuel Merello$^{1,3}$, Leonardo Bronfman$^1$, Guido Garay$^1$, Lars-\AA ke Nyman$^2$, \\Neal J. Evans II$^3$, C. Malcolm Walmsley$^{4,5}$ }

\affil{$^1$Departamento de Astronom\'ia, Universidad de Chile, Casilla 36-D, Santiago, Chile}


\affil{$^2$Joint ALMA Observatory (JAO), Alonso de Cordova 3107, Vitacura, Santiago, Chile}


\affil{$^3$University of Texas at Austin, 1 University Station, Austin, Texas, 78712, USA}


\affil{$^4$Osservatorio Astrofisico di Arcetri, Largo E. Fermi 5, 50125 Firenze, Italy}
\affil{$^5$Dublin Institute of Advanced Studies, Fitzwilliam Place 31, Dublin 2, Ireland}



\begin{abstract}
We report molecular line and dust continuum observations toward the high-mass star forming region G331.5-0.1, one of the most luminous regions of massive star-formation in the Milky Way, located at the tangent region of the Norma spiral arm, at a distance of 7.5 kpc. Molecular emission was mapped toward the G331.5-0.1 GMC in the CO$(J = 1\rightarrow$0) and C$^{18}$O$(J = 1\rightarrow$0) lines with NANTEN, while its central region was mapped in CS($J = 2\rightarrow1$ and $J = 5\rightarrow4$) with SEST, and in CS$(J = 7\rightarrow$6) and $^{13}$CO$(J = 3\rightarrow$2) with ASTE. Continuum emission mapped at 1.2 mm with SIMBA and at 0.87 mm with LABOCA reveal the presence of six compact and luminous dust clumps, making this source one of the most densely populated central regions of a GMC in the Galaxy. The dust clumps are associated with molecular gas and they have the following average properties: size of 1.6 pc, mass of $3.2\times10^3$ $\mathrm{M}_{\sun}$, molecular hydrogen density of $3.7\times10^4$ cm$^{-3}$, dust temperature of 32 K, and integrated luminosity of $5.7\times10^5$ $\mathrm{L}_{\sun}$, consistent with values found toward other massive star forming dust clumps. The CS and $^{13}$CO spectra show the presence of two velocity components: a high-velocity component at $\sim-89$ km s$^{-1}$, seen toward four of the clumps, and a low-velocity component at $\sim-101$ km s$^{-1}$ seen toward the other two clumps. Radio continuum emission is present toward four of the molecular clumps, with spectral index estimated for two of them of 0.8$\pm$0.2 and 1.2$\pm$0.2. A high-velocity molecular outflow is found at the center of the brightest clump, with a line width of 26 km s$^{-1}$ (FWHM) in CS$(J = 7\rightarrow$6) . Observations of SiO($J = 7\rightarrow6$ and $J = 8\rightarrow7$), and SO($J_K = 8_8\rightarrow7_7$ and $J_K = 8_7\rightarrow7_6$) lines provide estimates of the gas rotational temperature toward this outflow $>$120 K and $>75$ K, respectively.

\end{abstract}




\section{Introduction}

Massive stars play a key role in the evolution of the Galaxy; hence they are important objects of study in astrophysics.
Although their number is much smaller than the low mass stars, they are the principal source of heavy elements 
and UV radiation, affecting the process of formation of stars and planets \citep{Bally2005} and the physical, chemical and morphological structure of 
galaxies \citep[e.g.][]{Kennicutt98,Kennicutt2005}.

Diverse studies show that massive stars are formed in the denser regions of giant molecular clouds (GMC) \citep{Churchwell90, Cesaroni91, Plume92, Bronfman96}. The formation process of high-mass stars is, however, still under debate \citep{Garay99}. Observationally, recently formed massive stars cannot be detected at optical wavelengths because they are embedded 
within regions with high dust extinction. Another problem arises from theoretical considerations: a star with a mass larger than eight solar masses starts burning hydrogen during the accretion phase and the radiation pressure may halt or even reverse the infall process \citep[e.g.][]{Larson71,Kahn74,Yorke77,Wolfire87,Edgar04,Kuiper10}. Due to these difficulties, a paradigm that explains the formation of massive stars does not exist \citep{Zinnecker07, McKee07}, contrary to the successful model accepted for low mass stars. Understanding the physical process that dominates during the early stages of the formation of massive stars and its influence back on its harboring molecular gas requires a study of the physical conditions of the environment previous to the star formation or at very early stages.

The object of our study is one of the most massive GMCs in the Southern Galaxy, at  $l=331.5\arcdeg$ and $b=-0.1$, in the tangent region of the Norma spiral arm \citep{Bronfman85, Bronfman89}.
Several observations have shown that the central region of this GMC
harbors one of the most extended and luminous regions of massive star formation
in the Galactic disk. Single dish radio continuum observations show the
presence of an extended region of ionized gas (\gtres), with an angular size of
$\sim3\arcmin-4$\arcmin (projected size 6.5-8.7 pc at a distance of 7.5 kpc), close to the peak CO position of the GMC \citep{Shaver70, Caswell87}. From observations of the H109$\alpha$ and 
H110$\alpha$ hydrogen 
recombination lines,~\citet{Caswell87} determined that the line central velocity of 
the ionized gas is $\sim$ -89 \kms, similar to the peak velocity of the GMC. The dynamical timescale of the region of ionized gas is 
$\sim 8.5\times10^5$ years, and the rate of ionizing photons required to excite the \hii\ region is $1.1\times10^{50}$ s$^{-1}$.
In addition, IRAS observations show the presence of an extended source, with an
angular size similar to that of the ionized region and a FIR luminosity of
3.6$\times10^6$ $\mathrm{L}_{\sun}$.

While both the ionizing photon rate and the FIR luminosity could be provided
by a single O4 ZAMS star, it is most likely that the ionization and the heating of
the region  is provided by a cluster of massive OB stars. This explanation is
supported by {\sl Midcourse Space Experiment} and {\sl Spitzer Space
Telescope}-GLIMPSE survey data, which show that there are several sites of massive
star formation spread over a region of $\sim6'$ in diameter.
\citet{Bronfman2000} showed that toward the $l=331\arcdeg$ galactic longitude, 
there is a peak in the number density of IRAS sources with colors of ultracompact \ion{H}{2} (UCHII) regions and associated with \csdu\ emission, which are thought to correspond to massive star 
forming regions. 

Evidence of active star formation in the \gtres\ GMC is further provided by 
the detection of OH and methanol maser emission \citep{Goss70, Caswell80, Caswell97,
Caswell98, Pestalozzi2005}, as well as H$_2$O maser
emission (Caswell, private communication).  Hydroxyl masers are known to 
be excellent indicators of UCHII regions, whereas methanol maser emission is a 
common phenomenon in regions of massive star formation \citep{Menten86} and
H$_2$O maser emission is thought to be a probe of the 
earliest stages of massive star formation \citep{Furuya03}. 
Also, within a radius of 4.5' centered on the peak position of \gtres\  
there are three objects from the Red MSX Source (RMS) catalog 
of Massive Young Stellar Objects (MYSOs) candidates  \citep{Urquhart08}, further indicating that the \gtres\ central region 
is an active region of massive star formation. In addition, \cite{Bronfman2008} showed the presence of a compact, extremely high velocity molecular outflow in the \gtres\ region, suggesting that it is driven by one of the most luminous and massive objects presently known.
The \gtres\ region is thus an ideal region to study extremely active massive star formation in GMCs.

Here we present observations and a thorough analysis of the \gtres\ region with dust continuum and molecular line emission data. Section 2 describes the telescopes used and methods of data acquisition. Section 3 describes the results of our study at several scales, starting from GMCs ($\sim$100 pc), to molecular clumps sizes ($\sim$1 pc) and reaching sub-parsec scales in our interferometric data. This analysis considers dust emission and different molecular line tracers, along with mid-infrared and far-infrared data from SPITZER, MSX and IRAS telescopes, and free-free emission at cm wavelengths. Section 4 describes the physical parameters derived from our observations, including mass, density and bolometric luminosity estimates. We also include a rotational temperature analysis for a massive and energetic molecular outflow in the \gtres\ GMC central region. Our results are summarized in section 5.

 
 \section{Observations}
 
We observed emission in several molecular lines and continuum bands using several telescopes located in northern Chile: the NANTEN telescope of Nagoya University at the Las Campanas observatory, the 15-m Swedish-ESO Submillimetre Telescope (SEST) at La Silla, the 10-m Atacama Submillimeter Telescope Experiment (ASTE) at Pampa la Bola, and the 12-m Atacama Pathfinder Experiment (APEX) located in Llano Chajnantor. We also made observations using the Australia Telescope Compact Array (ATCA) located at Narrabri, Australia. 
The basic parameters of the molecular line observations are summarized in Table~\ref{tbl:observ_mol}. Column 1 gives the telescope used, Col. 2 and 3, the observed 
transition and line frequency, respectively. Col. 4 and 5 the half-power beam width and the main beam efficiency. Col. 6 to 10 give, respectively, the number of positions observed, the angular spacing, the integration time on 
source per position, the channel width and the resulting rms noise in antenna temperature, for each of the observed transitions.
The observing parameters for the continuum observations are summarized in Table~\ref{tbl:observ_cont}.

\subsection{NANTEN Telescope}

The NANTEN telescope is a 4-m diameter millimeter telescope designed for a large-scale survey of molecular clouds. With this telescope we observed the emission in the \coucj\ ($\nu = 115271.202$ MHz) and \cDOoj\ ($\nu = 109782.173$ MHz) lines during several epochs between 1999 and 2003. The half-power beam width of the telescope at 115 GHz is 2.6\arcmin, providing a spatial resolution of 5.7 pc at the assumed distance of the source of 7.5 kpc. The observed maps consisted in grids of 585 positions for the CO emission and 440 positions for the C$^{18}$O emission, with spacing between observed positions of 2.5\arcmin\ (nearly the beam size of the instrument at the observed frequencies). 
The front-end was a superconductor-isolator-superconductor (SIS) receiver cooled down to 4 K with a closed-cycle helium gas refrigerator, and the backend was an acousto-optical spectrometer (AOS) with 2048 channels. The total bandwidth was 250 MHz and the frequency resolution was 250 kHz, corresponding to a velocity resolution of $\sim$ 0.15 \kms. The typical system noise temperature was $\sim$ 250 K (SSB) at 115 GHz and $\sim$ 160 K (SSB) at 110 GHz. The main beam efficiency of the telescope is 0.89.
A detailed description of the telescope and its instruments are given by \cite{Ogawa90}, \citet{Fukui91} and \citet{Fukui92}. Data reduction were made using the NANTEN Data Reduction Software (NDRS).

\subsection{Swedish-ESO Submillimetre Telescope (SEST)}

\subsubsection{Millimeter continuum}
The 1.2 mm continuum observations were carried out using the 37-channel hexagonal
Sest IMaging Bolometer Array (SIMBA) during 2001. The passband of the
bolometer has an equivalent width of 90 GHz and is centered at 250
GHz (1200 \um). At the frequency band of SIMBA, the telescope has an effective
beamwidth of 24\arcsec\ (FWHM), that provides a spatial resolution of 0.87 pc at a distance of 7.5 kpc. The rms noise achieved is $\sim$50 mJy/beam. Data reduction was  made using the package MOPSI  (Mapping On-Off Pointing Skydip Infrared). The maps were calibrated from observations of Mars and Uranus and the calibration uncertainties are estimated to be lower than 20\%.

\subsubsection{Molecular lines}
With SEST we observed the emission in the \csccj\ ($\nu$ = 244935.644 MHz) and \csduj\ ($\nu$ = 97980.968 MHz) lines during May 2001. The half-power beam width and the main beam efficiency of the telescope were 22\arcsec\ and 0.48 at 245 GHz, and 52\arcsec\ and 0.73 at 98 GHz, respectively. The total bandwidth was 43 MHz and the frequency resolution was 43 kHz, corresponding to a velocity resolution of $\sim$ 0.05 \kms\ for the \cscc\ observations, and a velocity resolution of $\sim$ 0.13 \kms\ for the \csdu\ observations. The typical system noise temperature was 350 K for \cscc\ observations and 250 K for \csdu\ observations. The CLASS (Continuum and Line Analysis Single-dish Software) software, part of the GILDAS package, was used for data reduction.

Figures \ref{fig:grid_cs21} and \ref{fig:grid_cs54} on the online material present maps of the observed spectra toward the \gtres\ central region in the \csdu\ and  \cscc\ line emission over 81 positions with a spacing of 45\arcsec. The \csdu\ map then is well sampled, while the \cscc\ map is undersampled at this spacing, considering the smaller beam size at that frequency.

\subsection{Atacama Submillimeter Telescope Experiment}

With ASTE we observed the emission in the \csssj\ ($\nu$ = 342882.950 MHz) and \tcotdj\ ($\nu$ = 330587.957 MHz) lines during 2005 July and 2006 July. ASTE has a beam size at 350 GHz of 22\arcsec\ (FWHM). For both transitions 167 positions were sampled, within a $ \sim 360\arcsec \times 360\arcsec$ region with 22.5\arcsec\
spacing in position-switched mode, with the OFF position located at
$\alpha_{2000} = 16^h14^m08.4^s$ and $\delta_{2000} = -51\arcdeg
37\arcmin 23\arcsec$. The system temperatures ranged between 179 and
233 K, resulting in a rms noise antenna temperature of typically 0.1
K for an integration time of 4 minutes.

Main-beam temperatures, $T_{MB} = T^*_A/ \eta_{MB}$, obtained by dividing the antenna temperature
$T^*_A$ by the main-beam efficiency, $\eta_{MB}$, are used
throughout the analysis. During the first part of the observations, we mapped a grid of 88 positions, with
$\eta_{MB}=0.65$. On July, 2006. we completed the map observing 79 new
positions, with $\eta_{MB}=0.77$. The 1024 channels of the 350 GHz
band receiver provided a bandwidth of 128 MHz, and a channel width
of 0.125 MHz. The data reduction was carried out by use of the 
AIPS-based software package NEWSTAR developed at NRO.

Figures \ref{fig:grid_cs76} and \ref{fig:grid_13co32} on the online material present maps of the observed spectra toward the \gtres\ central region in the \csss\ and \tcotd\ line emission over 167 positions with a spacing of 22.5\arcsec\ (similar to the beam size of the instrument at this frequency). 

\subsection{Atacama Pathfinder Experiment}

\subsubsection{Millimeter continuum: the ATLASGAL survey}
The millimeter observations of the region were made as part of the APEX Telescope Large Area Survey of the Galaxy (ATLASGAL) project, which is a collaboration between the Max Planck Gesellschaft (MPG: Max Planck Institute f\"{u}r Radioastronomie, MPIfR Bonn, and Max Planck Institute f\"{u}r Astronomie, MPIA Heidelberg), the European Southern Observatory (ESO) and the Universidad de Chile. ATLASGAL is an unbiased survey of the inner region of the Galactic disk at 0.87 mm, made using the 295-channel hexagonal array Large Apex BOlometer CAmera (LABOCA), that operates in the 345 GHz atmospheric window, with a bandwidth of about 60 GHz. The angular resolution is 18.6\arcsec\ (HPBW) (corresponding to a resolution of 0.68 pc at a distance of 7.5 kpc) and the total field of view is 11.4\arcmin. The survey covers an area of $\pm$1.5\arcdeg\ in galactic latitude, and $\pm$60\arcdeg\ in longitude, with a sensitivity of 50 mJy/beam \citep{Schuller09}.

\subsubsection{Molecular lines}

Using the APEX-2a heterodyne receiver, we observed 
the emission in the \siossj\ ($\nu$ = 303926.809 MHz),  \sioosj\ ($\nu$ = 347330.635 MHz), \soos\ ($\nu$ = 304077.844 MHz) and \sooo\ ($\nu$ = 344310.612 MHz) lines during May 2007. The telescope
has a beam size of 17.5\arcsec\ at 345 GHz, corresponding to a spatial resolution of 0.63 pc at a distance of 7.5 kpc. The main beam efficiency of the telescope is $\eta_{MB}=0.73$ and the system temperature was $\sim$ 170 K.
The observations were made toward the peak position of the \coss\ outflow emission ($\alpha_{2000} = 16^h12^m10.09^s$ and $\delta_{2000} = -51\arcdeg 28\arcmin 38.4\arcsec$) reported by \cite{Bronfman2008}.
The data were reduced using the CLASS software of the GILDAS package.

\subsection{Australia Telescope Compact Array}

The ATCA radio continuum observations were made in two epochs: 2002 November using the 
6A configuration and 2005 November using the 1.5C configuration.
These configurations utilize all six antennas and cover east-west 
baselines from 0.5 to 6 km. Observations were made at the frequencies of 
4.80 and 8.64 GHz (6 and 3.6 cm), with a bandwidth of 128 MHz. The FWHM primary beam of 
ATCA at 4.8 GHz and 8.6 GHz are, respectively, 10\arcmin\ and 6\arcmin.
The total integration time in each frequency was 2.5 hr.

The calibrator PKS 1600-48 was observed before and after every on-source
scan in order to correct the amplitude and phase of the interferometer data
due to atmospheric and instrumental effects. The flux density was calibrated 
by observing PKS 1934-638 (3C84) for which values of 5.83 Jy at 4.8 GHz and
2.84 Jy at 8.6 GHz were adopted. Standard calibration and data reduction 
were performed using MIRIAD \citep{Sault95}. Maps were made by Fourier transformation of the uniformly weighted interferometer data using the AIPS task IMAGR. The noise level in the 4.8 and 8.6 GHz
images are, respectively, 0.41 and 0.37 mJy beam$^{-1}$.
The resulting synthesized (FWHM) beams are $2.7\arcsec\times1.8\arcsec$
at 4.8 GHz and $1.5\arcsec\times1.0\arcsec$ at 8.6 GHz.


\section{Results}

\subsection{The \gtres\ Giant Molecular Cloud}

\begin{figure}[h]
	\begin{center}
		\includegraphics[angle=-90,width=1\textwidth]{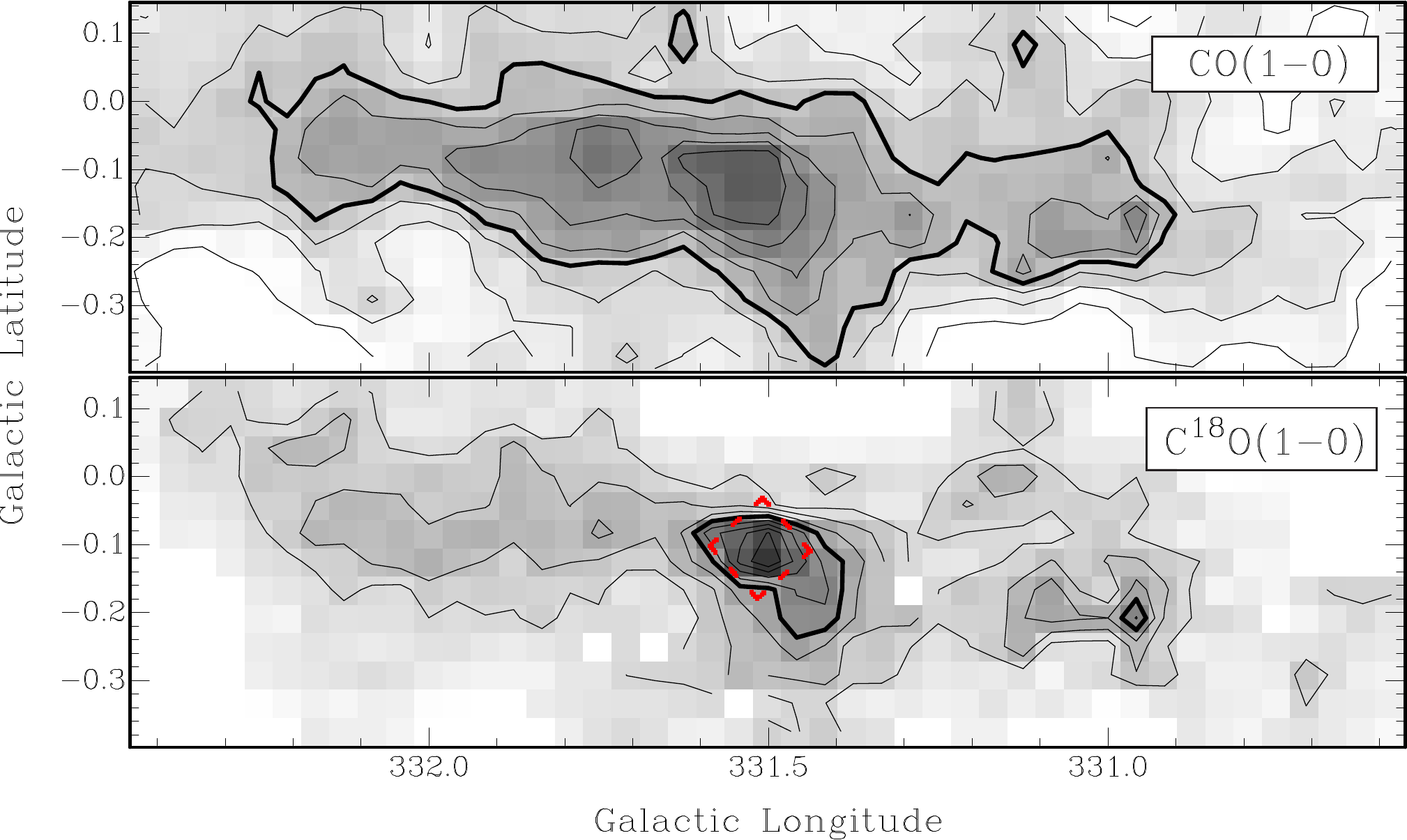}
	\end{center}
	\caption{Maps of velocity integrated emission, in the range -117.7 to -73.2. \kms, of the \gtres\ GMC. $Top:$ \couc. Contour levels are drawn from 20\% (3$\sigma$) to 90\%,
in steps of 10\%, of the peak intensity (297.6 K \kms.). $Bottom:$ \cDOo. Contour
levels are drawn from 20\% (4$\sigma$) to 90\%, in steps of 10\%,  of the peak 
intensity (19.6 K \kms). The 50\% emission in each map is drawn with a thicker contour. A 7\arcmin$\times$7\arcmin\ red dashed box, oriented in equatorial coordinates, is drawn toward the 70\% emission contour in the \cDOo\ map. This area is considered as the central region of this cloud.}
	\label{fig:map_co}
\end{figure}

\begin{figure}[h]
	\begin{center}
		\includegraphics[width=0.5\textwidth]{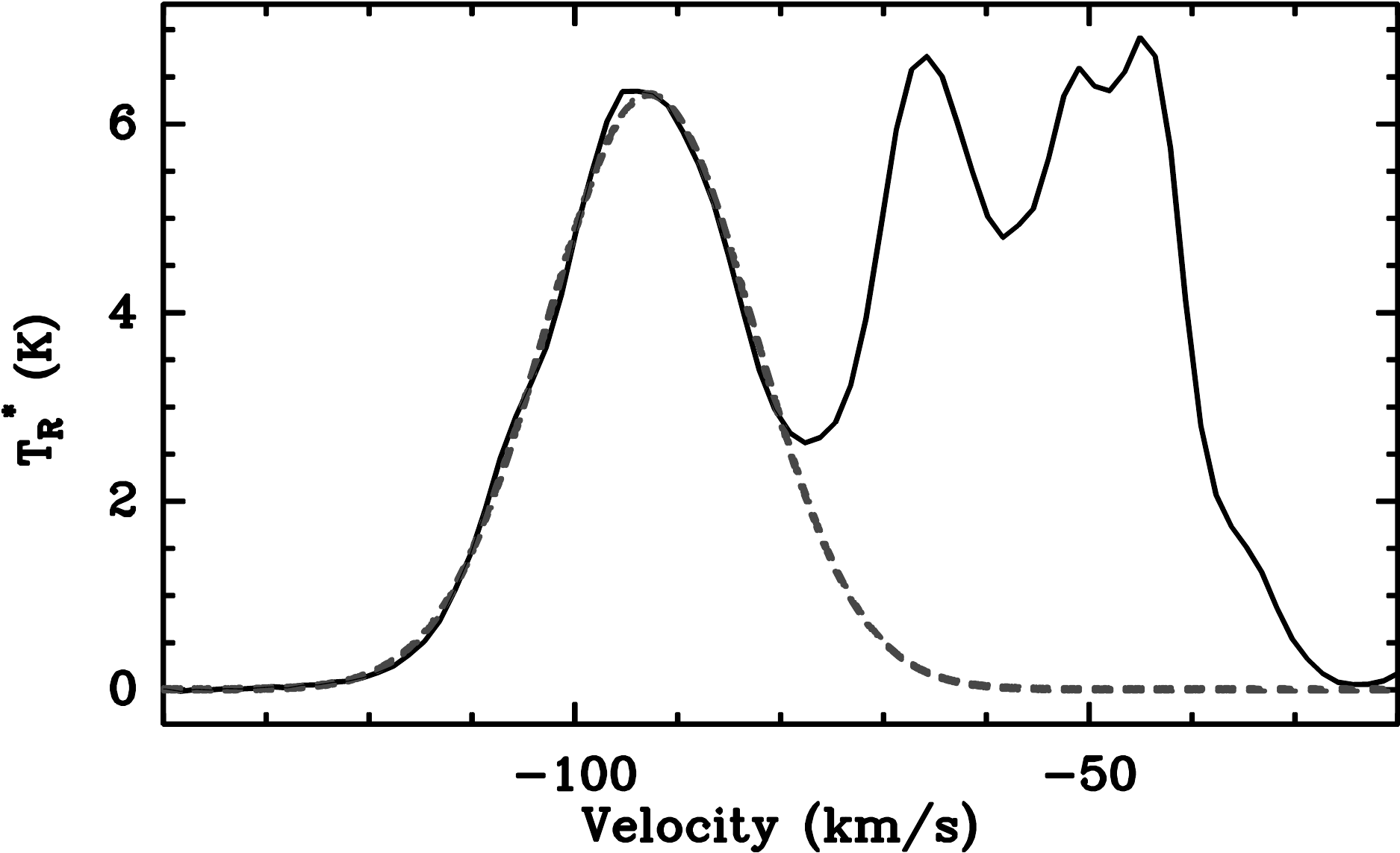}
	\end{center}
	\caption[Average profile of the CO emission over the whole GMC.]{Average profile of the \couc\ emission over the whole GMC. The dashed line corresponds to the gaussian fit of the spectrum on the velocity range of the \gtres\ GMC. The parameters of the gaussian fit are $v_0=-92.8\pm 0.006$ \kms, $\Delta v=23.8\pm0.017$ \kms\ and $T_{peak}=6.31$ K.}
	\label{fig:peaks_co}
\end{figure}

Figure \ref{fig:map_co} shows grey scale maps and contours of the velocity integrated CO and C$^{18}$O emission between 330.9\arcdeg\ $<$ \textit{l} $<$ 332.3\arcdeg\ and -0.42\arcdeg\ $<$ \textit{b} $<$ 0.14\arcdeg. The range of the velocity integration is from -117.7 to -73.2 \kms, chosen from the average spectrum of the region (shown in Fig. \ref{fig:peaks_co}). The contours represent 20\% to 90\%, in steps of 10\%, of the peak intensity of each map (297.6 K \kms\ for CO, 19.6 K \kms\ for C$^{18}$O). The rms noise of the CO and C$^{18}$O maps are 19.8 K \kms\ and 0.98 K \kms, respectively. The emission at 50\% of the peak is shown with a thicker contour in each map. The CO emission arises from an extended cloud elongated along the galactic longitude, with major and minor axes of 81.7\arcmin\ and 18.9\arcmin\ (full width at half-power), which imply linear sizes of 178 by 41 pc at the assumed distance of 7.5 kpc. The peak position on the CO map is located at $l=331.5\arcdeg ,\ b =-0.125\arcdeg$, with an intensity of 297.6 K \kms.

To estimate the kinematic distance of the \gtres\ cloud, we locate it in the subcentral point at the source longitude, for the following reasons. Using the Galactic rotation curve derived by \cite{alvarez90}, with Ro = 8.5 kpc and Vo = 220 \kms, and a mean average velocity of -92.8 \kms\ for the integrated GMC spectrum, the near kinematic distance would be 5.7 kpc and the far kinematic distance 9.2 kpc. While the near distance has been preferred by \cite{Goss72}, the far distance has been considered by \cite{Kerr70}.
However, the GMC has two velocity components, at -100 \kms\ and -90 \kms\ (see section~\ref{section:mle}), the lowest one almost reaching the terminal velocity due to pure rotation of -110 \kms.  \cite{Caswell87} reported hydrogen recombination line absorption at $l=331.515\arcdeg, b=-0.069$, with a $V_{LSR}$ of -89 \kms, as well as H$_2$CO absorption lines at -99.8 and -89.3 \kms, favoring the far distance, even when they adopted the near distance following \cite{Goss72}.  Nevertheless, \cite{Bronfman96} reported emission in \csdu\ from the IRAS point source 16086-5119,  at $l=331.552, b=-0.115$, with a $V_{LSR}$ of -100.7, corresponding to the higher velocity component, so the continuum source for the H$_2$CO absorption lines at both velocities is most probably the IRAS point source at -100 \kms, setting them both at the near distance.

Given the large spread in velocity of the integrated spectrum; the presence of two velocity components for the GMC, the lowest one almost reaching the terminal velocity; and the conflicting evidence for locating the GMC at the far or near distance, we will consider in the estimation of its physical parameters that the \gtres\ cloud is located in the tangent of the Norma spiral arm, at a distance of $\sim$7.5 kpc, corresponding to the subcentral point. With this assumption, the distance of the GMC will have an estimated uncertainty of $\sim$ 30$\%$, given by the near and far kinematic positions, leading to an uncertainty in the estimation of masses and luminosities of a factor of $\sim$2. A recent study of kinematic distances using HI absorption done by \cite{Jones12} puts the region G331.552-0.078, located toward the peak position of our CO emission maps, at the Norma spiral arm tangent point distance, at 7.47 kpc, a similar value to the one used in the present work.

The emission in the \cDOo\ line has a position angle of approximately $\sim$30\arcdeg\ with respect to the galactic longitude, with observed major and minor axes of 18.2\arcmin\ and 11.6\arcmin, respectively (39.7 $\times$ 25.3 pc at a distance of 7.5 kpc). The size of the C$^{18}$O structure is four times smaller than the CO cloud. It is necessary to keep in mind that the appearance and dimension of a molecular cloud depend strongly on the tracer used to observe it \citep{Myers95}. The CO observations most likely trace all the gas within the GMC, including low and high column density gas, while C$^{18}$O is tracing only the high column density gas.
The C$^{18}$O peak position coincides with the peak emission in the \couc\ line. Fitting gaussian profiles to the spectra observed at $l$=331.5\arcdeg, $b$=-0.125\arcdeg, we determine $V_{lsr}=-89.7$ \kms\ and $\Delta V$ (FWHM) = 3.8 \kms\ for CO, and $V_{lsr}=-89$ \kms\ and $\Delta V$ (FWHM) = 2.9 \kms\ for the C$^{18}$O emission.

\begin{figure}[p]
	\begin{center}
		\includegraphics[width=0.5\textwidth]{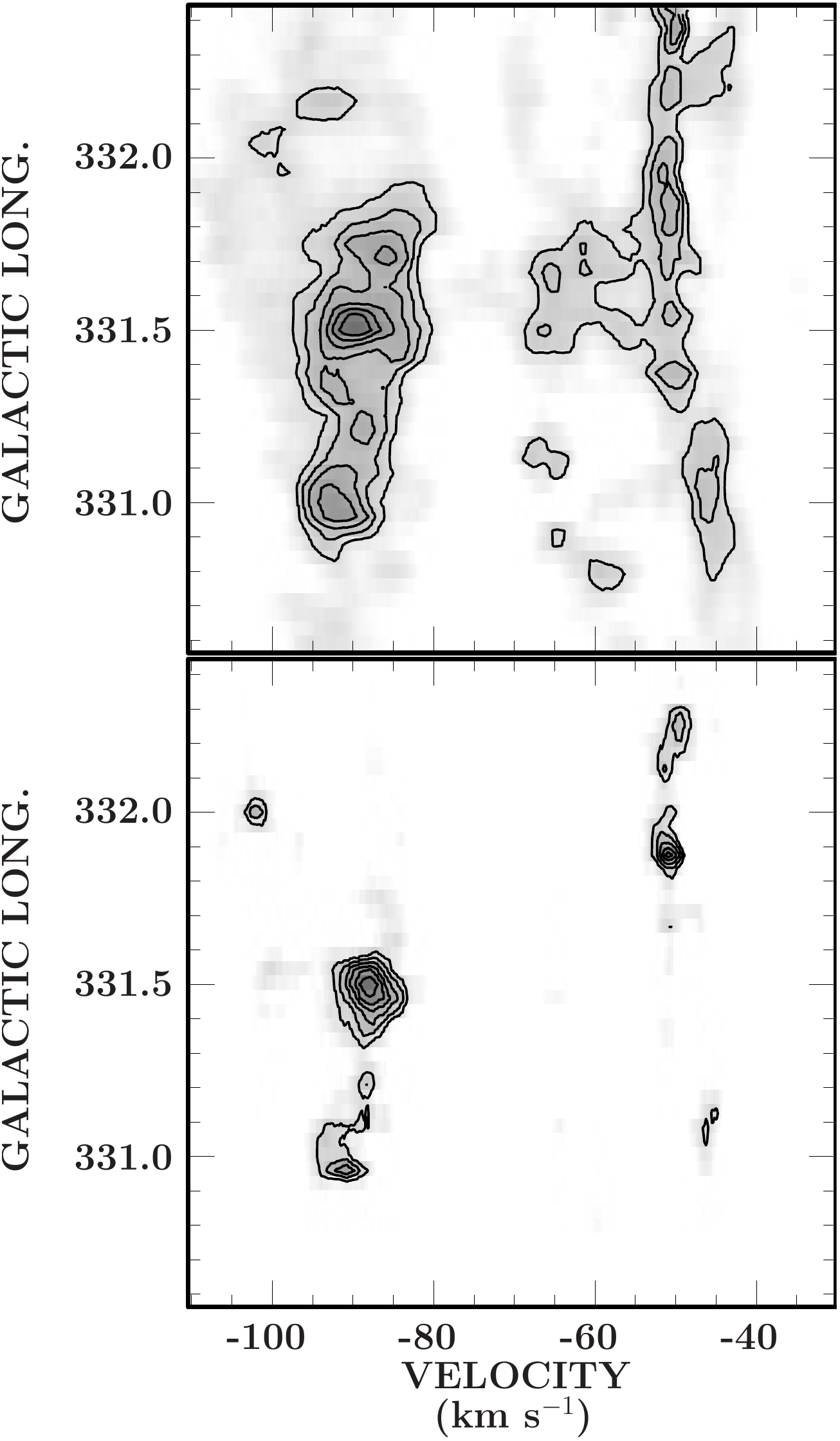}
	\end{center}
	\caption[PV map]{Position-velocity maps of the \couc\ ($top$) and \cDOo\ emission ($bottom$) integrated between -0.21 and -0.04 in Galactic Latitude. Contour 
levels are drawn at 40\% to 90\%, in steps of 10\%, of the peak intensity 
(11.5 K \kms\ for CO and 0.78 K \kms\ for C$^{18}$O).  The C$^{18}$O shows a clear 
component between  $l=331.3\arcdeg$ and $l=331.6\arcdeg$ at $\sim-90$ \kms.}
	\label{fig:pvmap}
\end{figure}

Using the \cDOo\ integrated map, we define the central region of the \gtres\ GMC as the region that emits more than seventy percent of the peak integrated emission (red dashed box in Figure~\ref{fig:map_co}). This area is 7\arcmin\ in size ($\sim$15 pc at the source distance) and it is centered at $l$ = 331.523\arcdeg, $b$ = -0.099\arcdeg\ ($\alpha_{2000} = 16^h12^m12^s$ and $\delta_{2000} = -51\arcdeg
28\arcmin 00\arcsec$). The central region is a single structure coherent in velocity, as it can be seen on the position-velocity maps for CO and C$^{18}$O (Fig. \ref{fig:pvmap}). Further observations with better angular resolution have shown that this region has two velocity components, with a difference of $\sim$12 \kms\ (see section~\ref{section:mle}).

\begin{figure}[p]
	\begin{center}
		\includegraphics[angle=-90,width=1\textwidth]{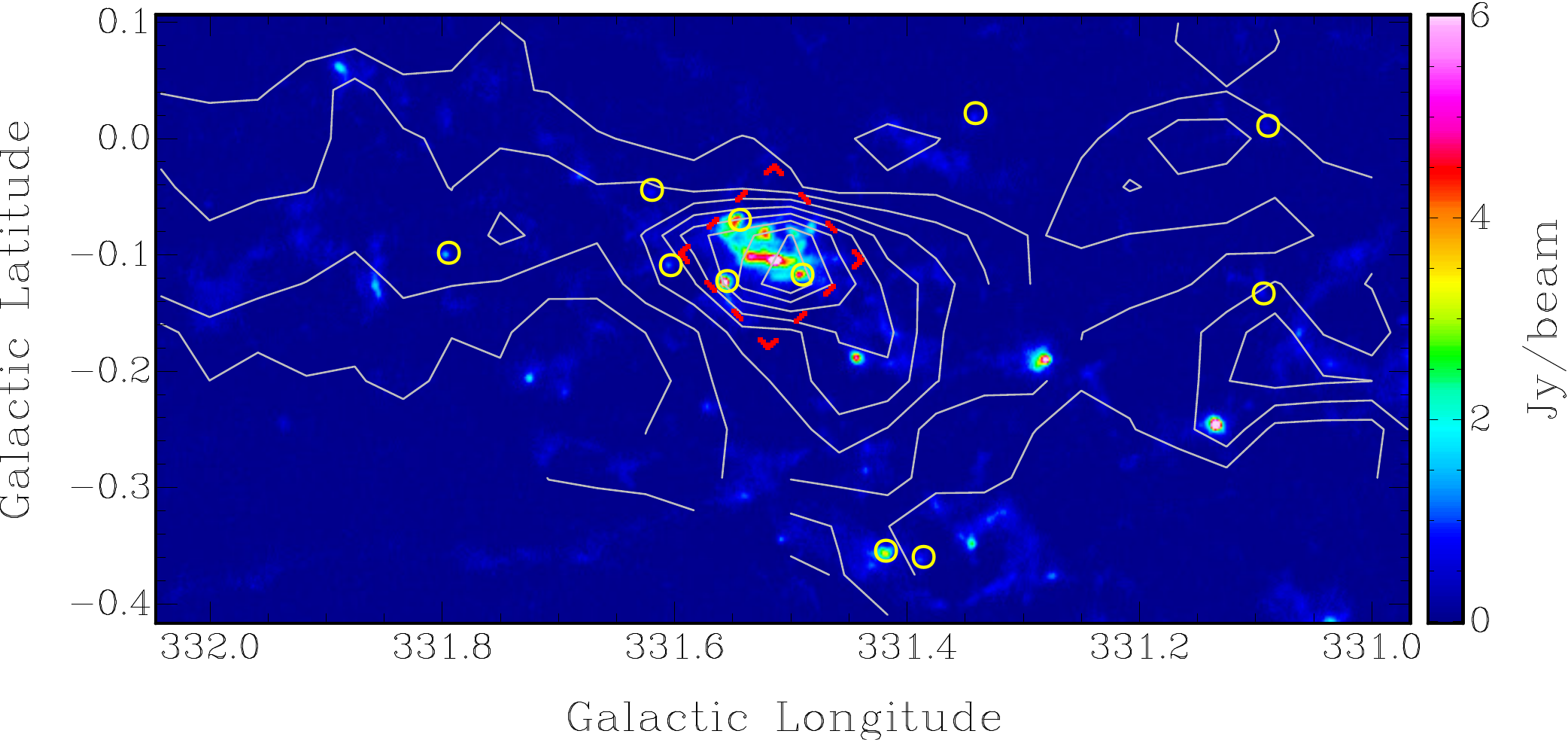}
	\end{center}
	\caption{Map of dust continuum emission at 0.87 mm from ATLASGAL towards the Norma spiral arm tangent region. Overlaid in grey is the contour map of the \cDOo\ 
emission integrated over the velocity range from -117.7 to -73.2 \kms. Contour 
levels are from 20\% to 90\%, in intervals of 10\%, of the peak intensity 19.6 K \kms. The red dashed box, 7\arcmin$\times$7\arcmin\ in size (15$\times$15 pc$^2$ at a distance of 7.5 kpc) and oriented in equatorial coordinates, shows the defined central region of the \gtres\ GMC.
The yellow circles show the position of the RMS sources in this region: G331.0890+00.0163A,
G331.0890+00.0163B,
G331.0931-00.1303,
G331.3865-00.3598,
G331.4117+00.0154,
G331.4181-00.3546,
G331.4904-00.1173,
G331.5414-00.0675,
G331.5582-00.1206,
G331.6035-00.1081,
G331.6191-00.0442A,
G331.6191-00.0442B,
and G331.7953-00.0979.
	\label{fig:atlasgal}
}
\end{figure}

Figure~\ref{fig:atlasgal} shows a map of the emission observed at 870 \um\ by ATLASGAL between 331\arcdeg\ $<$ \textit{l} $<$ 332\arcdeg\ and -0.4\arcdeg\ $<$ \textit{b} $<$ 0.15\arcdeg. Overlaid is a velocity integrated C$^{18}$O emission map. The range of velocity integration and the emission contours are the same as for Figure~\ref{fig:map_co}. The 0.87 mm map also shows the position of 13 sources (two of them are double sources) from the RMS catalog in the proximity of the tangent area of the Norma spiral arm. Clearly there is a concentration of dust continuum emission sources towards the defined central region of the cloud (red dashed box in Figure~\ref{fig:atlasgal}). Six millimeter-wave compact sources were identified inside the defined \gtres\ GMC central region, and  three of them are associated to RMS sources. This suggests an active formation of massive stars within the central region of \gtres\ GMC.

\subsection{The \gtres\ GMC central region}

In this section we present a detailed study of millimeter continuum and molecular line observation toward the 7\arcmin$\times$7\arcmin\ area defined in the previous section as the \gtres\ GMC central region. The beam sizes of the observations ($\sim$30\arcsec) are approximately five times better than the observations of \couc\ and \cDOo, and they give information in the $\sim$1 pc at a distance of 7.5 kpc. The maps of \gtres\ GMC have been presented so far in galactic coordinates, mainly to show the extension of this region in galactic longitude. Nevertheless, we present our observations here and following sections in a more traditional style using equatorial coordinates.

\subsubsection{Dust emission: Millimeter continuum observations}

\begin{figure}[h]
	\begin{center} 
		\includegraphics[width=0.5\textwidth]{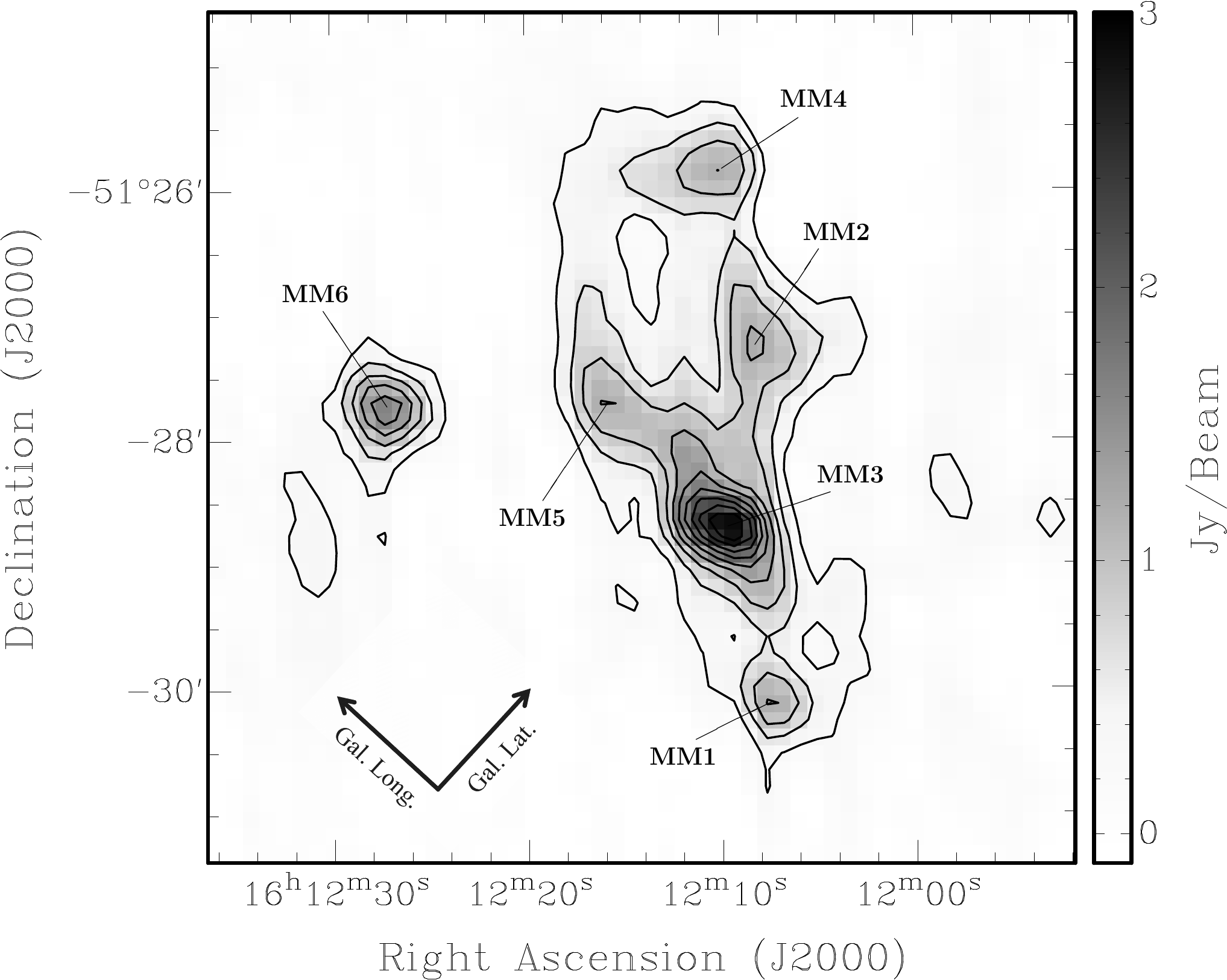}
		\includegraphics[width=0.5\textwidth]{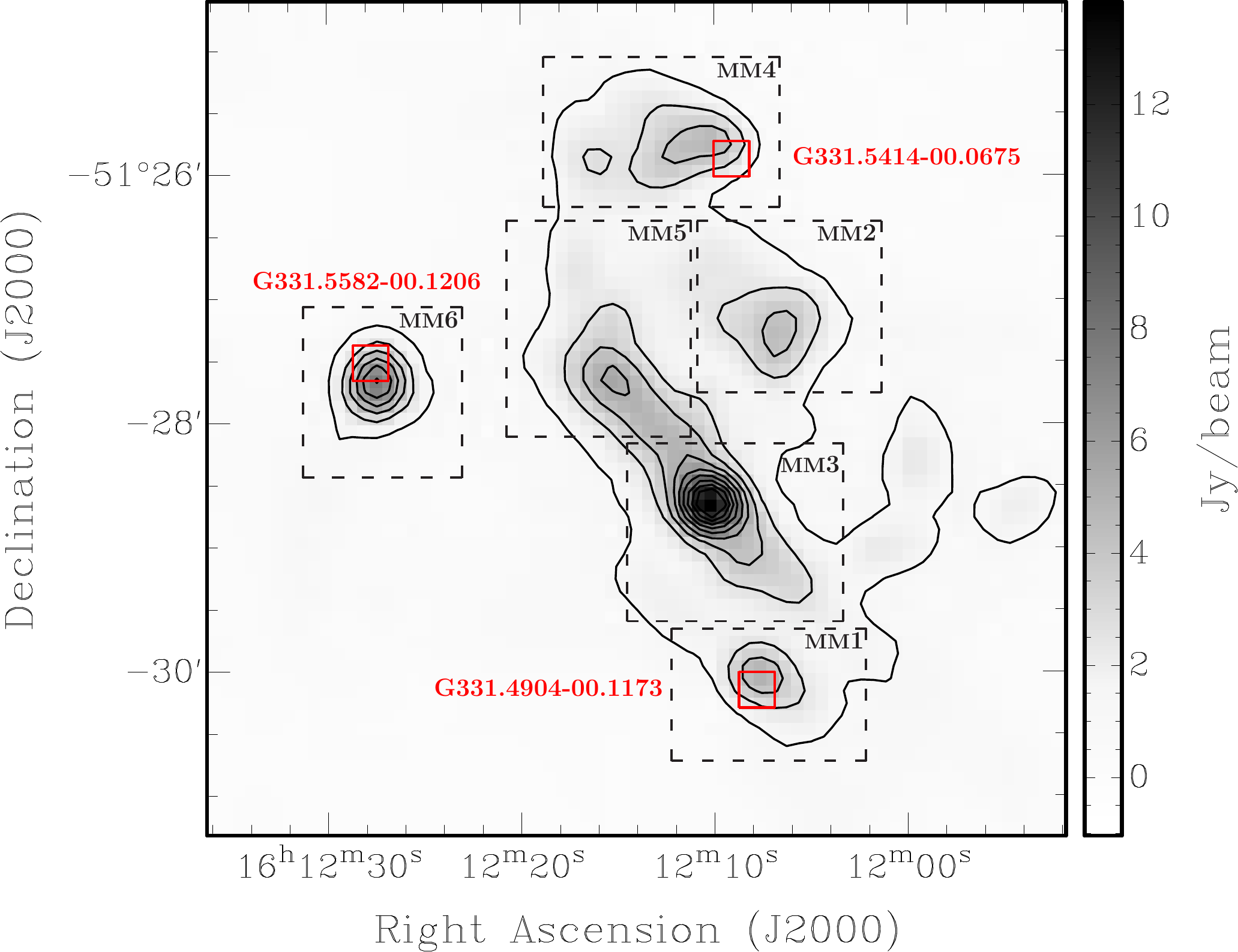}
	\end{center}
	\caption[Maps of 1.2 mm. (top) and 0.87 mm. (bottom) continuum emission 
observed with SIMBA and APEX.] { Maps of the dust continuum emission of the central region of the \gtres\ GMC.   
$Top$: Map of the 1.2 mm emission observed with SIMBA. The angular resolution is
24\arcsec.  Contour levels are from 10\% (3$\sigma$) to 90\%, in steps 
of 10\%, of the peak intensity of 2.9 Jy beam$^{-1}$. The peak position of millimeter clumps MM1 through MM6,
and the Galactic latitude and longitude directions are shown in this map. $Bottom$: Map of the 0.87 mm emission 
observed with LABOCA. The angular resolution is 18.6\arcsec. 
Contours levels are from 10\% (4$\sigma$) to 90\%, in steps 
of 10\%,  of the peak intensity of 13.8 Jy beam$^{-1}$. Dashed boxes in this map show the considered area of each millimeter clump for spectral energy distribution analysis. Red boxes mark the position of RMS sources in the region. The size of these red boxes is 18\arcsec$\times$18\arcsec\ to account for the beam size of the RMS catalog.}
\label{fig:dust}
\end{figure}

Figure~\ref{fig:dust} presents maps of the 1.2 mm and 0.87 mm dust continuum emission observed with the SEST and APEX telescopes respectively. They reveal the presence of six dust continuum substructures with strong emission. We labeled these sources as MM1, MM2, MM3, MM4, MM5 and MM6, according to increasing order in right ascension. 
The observed parameters of the millimeter sources are given in Table~\ref{tbl:dustpeaks}. Column 3 and 4 give their peak position. Column 5 and 6 give the peak flux density and the total flux density, respectively, the latter measured with the AIPS task IMEAN. Column 7 give the deconvolved major and minor FWHM angular sizes obtained from Gaussian fits to the observed spatial distribution. Column 8 gives the physical size of the sources, obtained from the geometric mean of the angular size obtained at 1.2 mm and 0.87 mm, and considering a distance to the sources of 7.5 kpc. The average size of the millimeter sources within the \gtres\ central region is 1.6 pc. Following the description of \citet{Williams2000}, we refer to these millimeter-wave structures as clumps.

\subsubsection{Molecular line emission} \label{section:mle}

The observed parameters of the line emission detected at the peak position of the millimeter clumps are given in Table~\ref{tbl:linepeaks}. Also given are the parameters of an average spectrum of each clump for the \csss\ and \tcotd\ lines, obtained from a three by three map centered at the peak position of each clump. For non-gaussian profiles, the velocity spread $\Delta V$ is estimated from $\Delta V\ =\ I(T_A^*)\, /\, T^*_{A}\, _{Peak}$. The equations for $I(T_A^*)$ and its associated error $\sigma^2_{I(T_A^*)}$ are obtained from \citep[e.g.][]{Shirley03}
\begin{eqnarray}
\nonumber
I(T_A^*)&=&\int_{v_1}^{v_2}{T_A^*\,dv} \\
\sigma^2_{I(T_A^*)}&=&\langle \sqrt{\Delta v_{line}\delta v_{chan}}\  \sigma_{T_A^*} \rangle^2_{map}+\sigma_{base}^2\ ,
\label{eqn:sigmalines}
\end{eqnarray}

\noindent with $\Delta v_{line}=v_2-v_1$ is the full velocity range of the line emission, $\delta v_{chan}$ is the velocity resolution of the spectrometer, and $\sigma_{T_A^*}$ is the uncertainty in the antenna temperature. The additional error contribution $\sigma_{base}$, from residual variation in the baseline due to linear baseline subtraction, was not considered. The error $\sigma_{I(T_A^*)}$ was propagated into the uncertainty in $\Delta V$. 

\begin{figure}[h]
	\begin{center} 
		\includegraphics[angle=0,scale=0.36]{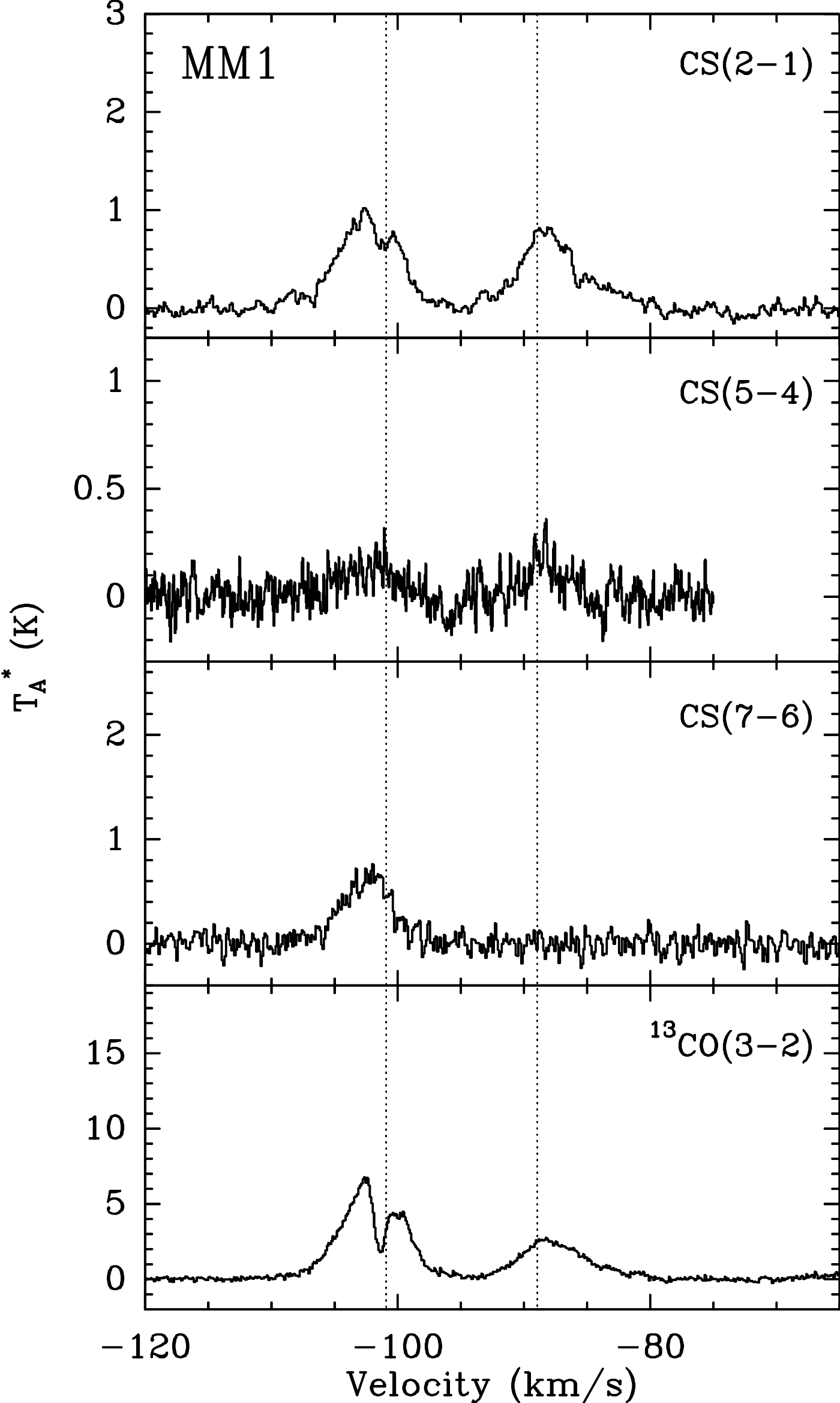}
		\includegraphics[angle=0,scale=0.36]{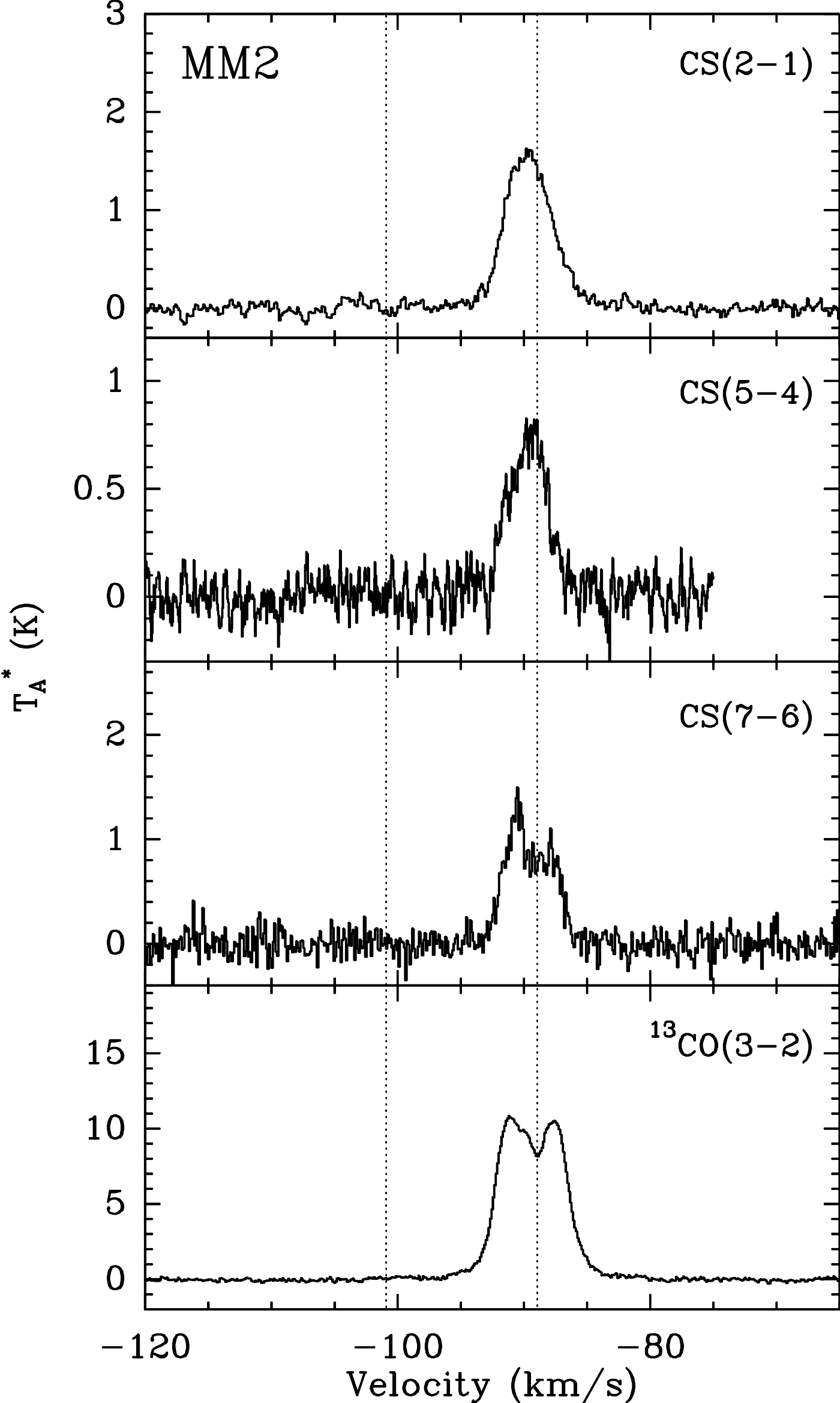}
		\includegraphics[angle=0,scale=0.36]{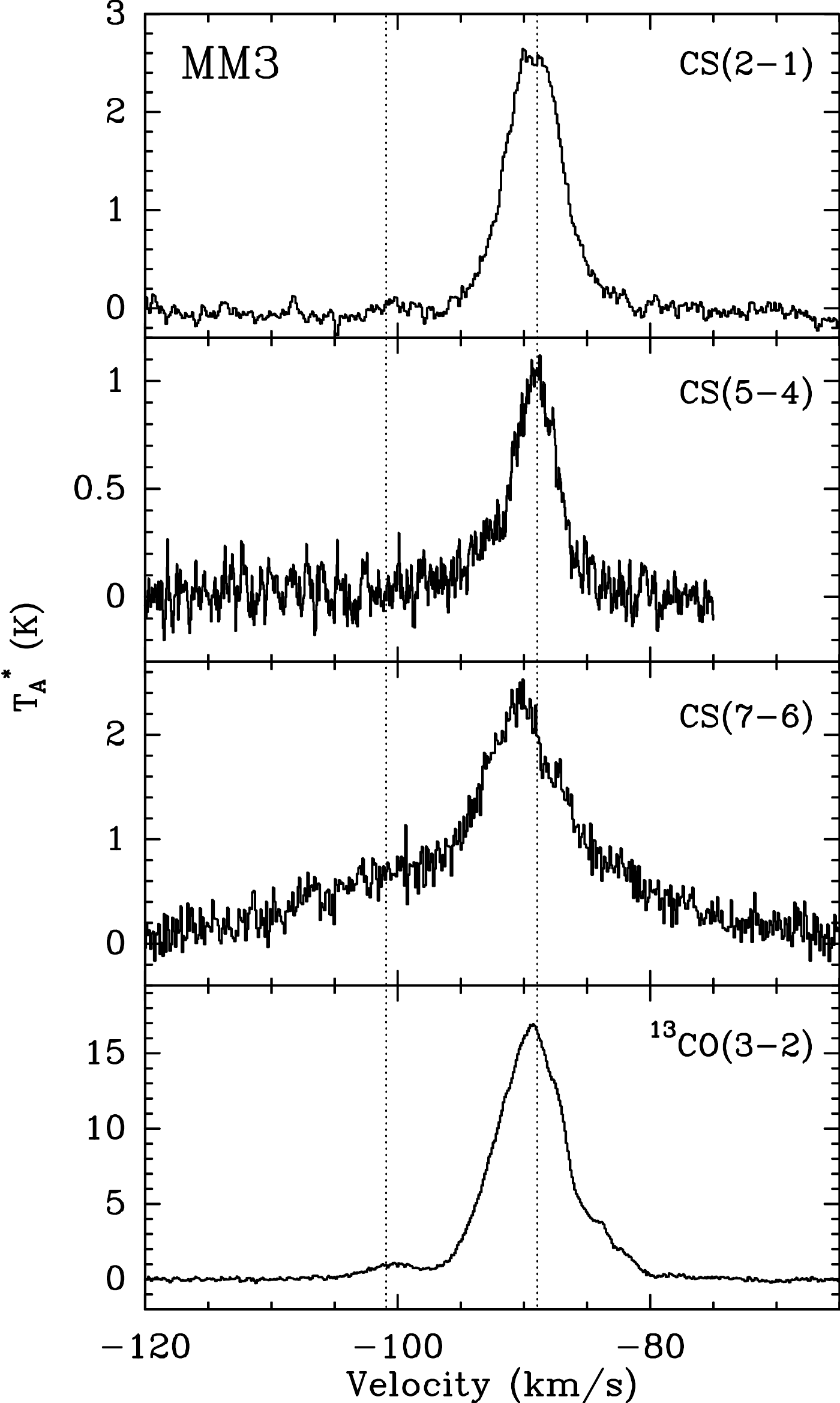}
		\includegraphics[angle=0,scale=0.36]{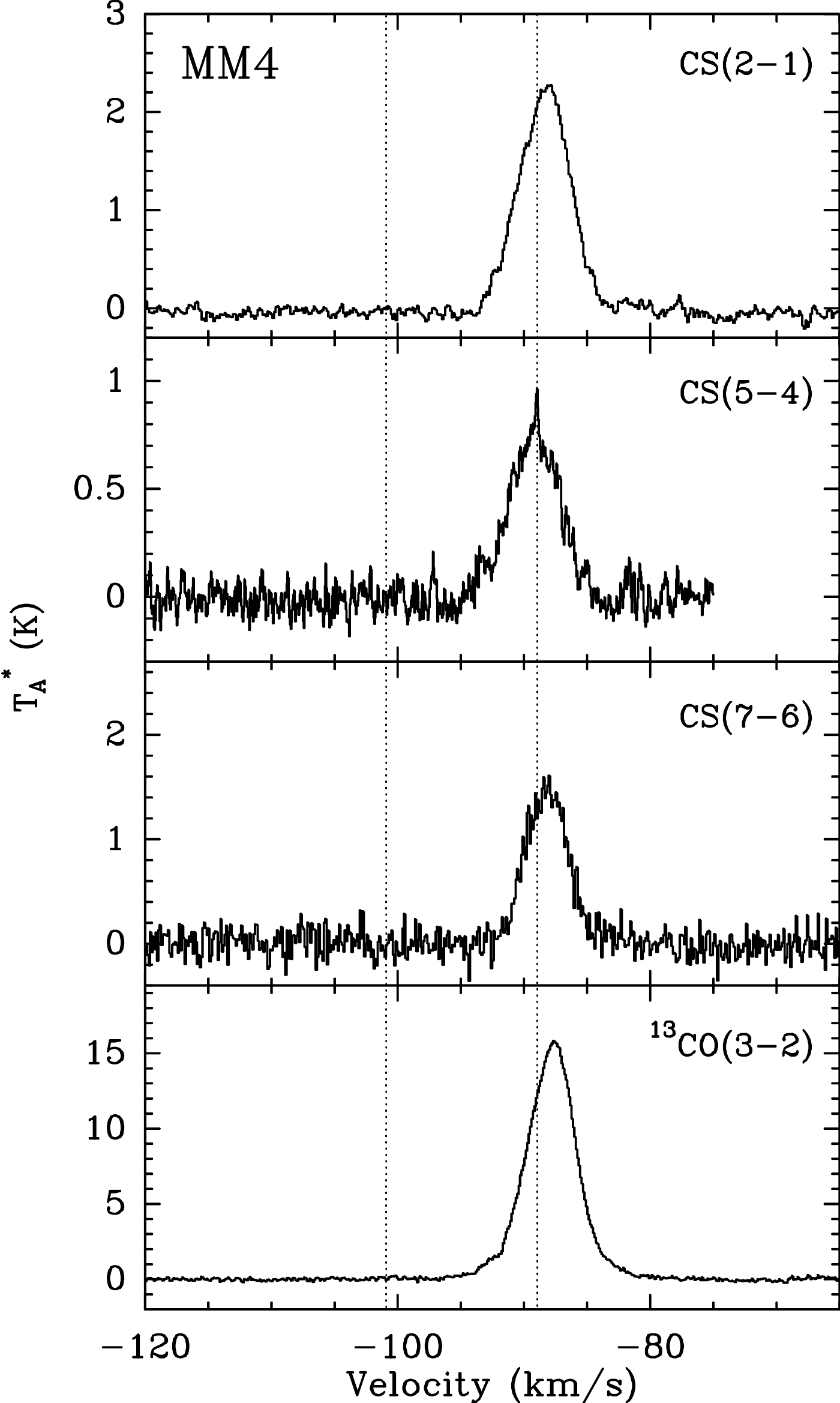}
		\includegraphics[angle=0,scale=0.36]{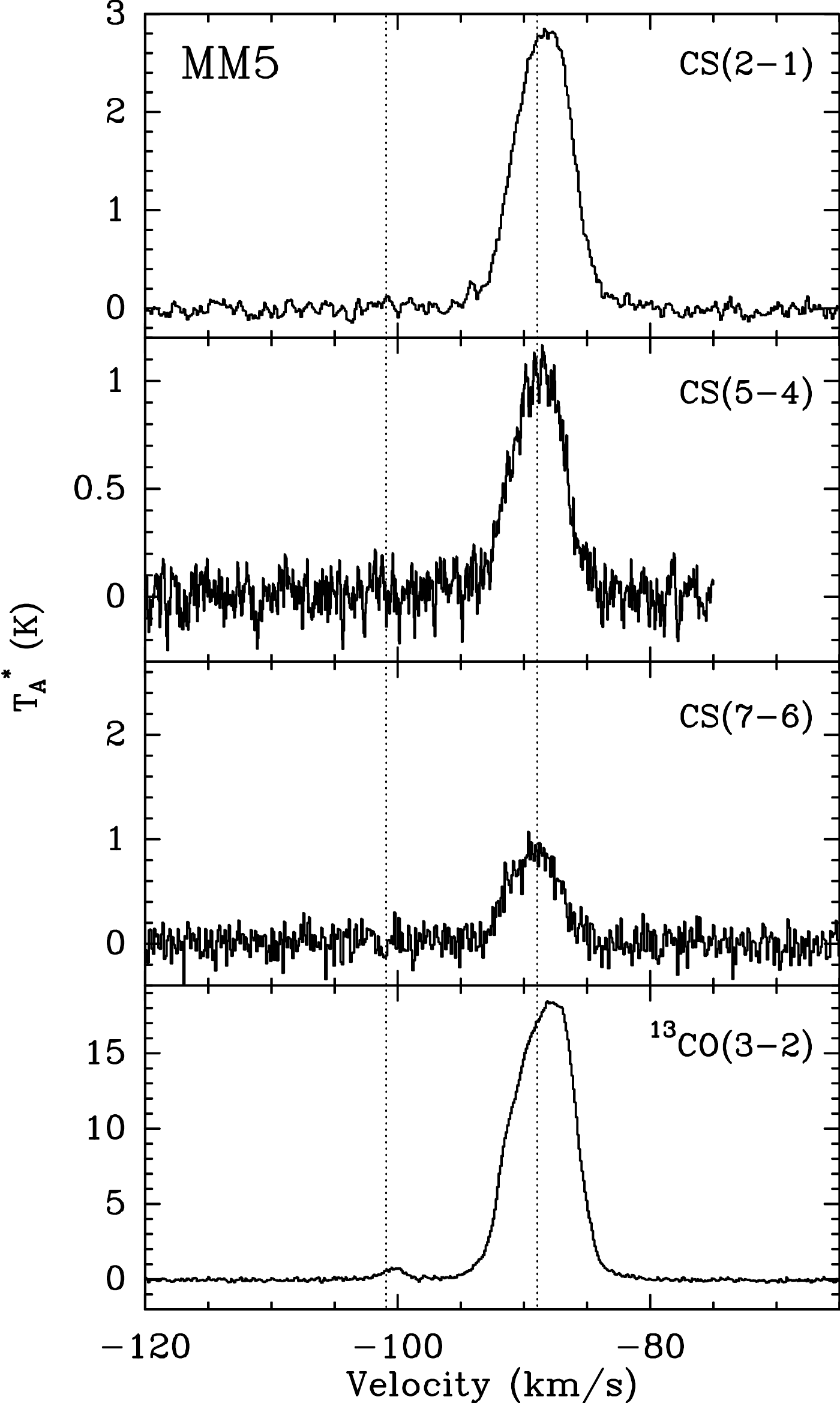}
		\includegraphics[angle=0,scale=0.36]{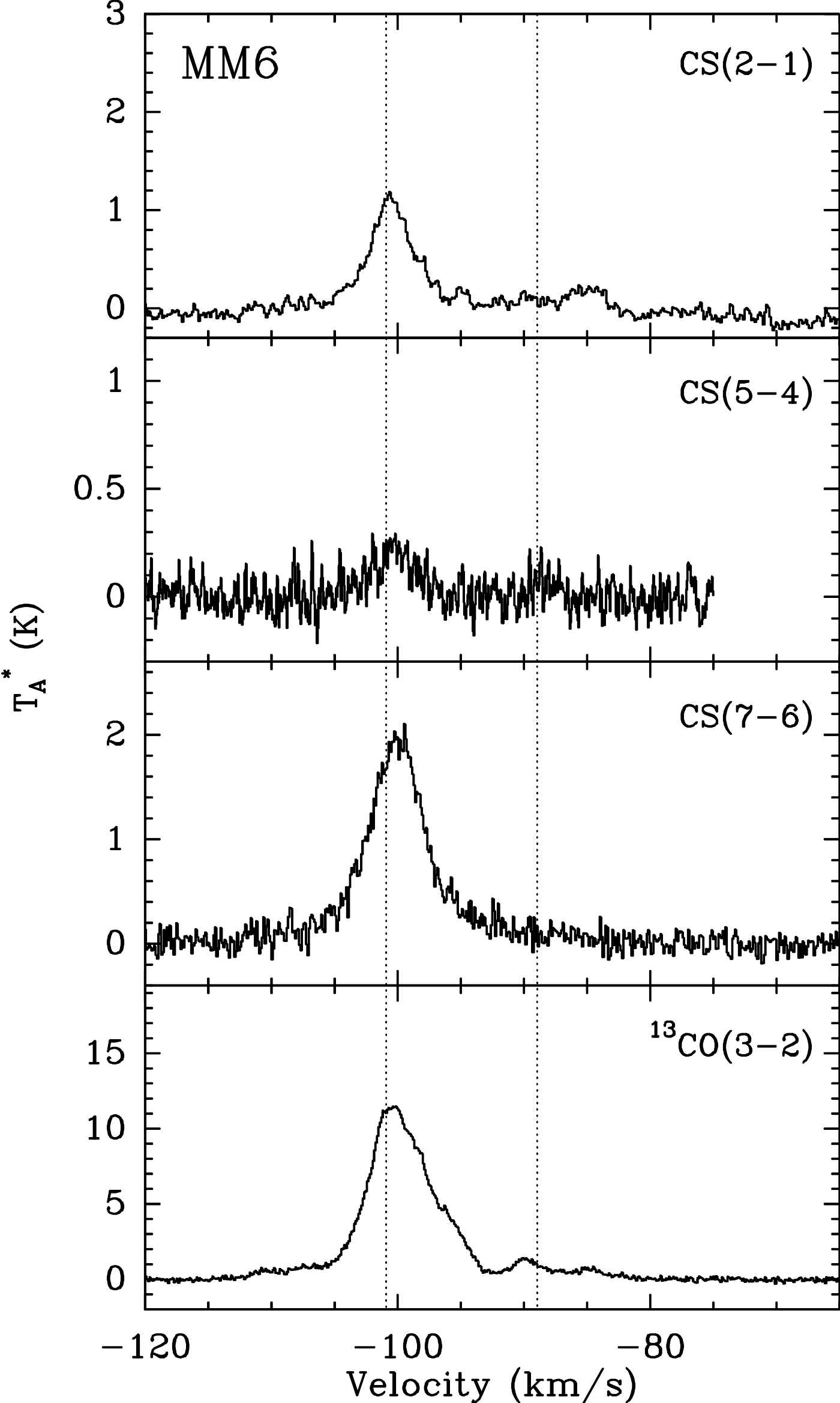}
	\end{center}
	\caption{Spectra observed at the peak position of the millimeter clumps 
MM1 through MM6. The transition is given in the upper right corner. These spectra 
indicate that the emission arises from two different components: a high velocity component at 
v=-88.9 \kms, and a low velocity component at v=-100.8 \kms, indicated with vertical dotted lines. Molecular sources associated with millimeter clumps MM2, MM3, MM4 and MM5 are part of a single structure at high velocity that we defined as the complex of clumps.}
	\label{fig:mm_spectra}
\end{figure}

Figure~\ref{fig:mm_spectra} shows the spectra toward the peak intensity position of the clumps MM1 through MM6 in the molecular lines \csdu, \cscc, \csss\ and \tcotd.

We discovered in the \csss\ map the presence of a broad and strong wing emission toward the peak position of the brightest 1.2 mm clump (MM3), at $\alpha_{2000} = 16^h12^m10.13^s$ and $\delta_{2000} = -51\arcdeg28\arcmin 37.5\arcsec$. The characteristics of this spectrum, along with other CO emission lines, were previously reported by \cite{Bronfman2008}. Their analysis indicates that it corresponds to an unresolved, energetic and massive molecular outflow (flow mass of $\sim55\ \Msun$; momentum of $\sim2.4\times10^3\ \Msun$ km s$^{-1}$; kinetic energy of $\sim 1.4\times10^{48}$ ergs).

\begin{figure}[h]
	\begin{center} 
		\includegraphics[width=0.5\textwidth]{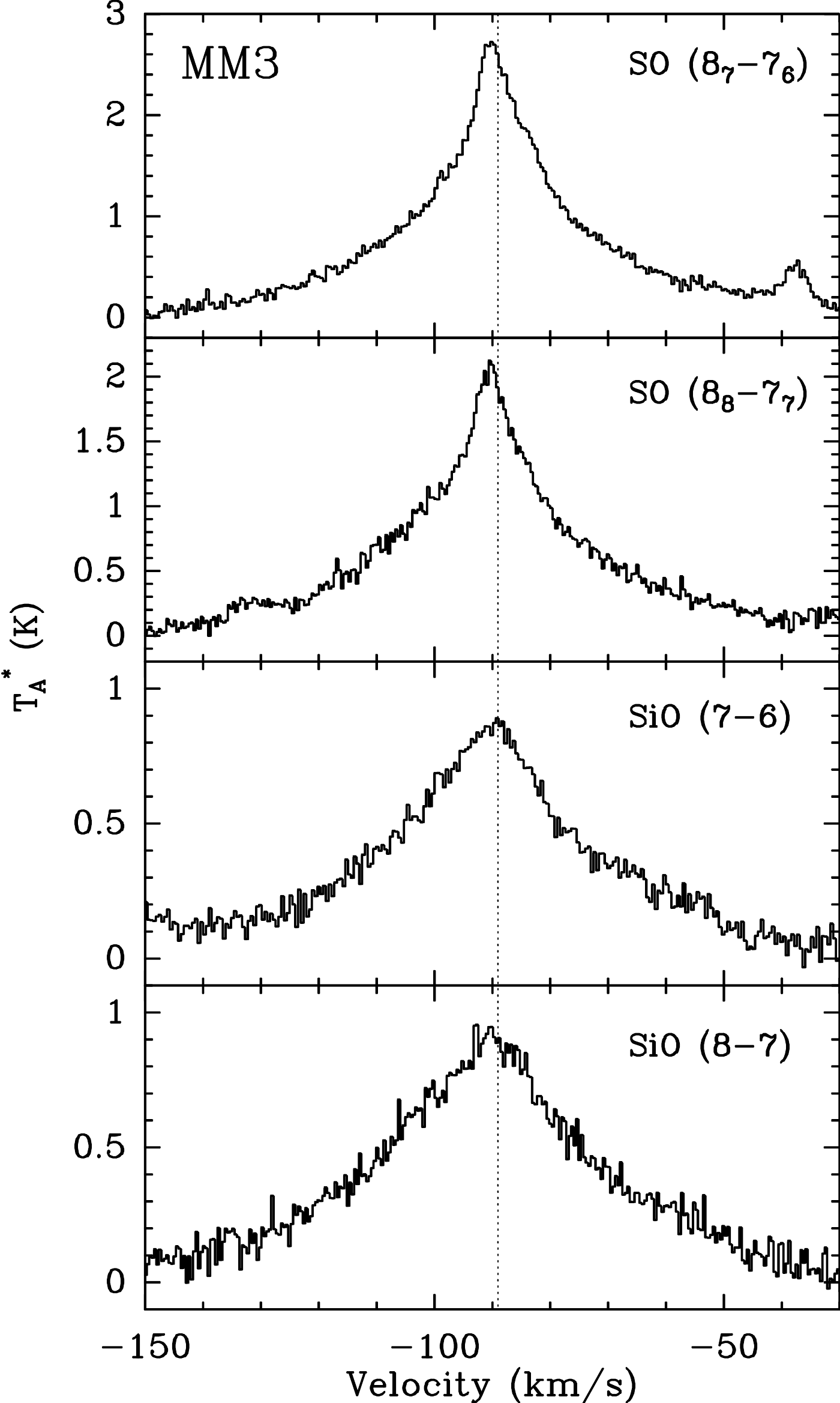}
	\end{center}
	\caption{SiO and SO spectra observed toward the peak position of clump MM3, 
which is associated with the high-velocity molecular outflow \outflow. Transitions 
are given in the upper right corner. The vertical dotted line indicates the ambient gas velocity.  }
	\label{fig:sio_so}
\end{figure}

The molecular transitions \sioss, \sioos, \soos\ and \sooo\ were observed toward the position of the molecular outflow. SO and SiO trace shocked gas towards dense cores (see 
\citealt{Gottlieb78, Rydbeck80, Swade89, Miettinen06}) where the density is high enough to excite it. High-velocity SiO and SO has been observed in some of the most powerful molecular outflows (e.g. \citealt{Welch81, Plambeck82, MPintado92}), and our current understanding is that SiO and SO are evaporated from the dust grains when the shock velocity is greater than about 20 \kms, depending on the composition of the grain-mantle in the preshock gas (e.g. \citealt{Schilke97}). Figure~\ref{fig:sio_so} shows the spectra of the SO and SiO lines.

\begin{figure}[h]
	\begin{center}
		\includegraphics[width=0.49\textwidth]{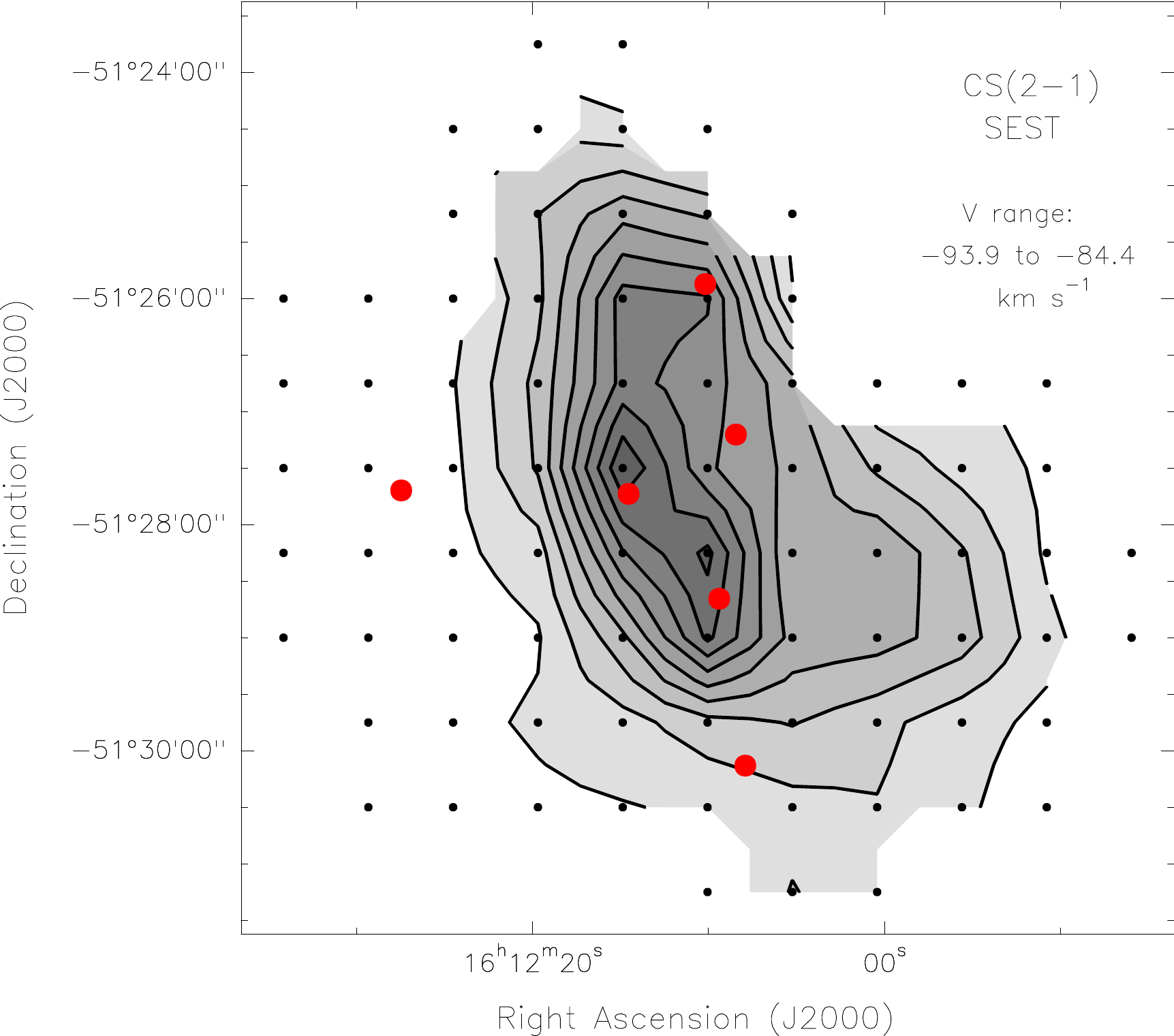}
		\includegraphics[width=0.49\textwidth]{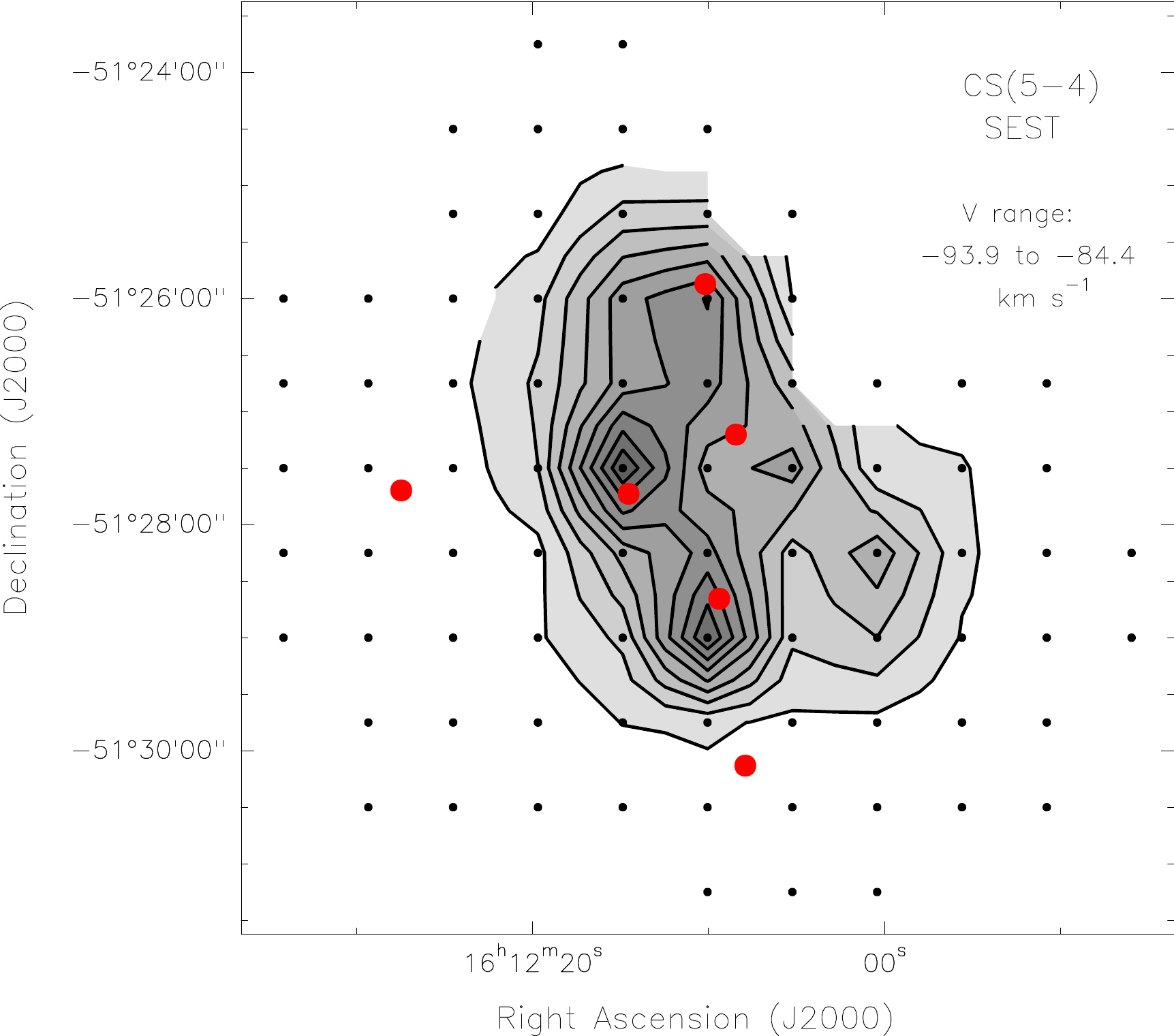}
		\includegraphics[width=0.49\textwidth]{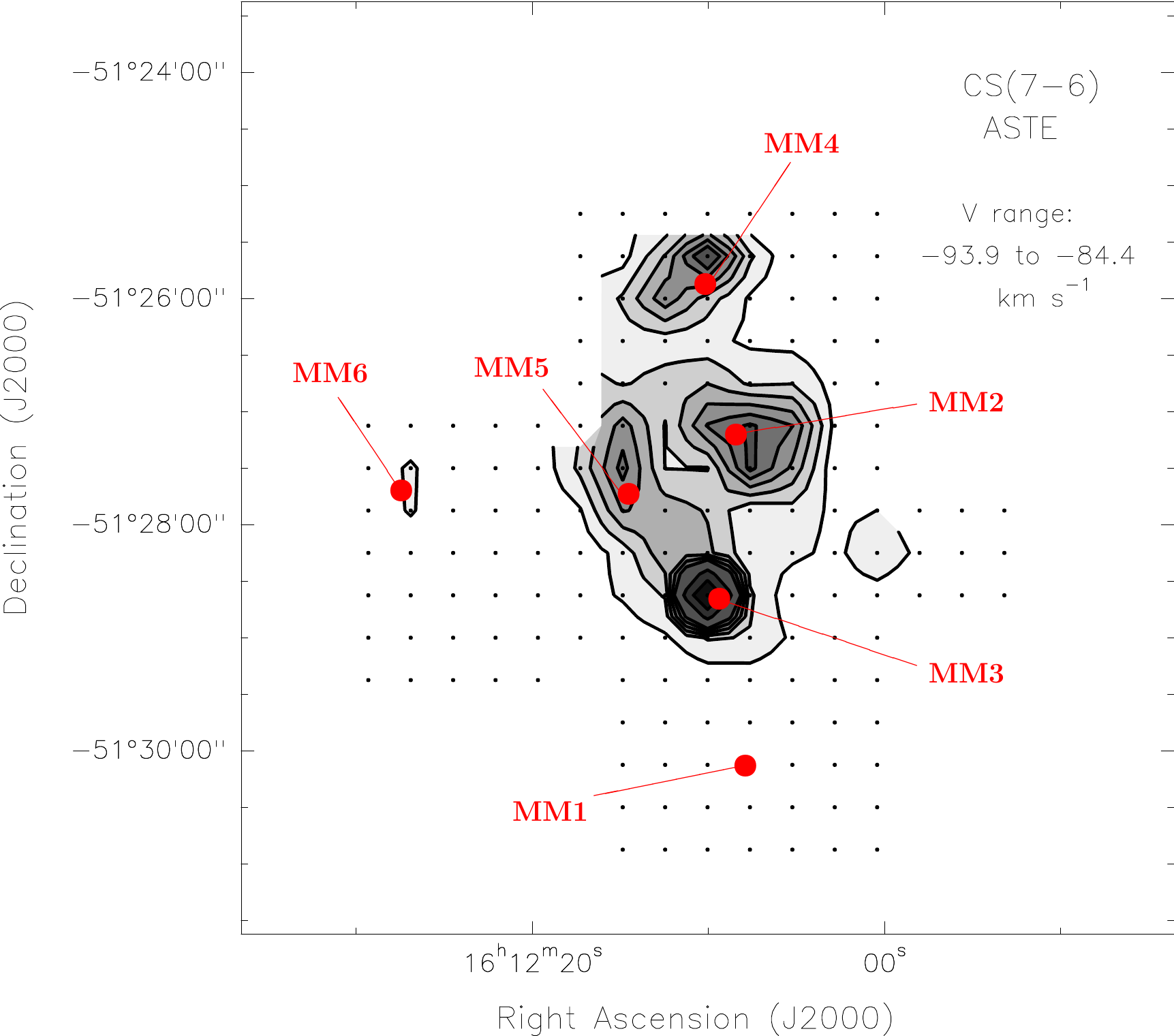}
		\includegraphics[width=0.49\textwidth]{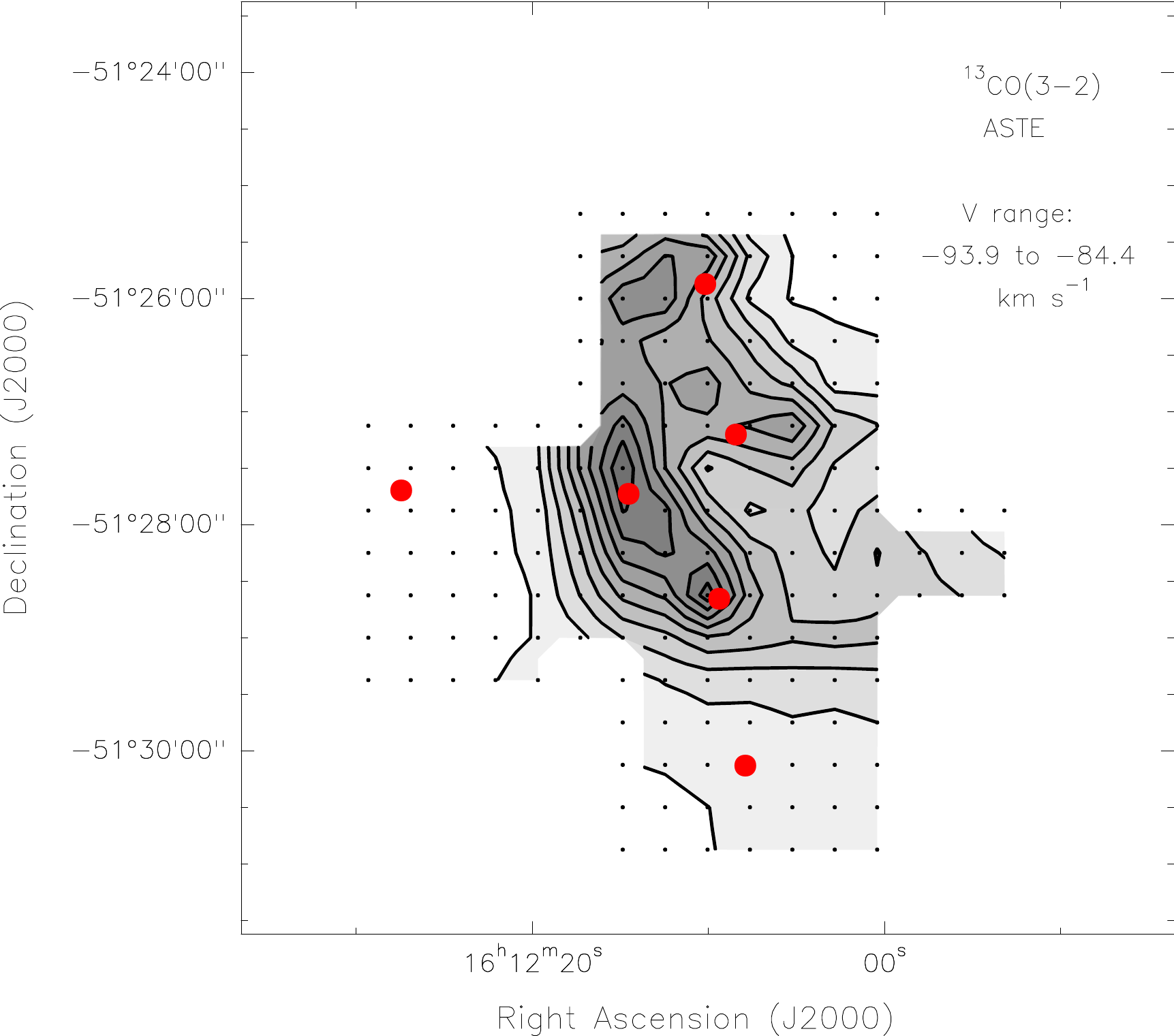}
	\end{center}
	\caption{Contour maps of velocity-integrated emission,
in the range from -93.9 to -84.4 \kms\  from the central region of the \gtres\ 
molecular cloud in four different transitions. 
For the \csss\ map, the contour levels are drawn at 5\%, 10\%, 15\%, 20\%, 25\%, 30\%, 50\% and 70\% of the peak flux density 25.6 K \kms. For \csdu, \cscc\ and \tcotd\ maps, the contours 
are drawn at 10\% to 90\%, with interval of 10\%, of the peak intensity. For \csdu, 
\cscc\ and \tcotd, the peak intensities are 15.3, 5.3 and 161.7 K \kms, respectively. The red points show the peak positions of the millimeter clumps MM1 through MM6, as is labeled in the \csss\ contour map. Molecular sources associated with millimeter clumps MM2, MM3, MM4 and MM5 are part of a single structure at high velocity that we defined as the complex of clumps.}
	\label{fig:map_mol_lowv}
\end{figure}

\begin{figure}[h]
	\begin{center}
		\includegraphics[width=0.49\textwidth]{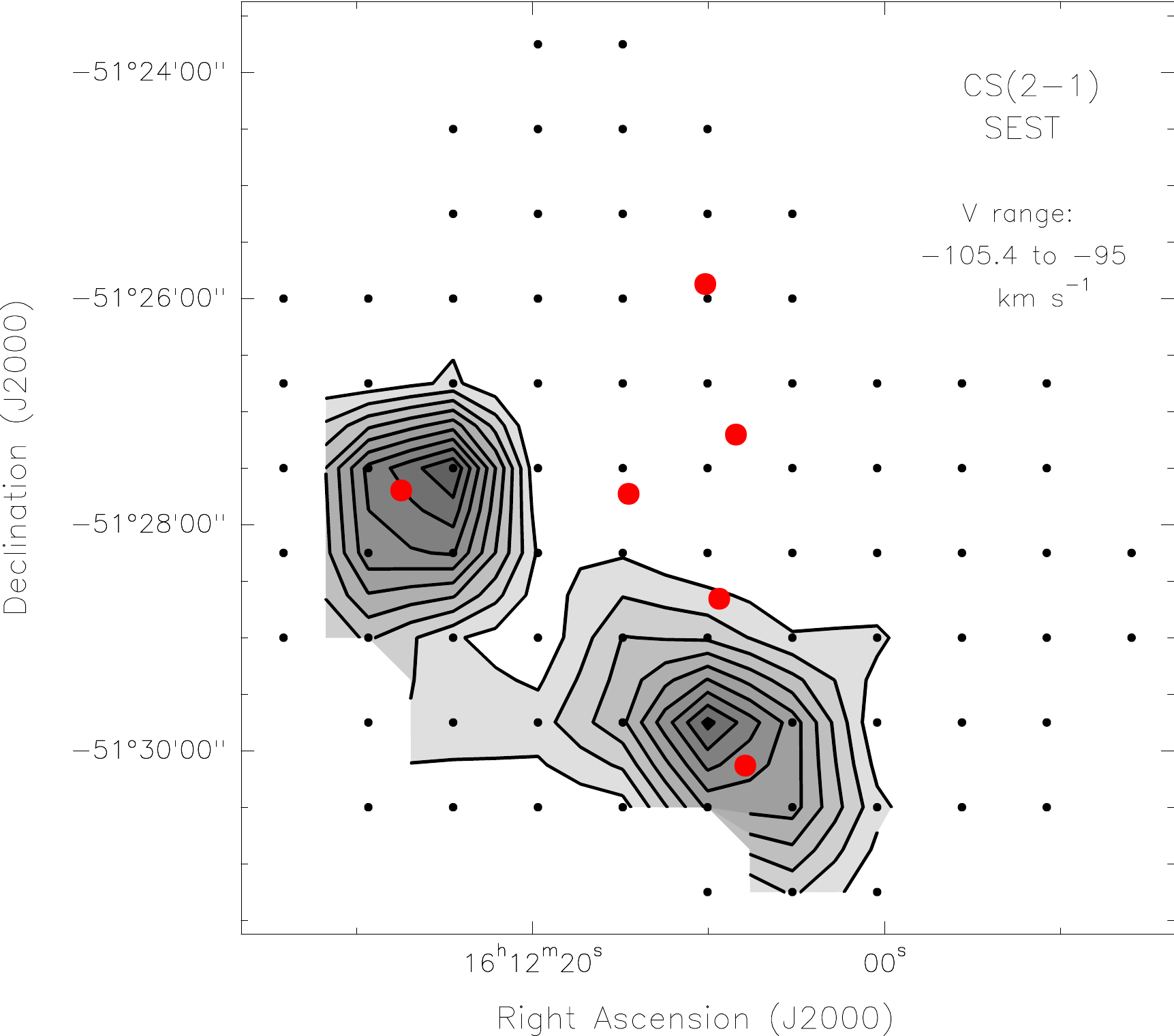}
		\includegraphics[width=0.49\textwidth]{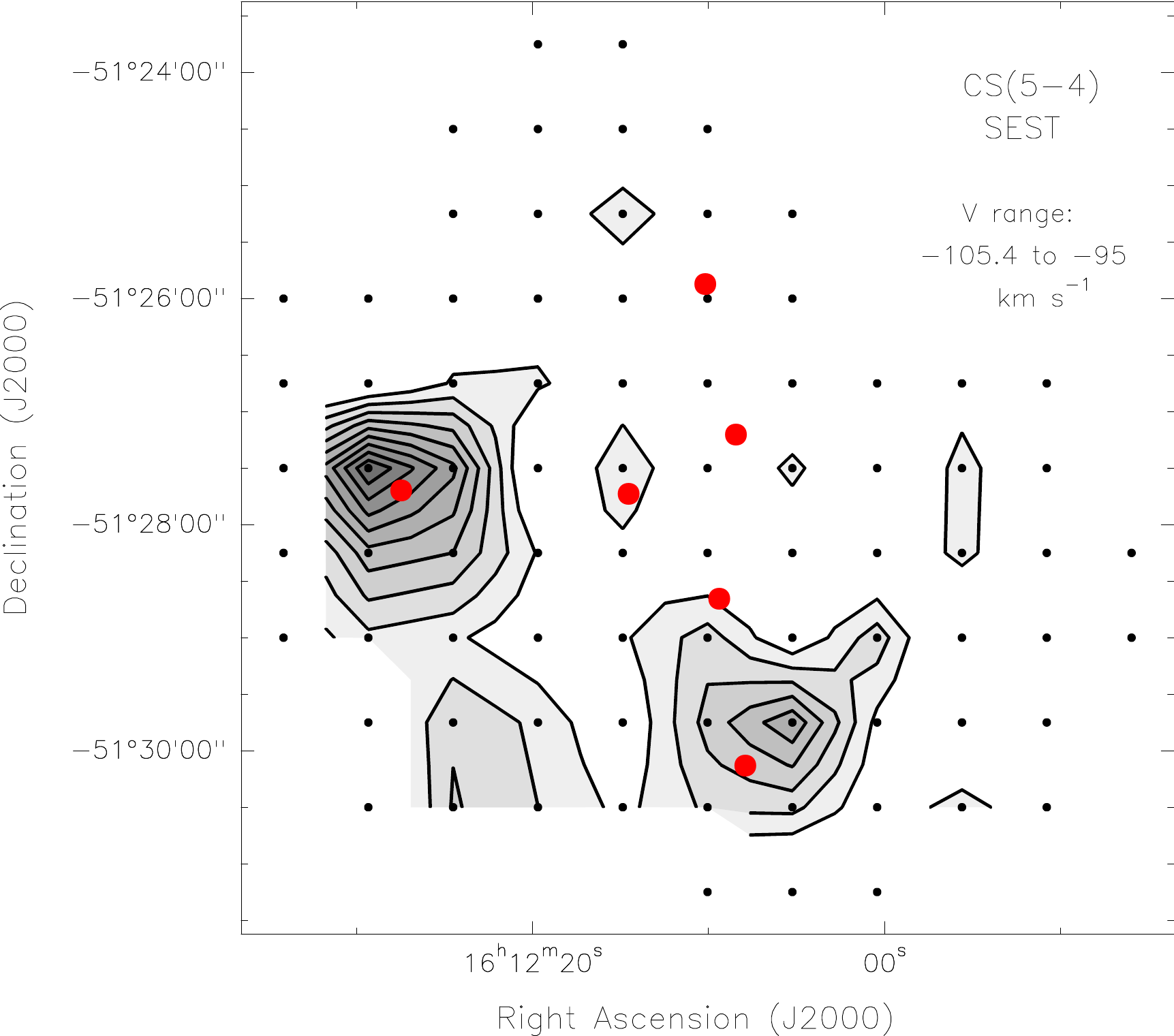}
		\includegraphics[width=0.49\textwidth]{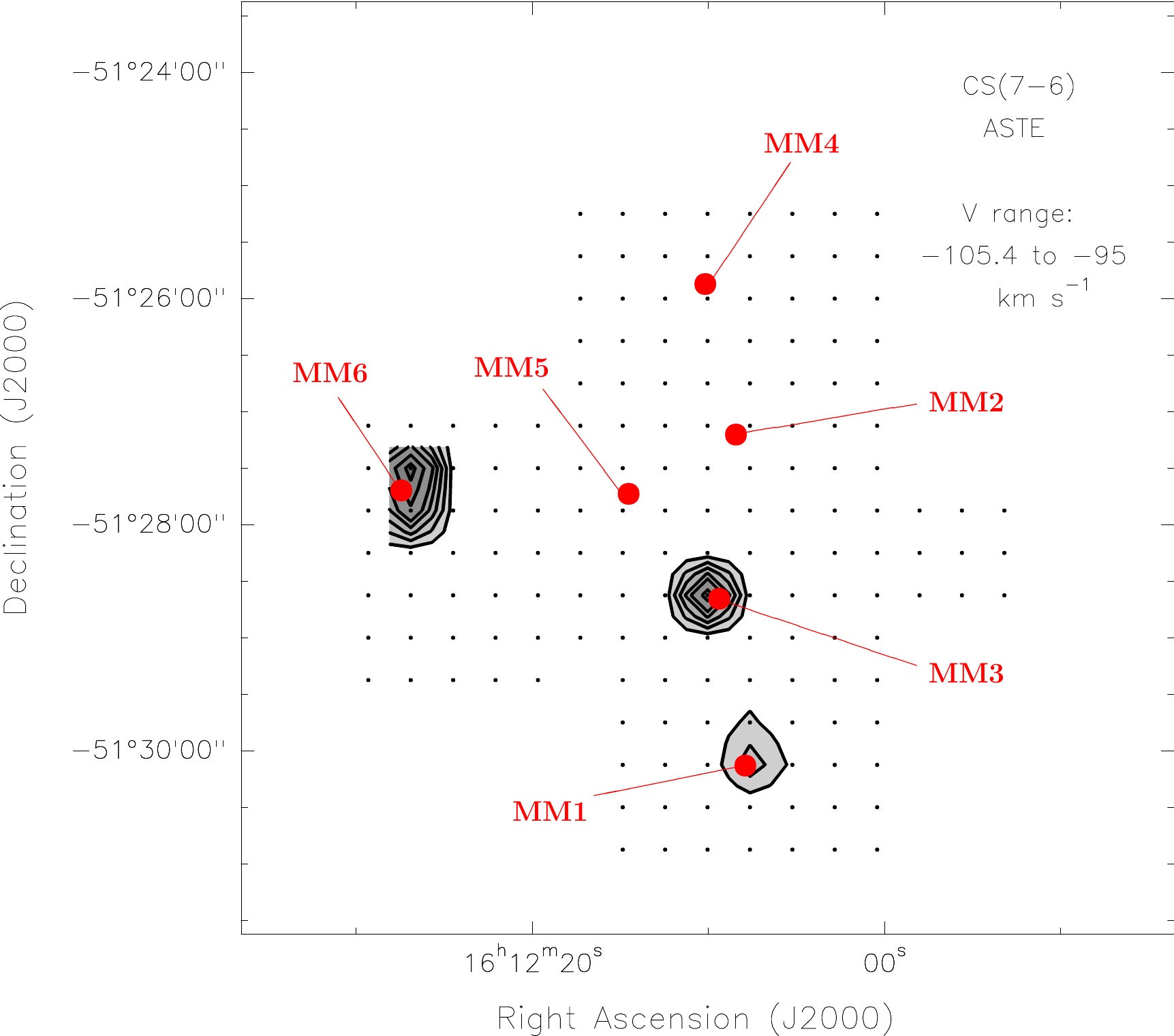}
		\includegraphics[width=0.49\textwidth]{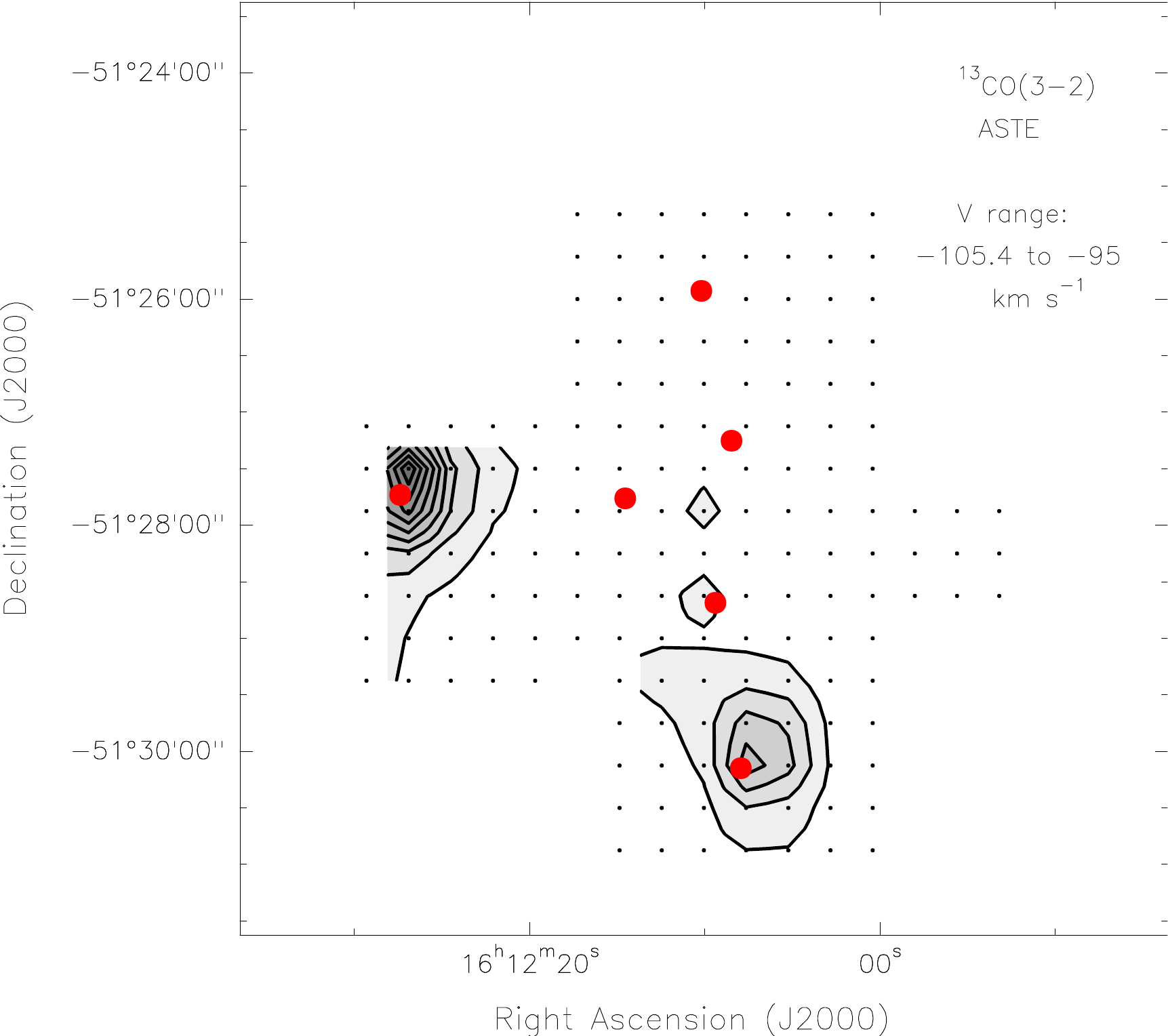}
	\end{center}
	\caption{Contour maps of velocity-integrated emission,
in the range between -105.4 and -95 \kms, from the central region of the \gtres\ 
molecular cloud in four different transitions.  For the \csss\ map, the contour 
levels are drawn at 10\%, 20\%, 30\%, 40\%, 50\%, 70\% and 90\% of the peak flux 
density 14.0 K \kms. For  \csdu, \cscc\ and \tcotd maps, the contours 
are drawn at 10\% to 90\%, with interval of 10\%, of the peak intensity. For \csdu, the peak intensity is 5.2 K \kms. For \cscc, the peak intensity is 1.4 K \kms. For \tcotd, the peak intensity is 86.7 K \kms. The red points show the peak positions of the millimeter clumps MM1 through MM6, as is labeled in the \csss\ contour map. The emission observed in the \csss\ map at $\alpha_{2000}=16^h12^m10.13^s$ and $\delta_{2000} = -51\arcdeg28\arcmin 37.5\arcsec$ (toward the MM3 point) is related with the emission integration on the blue wing of the \outflow\ outflow.}
	\label{fig:map_mol_highv}
\end{figure}

\begin{figure}[h]
	\begin{center}
		\includegraphics[width=0.5\textwidth]{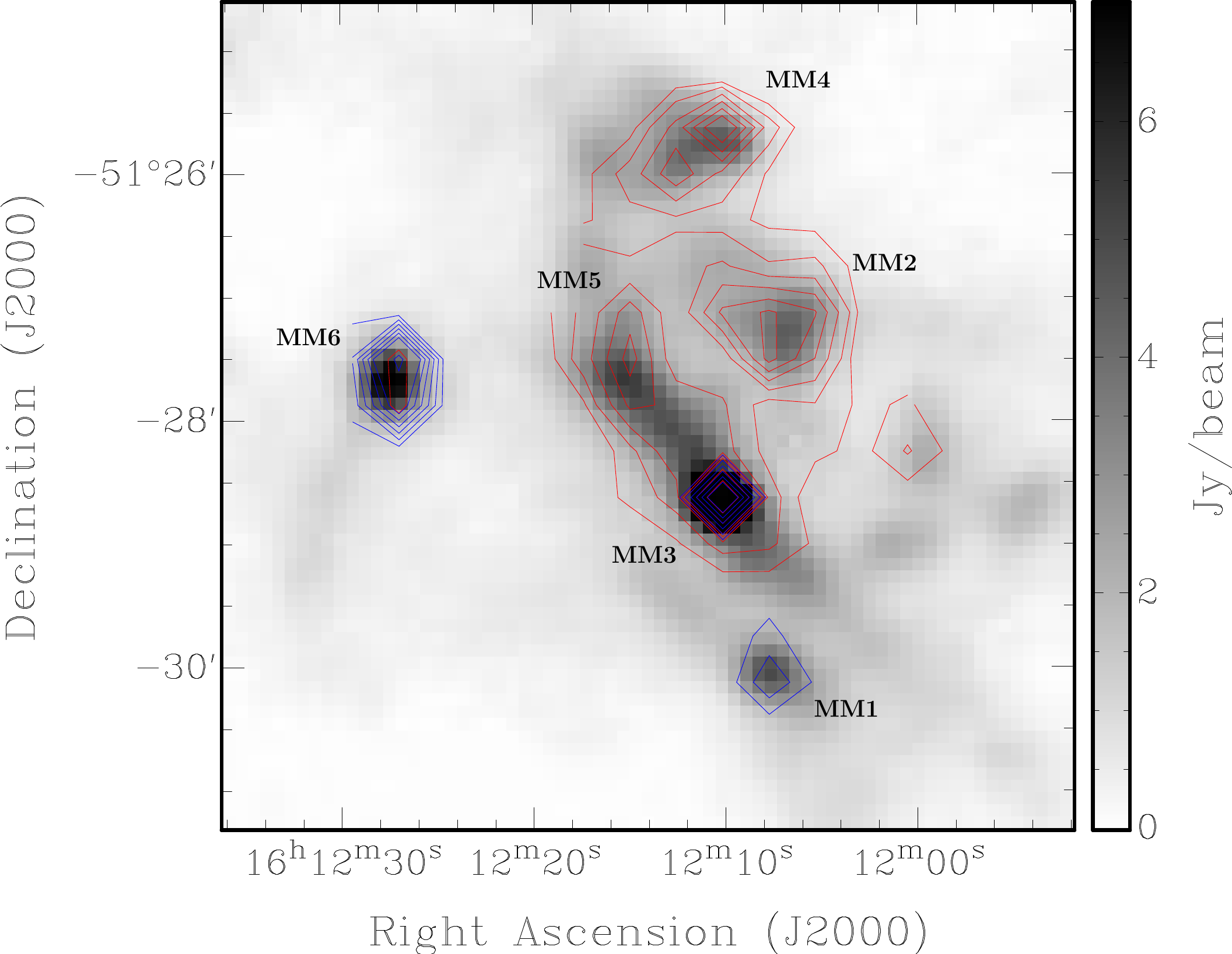}
	\end{center}
	\caption{0.87 mm greyscale image of the central region of the \gtres\ GMC, overlaid with CS velocity integrated contour map. The contour levels are drawn at 5\%, 10\%, 15\%, 20\%, 25\%, 30\%, 50\% and 70\% of the peak flux density 25.6 K \kms. The red contours show the integrated emission in the high velocity component of the spectra, integrated between -93.9 and -84.4 \kms. The blue contours show the low velocity component, integrated between -105.4 and -95 \kms. Labels MM1 through MM6 show the associated millimeter clump for each CS source. Molecular sources associated with millimeter clumps MM2, MM3, MM4 and MM5 are part of a single structure at high velocity that we defined as the complex of clumps.}
	\label{fig:laboca_cs}
\end{figure}

Line profiles of the CS and CO emission show the presence of two distinct velocity components centered at $\sim-88.9$ and $\sim-100.8$ \kms\ (see Fig.~\ref{fig:mm_spectra}). Maps of the velocity integrated emission in the ranges of -93.9 to -84.4 \kms\ and -105.4 to -95 \kms\ are presented in Figures~\ref{fig:map_mol_lowv} and \ref{fig:map_mol_highv}. Clumps MM1 and MM6 are associated with the lowest velocity component ($\sim-100.8$ \kms), while clumps MM2, MM3, MM4 and MM5 are associated with the high velocity component ($\sim-88.9$ \kms). There is a good correspondence between \csss\ map and the dust continuum emissions (see Fig.~\ref{fig:laboca_cs}), indicating that these two tracers are probes of regions with similar physical conditions.
We will refer to the region including the four high-velocity clumps as the ``{\bf complex of clumps}" in the \gtres\ GMC central region.

The angular sizes of the structures defined by the integration of the \csss\ line and associated to the clumps, and the nominal radii resulting from the deconvolution of the telescope beam size, are summarized in Table~\ref{tbl:cs_sizes}. For MM3, the \csss\ peak position and average spectra are well fitted with two gaussian components. The first component corresponds to the broad wind emission, associated with the high-velocity molecular outflow, while the second gaussian component corresponds to the ambient gas emission. For the peak position spectrum, the gaussian fit of the wing emission has $T_A^*=0.92\pm0.12$ K and $\Delta v=31.1\pm2.8$ \kms\ (FWHM), and the ambient gas gaussian fitting has $T_A^*=1.40\pm0.12$ K and $\Delta v=6.1\pm0.1$ \kms\ (FWHM).

\subsubsection{Mid-infrared and far-infrared emission}
Infrared observations are a useful tool to study the process of massive star formation. At these wavelengths, radiation is less affected by extinction than at visible wavelengths; therefore infrared observations can give us information of the nearby environment of newly formed massive stars that are surrounded and obscured by cool dust. The mid-infrared (range 5 - 25 \um) and far-infrared (range 25 - 400 \um) observations are dominated by the thermal emission from dust, that re-radiates the absorbed UV radiation from newly born OB stars and reprocessed photons from the ionized nebula. In the mid-infrared range the contribution from polycyclic aromatic hydrocarbons, from which emission is strong in photo-dissociation regions, is also significant.

\begin{figure}[h]
	\begin{center}
		\includegraphics[width=1\textwidth]{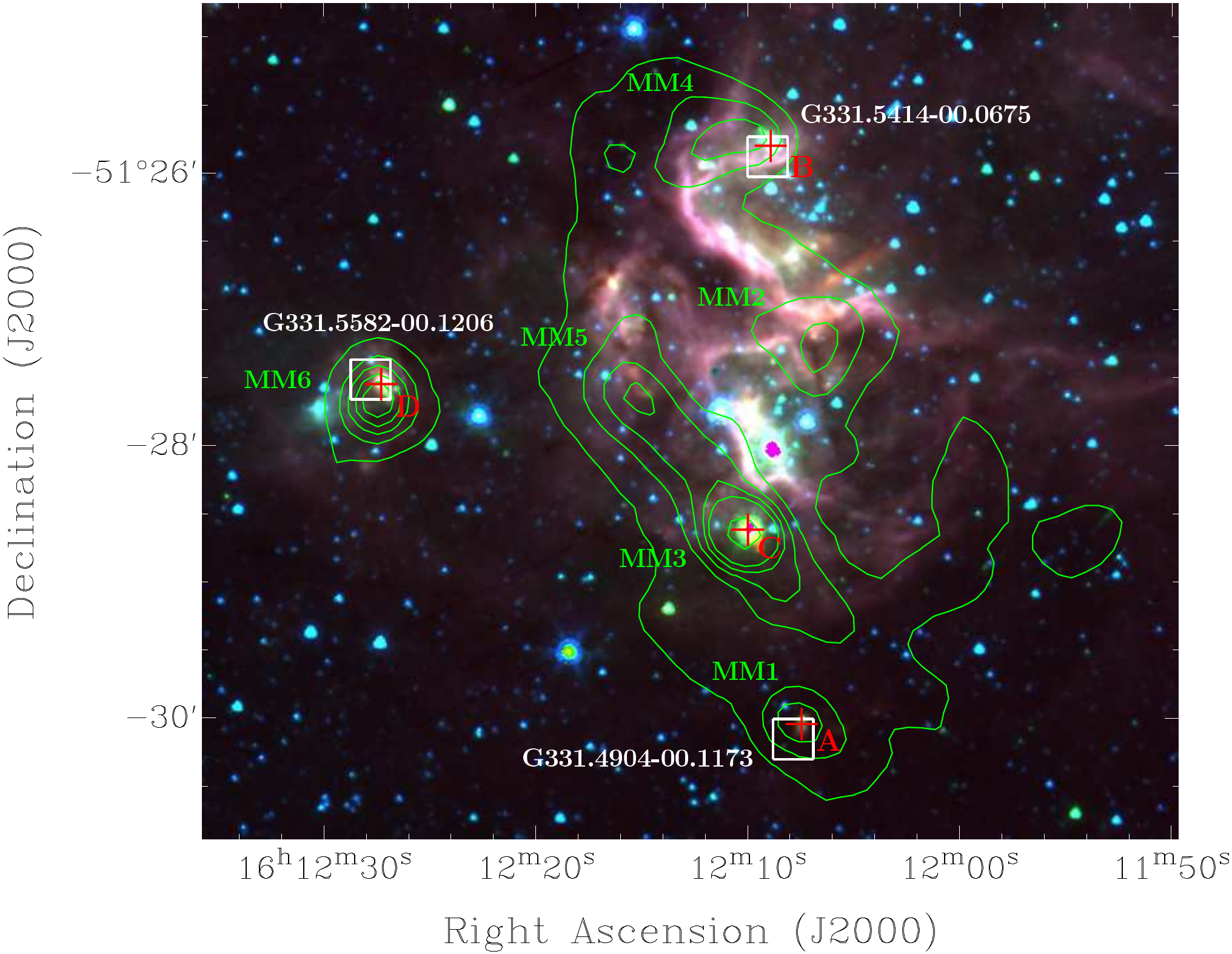}
	\end{center}
	\caption{Contour map of the 0.87 mm continuum emission from the G331.5-0.1 
region overlaid on a three color image of the mid infrared emission, made using the IRAC filters 1 (3.6 \um, blue), 2 (4.5 \um, green) and 4 (8 \um, red). Contour levels are drawn at 10\%, 20\%, 30\%, 40\%, 50\% and 90\% of the peak flux density 13.8 Jy beam$^{-1}$. Millimeter sources are labeled in the map with MM1, MM2, MM3, MM4, MM5 and MM6. Red crosses mark the positions of the four radio emission components in the central region of the \gtres\ GMC (A, B, C and D). Component B and D are associated with methanol and OH masers. Component C is associated only with OH maser emission. White boxes mark the position of the three RMS sources in the region (G331.4904-00.1173, G331.5414-00.0675 and G331.5582-00.1206). The RMS box sizes are 18\arcsec\ to account for the beam of MSX observation.}
	\label{fig:spitzer_laboca}
\end{figure}

Figure \ref{fig:spitzer_laboca} shows a three color infrared image made using data from the IRAC bands at 3.6 \um\ (blue), 4.5 \um\ (green) and 8.0 \um\ (red), overlaid with contours of the emission at 0.87 mm. While the infrared emission agrees with the millimeter emission in the global sense, the 8.0 \um\ emission seems to trace the envelope of a ionized region, with the millimeter emission of the complex of clumps, surrounding this structure. Also, we find strong and compact emission associated with the MM3 clump in the three IRAC bands, with an angular size of $\sim14\arcsec$.

For each millimeter clump, we measured the integrated flux densities in the four Spitzer IRAC bands (8.0 \um, 5.8 \um, 4.5 \um\ and 3.6 \um, \citealt{Fazio04}), on the MSX bands (21.3 \um, 14.7 \um, 12.1 \um\ and 8.3 \um,  \citealt{Mill94}) and IRAC bands (100 \um, 60 \um, 25 \um\ and 12 \um, \citealt{Neugebauer84}). The results are tabulated in Table~\ref{tbl:flujos}. 

\subsection{High angular resolution observations of the \gtres\ GMC central region.}
We present in this section interferometric observations of radio continuum emission toward the \gtres\ GMC central region. The beam size of the observations ($\sim$2\arcsec) allow us to study compact structures at sub-parsec scales. The analysis also considered association between the radio continuum components found in recent OH and methanol maser catalogs at high resolution.

\subsubsection{Radio continuum emission}

\begin{figure}[h]
	\begin{center}
		\includegraphics[width=0.5\textwidth]{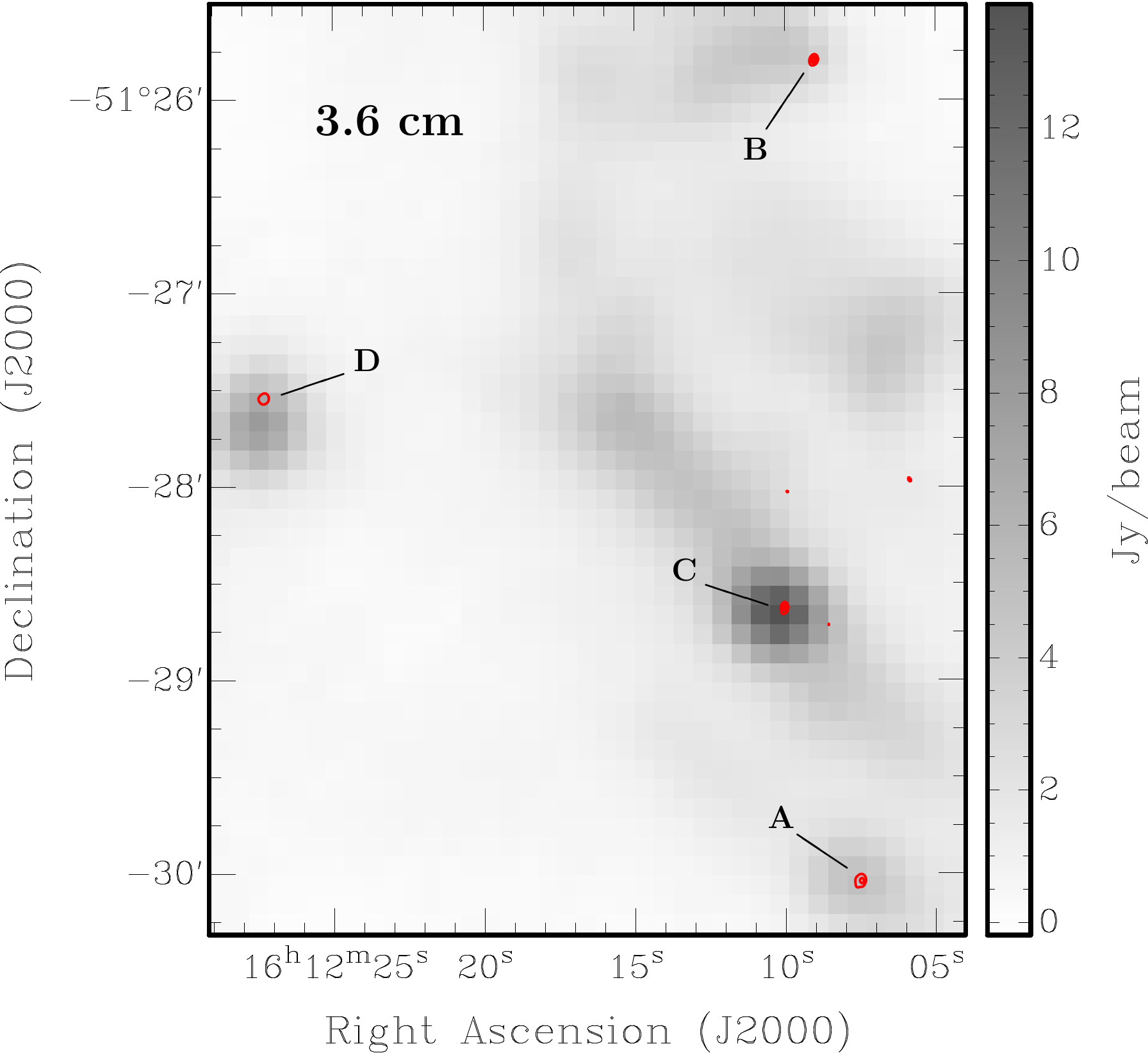}
		\includegraphics[width=0.5\textwidth]{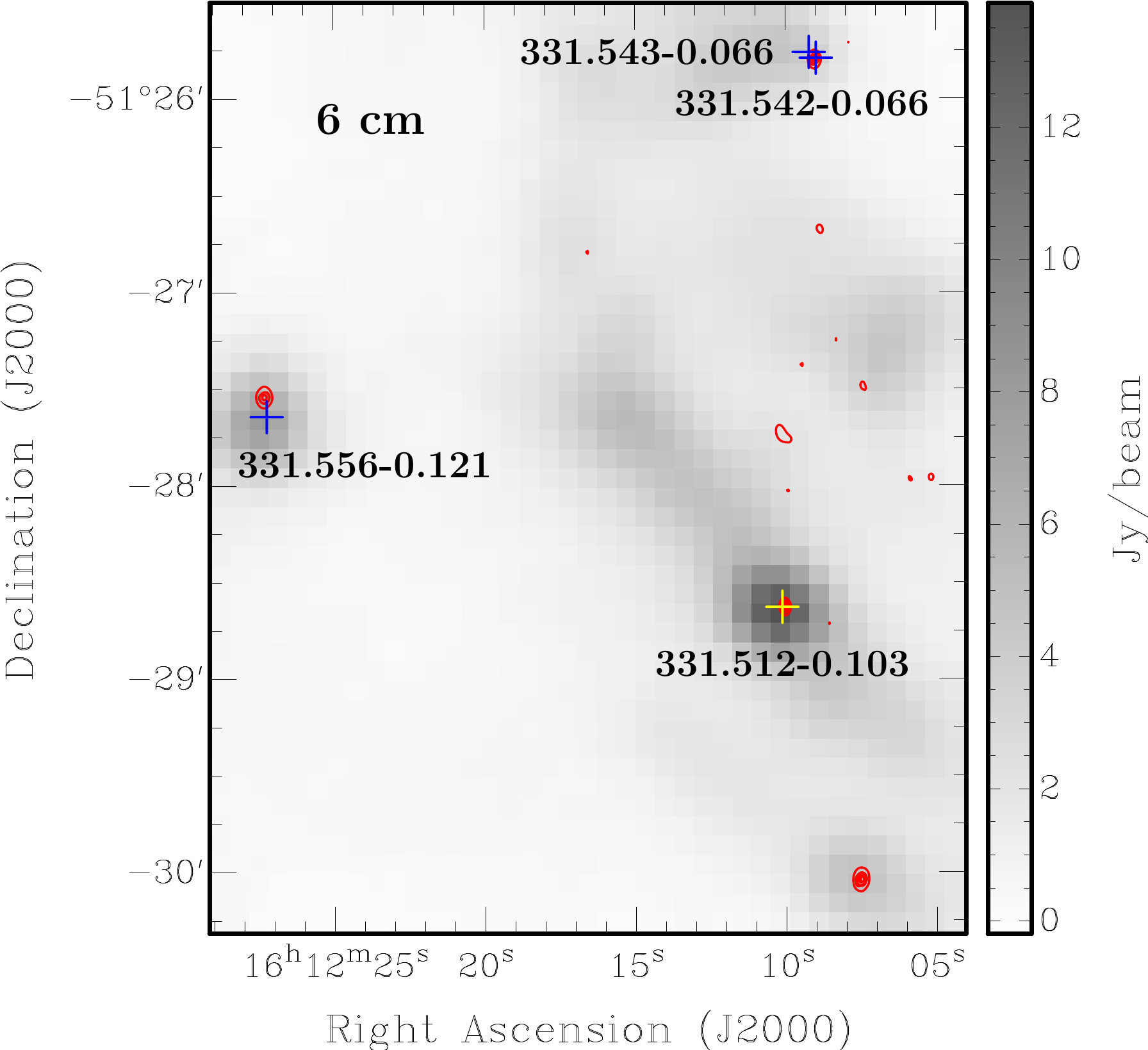}\\ 
	\end{center}
	\caption{Contour maps of the radio emission from the \gtres GMC central region at 3.6 cm (\textit{top}) and 6 cm (\textit{bottom}), overlaid on 0.87 mm continuum emission. The contours are drawn at 10\%, 30\%, 60\% and 90\% of the peak flux. For the 6 cm map, the peak intensity is 153.0 mJy beam$^{-1}$, rms noise is 1.04 mJy beam$^{-1}$. For the 3.6 cm map, the peak intensity is 159.5 mJy beam$^{-1}$, rms noise is 1.09 mJy beam$^{-1}$. The blue crosses in the 6 cm map show the position of methanol masers in the central region of the \gtres\ GMC. All methanol masers here are associated with OH masers. The yellow cross shows the position of the OH maser, without methanol counterpart, toward the \outflow\ outflow. }  
	\label{fig:radiomap}
\end{figure}

Figure~\ref{fig:radiomap} shows contour maps of the radio continuum emission at 6 and 3.6 cm towards the central region of \gtres\ GMC.
Our observations reveal the presence of four distinct compact radio sources within the
mapped field of 10\arcmin, labeled as components A,B,C and D, and associated with the clumps MM1, MM4, MM3 and MM6, respectively (see Fig.~\ref{fig:spitzer_laboca}). 
The positions, flux densities and beam deconvolved sizes (FWHM) of the sources, obtained with the AIPS task IMFIT, are given in Table~\ref{tbl:radio_param}.

\begin{figure}[h]
	\begin{center}
		\includegraphics[width=0.5\textwidth]{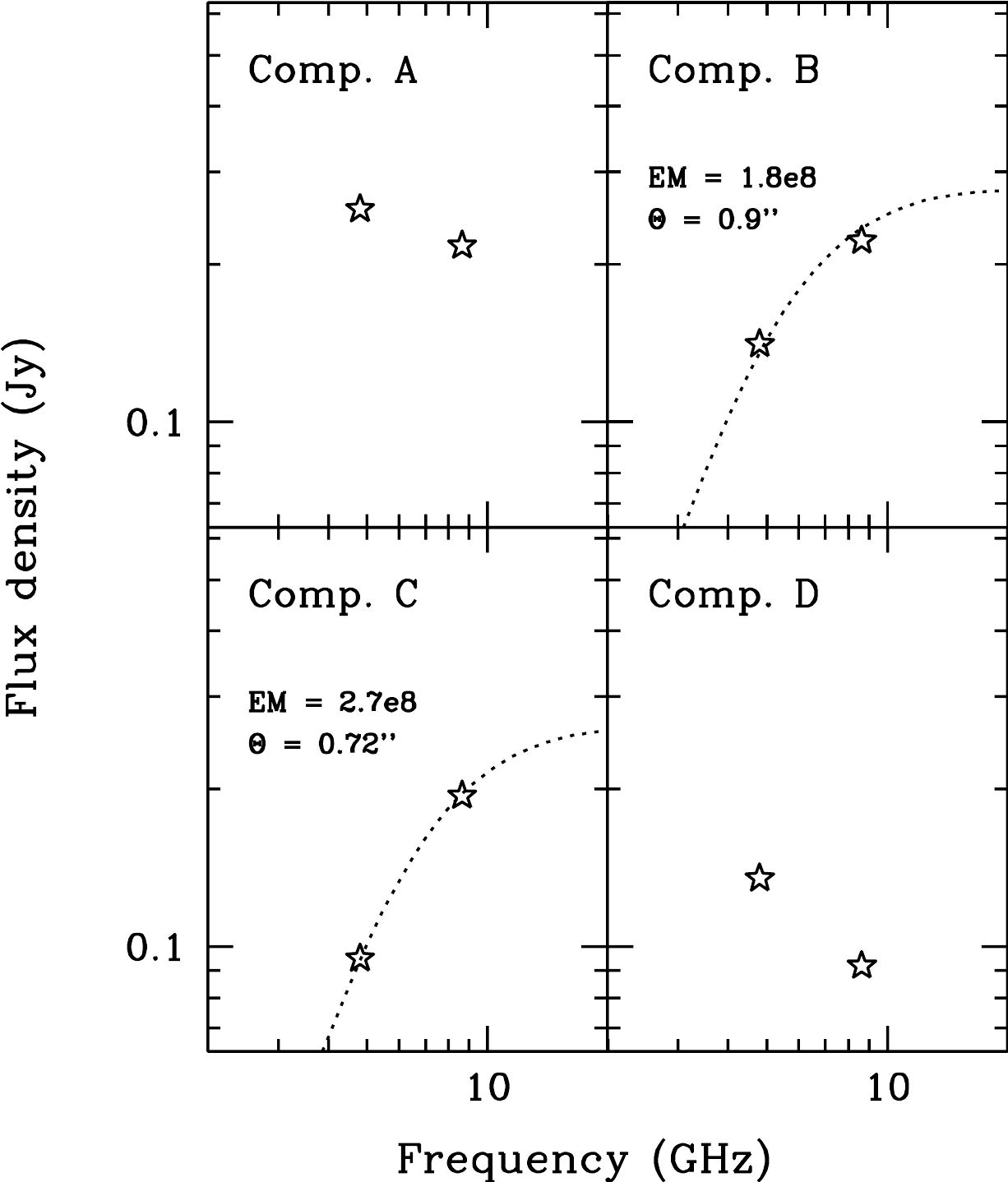}
	\end{center}
	\caption{Spectral energy distributions at short frequency for each radio component found in the central region of the \gtres\ GMC. For components B and C, the dotted line is a fit of the spectrum using modified blackbody function of the form $B_\nu(T_e)[1-exp(-\tau_\nu)]$, with $\tau_\nu\propto T_e^{-1.35} \nu^{-2.1} EM$. The emission measure $EM$ is in units of [pc cm$^{-6}]$. The $T_e$ considered is 10000 K. The spectral index of components B and C are 0.8$\pm$0.2 and 1.2$\pm$0.2, respectively.}
	\label{fig:radio_spectra}
\end{figure}

Components A, B and D are also associated, respectively, with the G331.4904-00.1173, 
G331.5414-00.0675, and G331.5582-00.1206 RMS sources candidates to MYSOs 
\citep{Urquhart08}. The detection of radio emission indicate that these 
three candidates are UCHII regions and therefore related with a more evolved state than a genuine MYSOs. The brightest radio 
source (component C) is associated with the very energetic outflow detected towards 
the \gtres\ central region \citep{Bronfman2008}. Figure~\ref{fig:radio_spectra} shows the radio spectra of all four components.
Components B and C exhibit an increasing flux density with frequency, thus 
they could be optically thick ultracompact regions or stellar wind sources.
The dotted line corresponds to the best fit of the spectral energy distribution (SED) assuming that they correspond to 
regions of ionized gas with uniform density, and that the emission can be modeled as a modified blackbody, with $I_\nu=B_\nu(T_e)[1-\exp(-\tau_\nu)]$ and the opacity $\tau$ estimated from \citep{Wilson09}:

\begin{equation}
\tau_\nu=8.235\times10^{-2}\left(\frac{T_e}{\mathrm{K}}\right)^{-1.35}\left(\frac{\nu}{\mathrm{GHz}}\right)^{-2.1}\left(\frac{EM}{\mathrm{pc\ cm^{-6}}}\right)\ ,
\end{equation}

\noindent with $T_e$ the electron temperature and $EM$ the emission measure. We considered $T_e=10^4$ K. The derived spectral indexes of component B and C are given in Column 9 of  Table~\ref{tbl:radio_param}. Component B has a spectral index of the radio emission between 4.8 and 8.6 GHz of 0.8$\pm$0.2, similar  the value expected for a spherical, isothermal, constant-velocity model of stellar wind \citep{Reynolds1986}. The error in the determination of spectral indexes 
include the 5\% uncertainty in the absolute flux density scale. Component C spectral index has a value of 1.2$\pm$0.2 already reported by \cite{Bronfman2008}. This index suggests that source 
C corresponds to an ionized jet which is likely to drive the associated energetic 
molecular outflow.

Components A and D are more extended than B and C, and they present a decrease in the integrated flux density from 4.8 to 8.6 GHz. We attribute this particular behavior to an observational artifact, that could be related with a less complete uv-coverage at higher frequency. An estimation of the flux density inside an area comparable with the beam size ($\sim$2.5\arcsec) gives an almost flat spectrum for component A, and even though the flux still decreases with frequency for component D, the flux difference was reduced from 47\% to 30\%. We expected for both sources, A and D, a flat spectral energy distribution in the radio continuum, consistent with the free-free emission from a thermal source.

\subsubsection{Maser emission}

Earlier studies of masers across our Galaxy have shown that regions of high-mass star formation can be associated with OH, H$_2$O and Class II methanol masers \citep{Caswell1995, Szymczak2002}. In particular, OH masers are usually considered signposts of UCHII regions, while methanol masers are found associated to hot molecular cores, UCHII regions and near-IR sources \citep{Garay99, Bartkiewicz2012}. Therefore, OH and methanol masers often coincide, and they are found in regions with active star formation. With this in mind, we searched in methanol and OH surveys toward the Norma spiral arm region. We found three sources from the 6-GHz methanol multibeam maser catalog from \cite{Caswell2011}. Two of them, 331.542-0.066 and 331.543-0.066, are toward the MM4 millimeter clump and they are coincident, within the errors, with the radio component B. The velocity of the peak reported for these maser sources are -86 \kms\ and -84 \kms\ respectively, within 5 \kms\ difference from the molecular line velocities of MM4 shown in Table~\ref{tbl:linepeaks}, and therefore consistent with the ambient velocity of the complex of clumps. The third 6668 MHz methanol maser spot, 331.556-0.121 ($\alpha_{2000} = 16^h12^m27.21^s$ and $\delta_{2000} = -51\arcdeg27\arcmin 38.2\arcsec$), is associated with the MM6 millimeter clump, it has a velocity at peak emission of -97 \kms, similar to values found with molecular lines for MM6, and it is located 5.8\arcsec\ ($\sim$3 times the beamsize in the maser catalog) away from the peak position of radio component D. Each of these three maser spots has a counterpart OH maser \citep{Caswell09,Caswell98}. 
The peak position of radio component C ($\alpha_{2000}=16^h12^m10.04^s$,
$\delta_{2000}= -51\arcdeg28\arcmin37.7\arcsec$) is coincident, within the
errors, with the interferometrically derived position of an
1665 MHz OH maser spot (G331.512-0.103; Caswell, 1998). No methanol
class II emission was reported towards the outflow up to a detection limit
of 0.6 Jy \citep{Caswell2011}. The position of masers in the  \gtres\ GMC central region is shown in Fig.~\ref{fig:radiomap}, along their association with the radio continuum sources. 

No methanol or OH emission is associated with the MM1, MM2 and MM5 clumps. While MM1 is associated with the radio component A and with a RMS source, no other signposts of star formation activity are found in MM2 and MM5, suggesting that these clumps are in a state prior to the formation of an UCHII region.

\section{Discussion}

\subsection{Physical properties of the GMC}

The mass of the \gtres\ GMC can be derived from the CO NANTEN observations considering two approaches: a) from the relation of the integrated intensity of CO line emission with H$_2$ mass, and b) using the local thermodynamic equilibrium (LTE) formalism.

In the first method, the column density N(H$_2$) of the GMC is considered proportional to the CO emission integrated in velocity [$I(\mathrm{CO})\;=\;\int T_{\mathrm{mb}} (^{12}\mathrm{CO})dv\ $(\Kkms)], 

\begin{equation}
N(\mathrm H_2)=X(\mathrm{CO})I(\mathrm{CO}),
\end{equation}

\noindent where 
$X$(CO) corresponds to the Galactic average $I$(CO)-to-$N$(H$_2$) conversion factor. With this approach, the luminosity of CO, $L_{\mathrm{CO}}=D^2\int\!\!\!\int T_{\mathrm{mb}} (^{12}\mathrm{CO})\, dv\, d\Omega$, is proportional to the total mass of the cloud, $M_{\mathrm{CO}}=\alpha_{\mathrm{CO}}L_{\mathrm{CO}}$. Here we assumed that $X\mathrm{(CO)} = 2.3\times10^{20}$ cm$^{-2}\ \left(\Kkms\right)^{-1}$ \citep{Kennicutt2012}, which implies $\alpha_{\mathrm{CO}}=4.6\ {\mathrm{M}_\odot}\ (\Kkms {\mathrm{pc}}^{2})^{-1}$ with correction for helium.

From the CO emission integrated in the velocity range between -117.7 and -73.3 \kms, we measured a CO luminosity of $1.9\times10^6\;\left(\mathrm K\ \kms\,\mathrm{pc}^2\right)$, giving a total mass of $8.7\times10^6\ {\mathrm M}_\odot$ and an average column density of $3.2\times10^{22}$ cm$^{-2}$. 

We also estimate the column density and mass from the C$^{18}$O observation following the derivation of physical parameters presented by \cite{Bourke97}. Assuming that the emission is optically thin,
\begin{eqnarray} \nonumber
N(\mathrm{C^{18}O})&\ =\ &2.42\times10^{14}\frac{T_{\mathrm{ex}}+0.88}{1-\exp(-5.27/T_{\mathrm{ex}})}\\ &&\times \frac{1}{J(T_{\mathrm{ex}})-J(T_{\mathrm{bg}})} \int T_{\mathrm{mb}}(\mathrm{C^{18}O})\,dv,
\end{eqnarray}
\noindent\ and
\begin{eqnarray} \nonumber
\left( \frac{M_{\mathrm{LTE}}}{M_\odot}\right)&\ =\ &1.44 \left(\frac{\mu_m}{2.72 m_\mathrm{H}}\right) \!\!\left(\frac{[\mathrm{H_2/C^{18}O}]}{3.3\times10^6}\right)\!\!
\left(\frac{D}{\mathrm{kpc}}\right)^2\\ \nonumber&&\times\frac{T_{\mathrm{ex}}+0.88}{1-\exp(-5.27/T_{\mathrm{ex}})}\ \frac{1}{J(T_{\mathrm{ex}})-J(T_{\mathrm{bg}})}\\&&\times\int\!\!\!\int T_{\mathrm{mb}} (\mathrm{C^{18}O})\, dv d\Omega\ ,
\end{eqnarray}
\noindent where $T_{\mathrm {ex}}$ is the excitation temperature of the transition, $T_{\mathrm{bg}}$ is the background temperature, $v$ is in \kms, $\Omega$ is in arcmin$^2$, and

\begin{displaymath}
J(T)=\frac{h\nu}{k}\frac{1}{\exp(h\nu/kT)-1}\ . \nonumber
\end{displaymath}

We assumed an abundance ratio $[\mathrm{CO/C^{18}O}]=330$, which is the suggested value for GMCs inside the 4 kpc Galactic molecular ring \citep{Wilson94}. Considering an average temperature of 15 K, the total mass value for this region is then $3.5\times10^6$ \Msun, with an average column density $N(\mathrm{C^{18}O}) = 4.0\times10^{15}\;$cm$^{-2}$. The mass estimated is less than half that the mass obtained from CO, but the denser material traced by C$^{18}$O is emitting in a smaller area. Since the continuum millimeter sources are seen in the C$^{18}$O map concentrated within the area defined for the \gtres\ central region, we estimated the column density towards the position of the peak emission on this map, considering a higher temperature than the average value used for the GMC. For $T=20$ K, we obtained $N(\mathrm{C^{18}O}) = 2.6\times10^{16}\;$cm$^{-2}$.


\subsection{Physical properties of the \gtres\ central region.}

\subsubsection{Spectral energy distributions}

To determine the dust temperature and bolometric luminosity of the GMC central region and each clump, we fit the spectral energy distribution (SED) using a modified blackbody model,

\begin{equation}
S_\nu=\Omega B_\nu(T_d)\{1-exp(-\tau_\nu)\},
\end{equation}

\noindent where $S_\nu$ is the flux density of the source, $B_\nu$ is the Planck function, $\Omega$ is the solid angle subtended by the emitting region, and $T_d$ is the dust temperature. The optical depth, $\tau_\nu$, is assumed to depend on frequencies, $\tau_\nu=(\nu /\nu_0)^\beta$, 
with $\beta$ the power law spectral index and $\nu_0$ the frequency at which the optical depth is unity.

\begin{figure}
\begin{center}
	\includegraphics[width=0.5\textwidth]{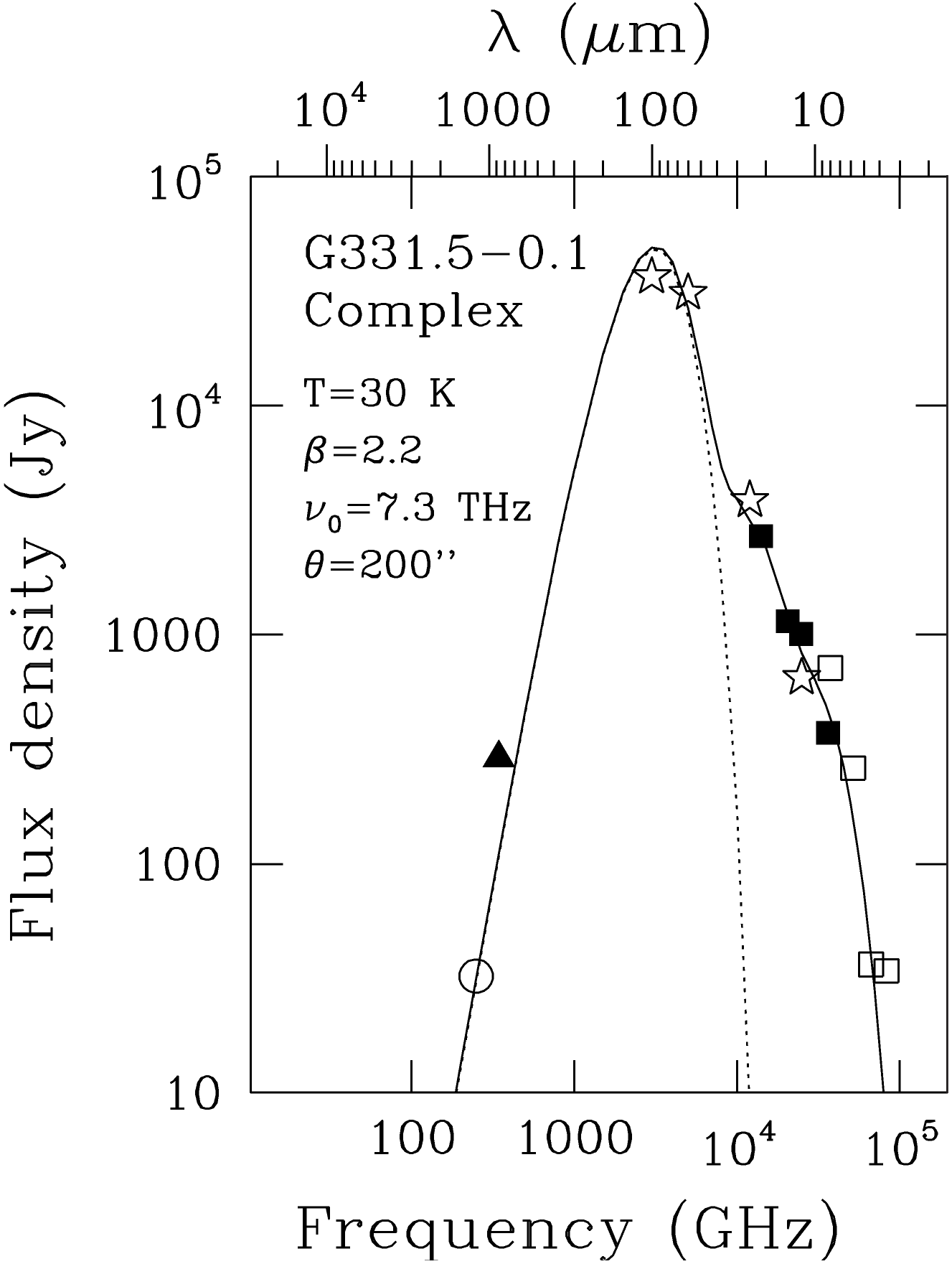}
\epsscale{0.4}
\end{center}
\caption{Spectral energy distribution of the region harboring the \gtres\ complex of clumps. From left to right, the circle marks the SIMBA flux, the triangle the LABOCA, the stars mark IRAS fluxes, filled-squares MSX fluxes, and open-squares SPITZER fluxes. The solid curve is a fit to the spectrum using three modified blackbody functions of the form $B_\nu(T_e)[1 - exp\left( - (\nu/\nu_0)^\beta\right)]$, with different temperatures (cold, warm and hot components). The fitted parameters of the cold component (represented as the dotted line) are listed.}
\label{fig:sed_maincore}
\end{figure}

\begin{figure}[h]
\begin{center}
	\includegraphics[width=1\textwidth]{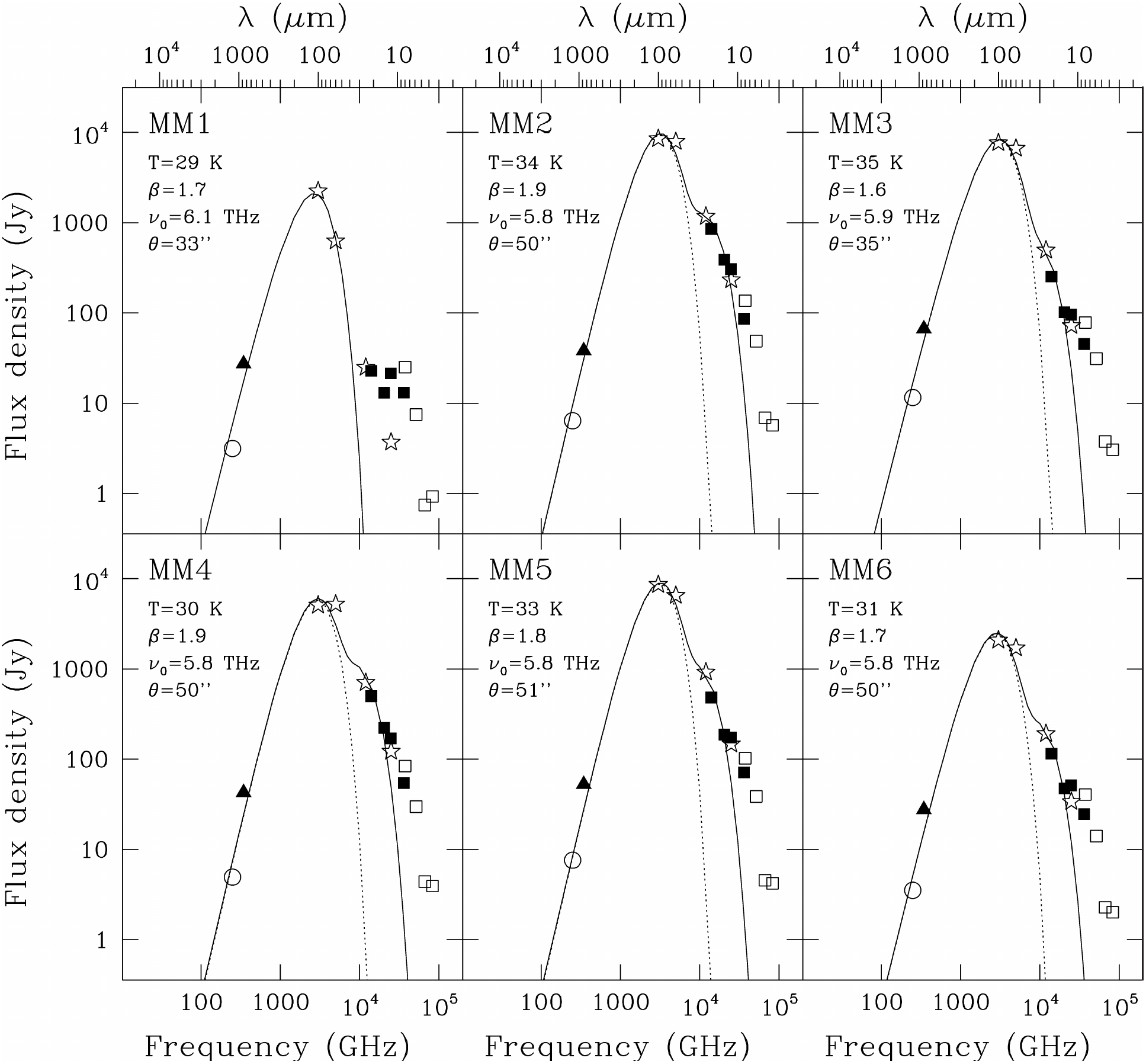}
\end{center}
\caption{Spectral energy distributions of each millimeter clump in the \gtres\ central region. The symbols are the same as in Fig.~\ref{fig:sed_maincore}. The solid curve is a fit to the spectrum using two modified blackbody functions of the form $B_\nu(T_e)[1 - exp\left( - (\nu/\nu_0)^\beta\right)]$, with different temperatures (cold and warm components). In the case of MM1, only on modified black function was considered. The fitted parameters of the cold component (represented as the dotted line) are listed on the upper left.}
\label{fig:seds_clumps}
\end{figure}

SEDs were fit for each clump and for the region harboring the complex of clumps at high velocity. The flux densities were measured in the areas shown in dashed boxes in figure~\ref{fig:dust}, using the data observed at 1.2 mm and 0.87 mm, and at four IRAS, MSX and SPITZER bands (see Table~\ref{tbl:flujos}). A good fit to the emission from the complex of clumps required three dust components at different temperatures (see Fig.~\ref{fig:sed_maincore}). For the individual clumps, we only considered a two component model, without considering the SPITZER data (Fig.~\ref{fig:seds_clumps}). In the case of MM1 clump, the source is only marginally detected at short wavelengths, and therefore we only considered a cold component for the model.  The parameters obtained (see Table~\ref{tbl:seds_clumps}) are typical of regions associated with high-mass star formation \citep{Faundez2004}. For the six clumps in the \gtres\ central region, the average temperature of the cold fit component is 32 K, and for the region containing the complex of clumps, the cold component has a fitted temperature of 30 K. The average value of the spectral indexes of the cold component for the six clumps is 1.8, which is consistent with the OH5 model from \cite{Ossenkopf94}. The steeper value of the spectral index found for the region containing the complex of clumps at high velocity could be due to low recovery of some extended flux in the 1.2 mm image. The bolometric luminosities were obtained according to:
\begin{equation}
L_{bol}=F_{int} \ 4\pi\  d\:^2\ ,
\end{equation}

\noindent with $d$ the distance of the source. The results are tabulated in Table~\ref{tbl:lumin}. The clumps have an average bolometric luminosity of $L_{bol}=5.7\times10^5\;\Lsun$, above the average value derived by \cite{Faundez2004} of 2.3$\times10^5\;\Lsun$. If the source of energy of each clump is due to a single object, then it will correspond to a star  with spectral type of O6 ZAMS or earlier \citep{Panagia73}. However, an inspection of Figure~\ref{fig:spitzer_laboca} shows that the clumps are unlikely to be associated with just one star. Most likely the \gtres\ central region has several early type stars, considering, for example, that a dozen of O9 ZAMS stars will produce the luminosity of a single O6 ZAMS star.

\subsubsection{Masses and densities}

Using the mm continuum observations we determined the mass of each clump from the expression:                 
\begin{equation} 
M_d=\frac{S_{\nu}D^2}{k_{\nu}B_{\nu}(T_d)}\ ,
\end{equation}

\noindent where $S_\nu$ is the flux density, $D$ is the distance to the source, $k_{\nu}$ is the dust mass coefficient, $B_\nu$ is the Planck function, and $T_d$ is the dust temperature.

We use a dust opacity of 1 cm$^2\:$g$^{-1}$ at 1.2 mm and 1.89 cm$^2\:$g$^{-1}$ at 0.87 mm, which are values computed for typical conditions of dense protostellar objects \citep{Ossenkopf94}. The opacity at 0.87 mm is obtained from a linear interpolation in the OH5 model of grains with thin ice mantles. Assuming a dust-to-gas mass ratio of $R=M_d/M_g=0.01$, the total mass of each clump is given by:
\begin{eqnarray}\nonumber
\left(\frac{M_{1.2 mm}}{M_\odot}\right)&\ =\ &20.82\left(\frac{S}{\mathrm{Jy}}\right)\left(\frac{\mathrm{1\: cm^2\: g^{-1}}}{\kappa_{1.2 mm}}\right)\left(\frac{D}{\mathrm{kpc}}\right)^2\\
&&\times\left\{\exp\left(\frac{11.990}{T_d}\right)-1\right\}\\ \nonumber
\left(\frac{M_{0.87 mm}}{M_\odot}\right)&\ =\ &4.29\left(\frac{S}{\mathrm{Jy}}\right)\left(\frac{\mathrm{1.89\: cm^2\: g^{-1}}}{\kappa_{0.87 mm}}\right)\left(\frac{D}{\mathrm{kpc}}\right)^2\\
&&\times\left\{\exp\left(\frac{16.538}{T_d}\right)-1\right\}\ \ .
\end{eqnarray}   

The mass of the clumps is estimated using the temperature obtained from the SED fitting. The masses estimated from the emission at 0.87 mm and 1.2 mm are given in Table~\ref{tbl:dust_cs_masses}. If we instead considered a fiducial temperature $T=20$, for example, the masses will be in general larger by a factor of two with respect to the masses estimated using the temperature from SED fitting. We note that the masses obtained from 1.2 mm are lower than the masses obtained from the 0.87 mm observations typically by a factor of two. Uncertainties in the dust opacities or in the dust temperatures could give other possible explanations. A better determination of the power law index $\beta$ of the dust optical depth could improve this value. Considering the results from the 1.2 mm observations, the average mass of the millimeter clumps in the central region is $3.2\times10^3\:\Msun$. The total mass and size of the complex of clumps are $\sim 2.1\times10^4\:\Msun$ and $\sim$ 15 pc. These values are bigger than the average masses and sizes of dust compact sources associated with high-mass star forming regions \citep[$\sim 5\times10^3\:\Msun$ and $\sim$1 pc,][]{Faundez2004}. However, each clump is by itself like one of these sources. 

Columns 4 and 5 in Table~\ref{tbl:dust_cs_masses} give the number density and surface density estimated from the emission at 1.2 mm for each clump. The density and the average surface density were computed using the expressions
\begin{equation} 
n=\frac{M}{(4/3) \pi R^3\mu\:m_H};\ \ \ \ \sum=\frac{M}{\pi R^2}\ \ \ \ ,
\end{equation}
\noindent assuming spherical morphologies and a mean mass per particle of $\mu=2.29$ (corrected for helium). The clumps have an average gas surface density of 0.4 \mbox{g cm$^{-2}$} ($\sim$1900 \Msun\ pc$^{-2}$). The values of the surface densities are similar to those of the densest clumps reported in a study of massive star-forming regions by \cite{Dunham11}, and they are above the surface density threshold of $\Sigma_{th}=129\pm14\ \Msun\:$pc$^{-2}$ found by \cite{Heiderman10} for ``efficient" star formation.  In an independent study \cite{Lada10} found a threshold of $\Sigma_{th}=116\ \Msun\:$pc$^{-2}$. Only one of the millimeter clumps in the \gtres\ central region have gas surface densities above 1g cm$^{-2}$, which is the theoretical lower limit determined by \cite{Krumholz08} to avoid excessive fragmentation and form massive stars. Nevertheless the inner and densest parts of each clump, mapped with a high-density gas tracer such as \csss, will exceed this requirement.

\subsubsection{Virial and LTE masses}

For a spherical, nonrotating cloud of radius R, and a gaussian velocity profile $\Delta v$ (FWHM), the virial mass is given by
\begin{equation}
\left(\frac{M_{virial}}{M_\odot}\right)=210\left(\frac{\Delta v}{\kms}\right)^2\left(\frac{R}{\mathrm{pc}}\right),
\end{equation}

\noindent assuming that the broadening due to optical depth is negligible. Column 6 in Table~\ref{tbl:dust_cs_masses} shows the virial mass obtained for each clump, determined from the \csss\ molecular line. The radius of the clumps was computed from the deconvolved angular size measured from the maps, and $\Delta v$ from the spatially average CS line emission (see Table~\ref{tbl:cs_sizes}). In the case of MM3, the outflow clump, we considered only the gaussian component corresponding to the ambient gas emission. The total mass obtained for the complex of clumps is $1.4\times10^4\:\Msun$.

We also estimate the mass of the complex of clumps from the \tcotd\ observations using the LTE formalism, assuming that this transition is optically thin. Using eq. [A4] from \cite{Bourke97}, the mass is estimated from the $^{13}$CO line in a similar way than the C$^{18}$O mass presented in section 4.1:

\begin{eqnarray} \nonumber
\left( \frac{M_{\mathrm{LTE}}}{M_\odot}\right)&\ =\ &0.082 \left(\frac{\mu_m}{2.72 m_\mathrm{H}}\right) \!\!\left(\frac{[\mathrm{H_2/^{13}CO}]}{5.3\times10^5}\right)\!\!\left(\frac{D}{\mathrm{kpc}}\right)^2\\
\nonumber&&\times\frac{T_{\mathrm{ex}}+0.88}{1-\exp(-15.87/T_{\mathrm{ex}})} \ \frac{\exp(15.87/T_{\mathrm{ex}})}{J(T_{\mathrm{ex}})-J(T_{\mathrm{bg}})}\\
&&\times\int\!\!\!\int T_{\mathrm{mb}} (\mathrm{^{13}CO})\, dv d\Omega\ ,
\end{eqnarray}

\noindent where $v$ is in \kms\ and $\Omega$ is in arcmin$^2$. We considered an excitation temperature of 30 K and a $[\mathrm{CO/^{13}CO}]$ abundance ratio of 53 \citep{Wilson94}. The mass obtained for the complex of clumps is $2.9\times10^4\:\Msun$, which agrees within a factor of two with the mass estimates from millimeter continuum emission and virial mass from CS.

Once the mass of the complex and of the individual clumps were obtained, we are able to investigate whether the clumps at low velocity (MM1 and MM6) and the complex of clumps at high velocity are bounded. Considering that the velocity difference $v_{dif}$ of the two systems is 11.9 \kms, and that the projected distance $r_p$ from their respective center of mass is $\sim$5 pc, a gravitationally bounded system will require a mass of $\sim2\times10^{5}$ \Msun, exceeding the total mass of the clumps by a factor of ten. Then, we discarded that the low velocity clumps are gravitationally bounded to the complex. We cannot rule out for the moment other kind of interaction between these two systems.   

\subsubsection{Rotational temperatures of the \outflow\ outflow \label{section:rot_diag}}

When observations of several transitions in a particular molecule are available, it is possible to estimate the rotational temperature, $T_{rot}$ and the total column density, $N_{T}$ from a rotational diagram analysis, assuming that the lines are optically thin and that excitation temperatures follow $T_{ex}>>T_{bg}$ \citep[e.g.][]{Linke79,Blake87,Garay2002}. We considered the equation

\begin{equation}
\frac{N_k}{g_k}\;=\;\frac{3k\int{T_{mb} dv}}{8\pi^3\mu^2\nu S}\ ,
\label{eqn:Ng}
\end{equation} 

\noindent with $S$ the intrinsic line strength, $\mu$ the permanent dipole moment, $N_k$ and $g_k$ the column density and degeneracy of the upper transition state. If we assume that a single rotational temperature is responsible for the thermalization of the population distribution, then

\begin{equation}
\frac{N_k}{g_k}\;=\;\frac{N_T}{Q(T_{rot})}\; \mathrm{exp}(-E_u/k\:T_{rot})\ ,
\label{eqn:rotation}
\end{equation}

\noindent with $N_T$ is the total molecular column density summed over all the levels, $E_u$ is the energy of the upper transition state, and $Q(T_{rot})$ is the rotational partition function at temperature $T_{rot}$. 

The previous analysis was made on the redshifted and blueshifted wing emission of the \outflow\ outflow, using the SiO transitions $J = 7\rightarrow$6 and $J = 8\rightarrow$7, and the SO transitions $J_K = 8_8\rightarrow7_7$ and $J_K = 8_7\rightarrow7_6$. We consider two ranges of  velocity for integration:  -144.1 to -95.6, and -79.9 to -16.3. The values computed for $T_{rot}$, $Q(T_{rot})$ and $N_T$ from the diagrams are tabulated in Table~\ref{tbl:rotation}. The uncertainty in the estimates of rotational temperatures are obtained using the error in the estimate of $I(T_{MB})=I(T_A^*)/ \eta_{MB}$ (see equation~\ref{eqn:sigmalines}), and propagating them through equations~\ref{eqn:Ng} and \ref{eqn:rotation}

For SiO transitions, the rotational temperatures of the blue and red wing are 123$\pm$15 K and 138$\pm$29 K respectively, with column densities ranging between $2.6\times10^{13}$ and $3.3\times10^{13}$ cm$^{-2}$. For the SO lines, the rotational temperatures of the blue wing (94$\pm$5 K) is higher than the temperature of the red wing, ($77\pm4$ K), with column densities varying between $3.8\times10^{14}$ and  $4.6\times10^{14}$ cm$^{-2}$. We note that since our estimation of temperatures has been done with only two transitions for each molecule, these values should be considered as a rough estimation and more transitions should be considered for a more robust estimation of the temperature in the wings of the molecular outflow. ALMA observations recently obtained toward this source (Merello et al., in prep.) shows several transitions of SO$_2$ that will give better determination of rotation temperatures.

\subsubsection{Comparison with other massive star-forming regions}
In this section we compare the characteristics of the clumps in \gtres\ with those of the sample of 42 massive star forming regions presented by \cite{Molinari2008}. In that study, a 8-1200 \um\ SED for each YSO is presented, using MSX, IRAS and sub-mm/mm data, aiming to relate their envelope mass with the bolometric luminosity. Those sources that were fitted with an embedded ZAMS envelope were named as ``IR", and those fitted only with a modified blackbody peaking at large wavelengths were named ``MM". A further classification, as ``P" or ``S" source, was made according if they were the primary (most massive) object in the field or not. The parameters fitted for MM sources gave temperatures with a median value of $\sim$20 K, and spectral indexes $\beta$ with median value $\sim1.5$. Our sample of millimeter clumps in the \gtres\ GMC central region seems to correspond to the MM sources in their sample.

\begin{figure}[h]
\begin{center}
	\includegraphics[width=0.5\textwidth]{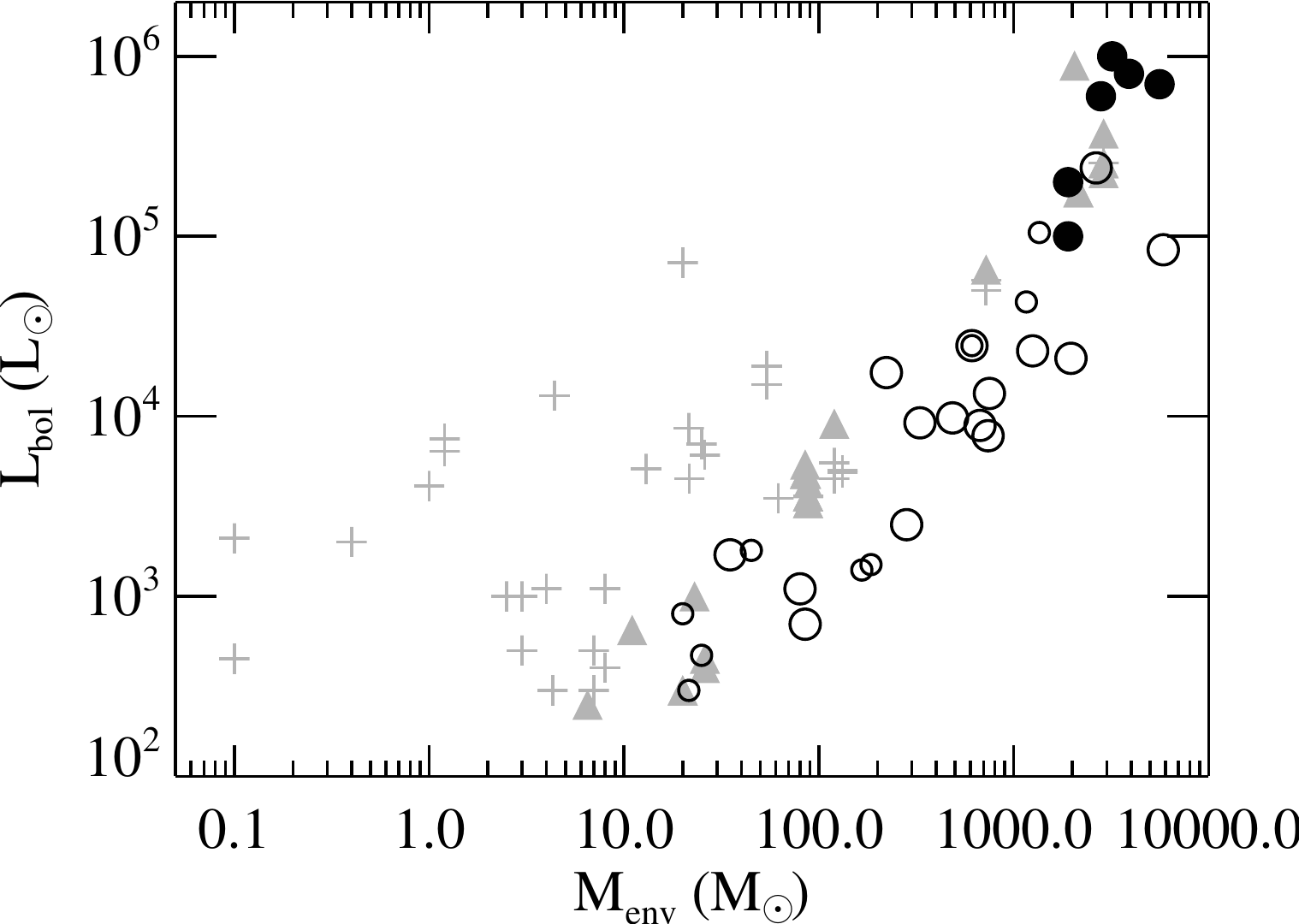}

\end{center}
\caption{$L_{bol}-M_{env}$ diagram for the six millimeter clumps (MM1-MM6) in the \gtres\ GMC central region (filled black circles), together with sources from \cite{Molinari2008}. From Molinari et al.\ sample, MM sources are represented as open circles (small representing MM-S sources), IR-P sources are shown in grey filled triangles and IR-S are indicated in crosses.}
\label{fig:lumin_mass}
\end{figure}

Figure~\ref{fig:lumin_mass} shows a $L_{bol}-M_{env}$ diagram for the sample of 42 sources from \cite{Molinari2008}, including the values derived for our six millimeter clumps (filled circles). Clearly our sources are more massive and luminous that most of the objects in that sample, with 
bolometric luminosities in general an order of magnitude larger than their MM sources. It is worth noticing that the six clumps in the \gtres\ GMC central region fall in the $L_{bol}-M_{env}$ diagram toward the prolongation of the isochrone of $\sim10^5$ yrs shown in Fig. 9 of \cite{Molinari2008}. This trend is associated with the end of the accelerated accretion phase, and therefore our sample of millimeter clumps MM1-MM6 may be in the phase of envelope clean-up, observationally identified with Hot Cores and UC \ion{H}{2} regions by Molinari et al.

Similar conclusions can be obtained when compared with the study with $Herschel$ of star-forming compact dust sources from \cite{Elia2010}, where modified blackbody SED fitting was made for sources extracted in two 2\arcdeg$\times$2\arcdeg\ Galactic plane fields centered at $l$=30\arcdeg, $b$=0\arcdeg, and $l$=59\arcdeg, $b$=0\arcdeg\ using the five Hi-GAL bands (70, 160, 250, 350 and 500 \um). Our sample of millimeter clumps remains in the extreme of most luminous and massive millimeter sources.


\section{Summary}

Observations at several wavelengths show that the \gtres\ GMC central region is one of the largest, most massive and most luminous regions of massive star formation in the Galactic disk. We have obtained a multi-tracer view of this source using radio, (sub)millimeter and infrared images, starting from an analysis of the global parameters of the parent GMC of this region at a scale of ~1\arcdeg, and continuing to smaller scales until identifying and resolving features at a scale inside \gtres\ of one arcsecond.

Millimeter maps reveal that the \gtres\ GMC central region harbors six massive clumps within a region of 15 pc in diameter. The average size, mass and molecular hydrogen density of these clumps are 1.6 pc, $3.2\times10^3\: \Msun$ and $3.7\times10^4$ cm$^{-3}$, respectively. From the SED analysis of the clumps, we estimated an average dust temperature and bolometric luminosity of 32 K and $5.7\times10^5\: \Lsun$, respectively. These values are similar to those of massive and dense clumps typically found in high-mass star forming regions. The high number of massive and dense clumps within G331.5-0.1 makes it one of the most densely populated GMC central regions in the Milky Way. At the center of the brightest, most massive and densest molecular clump within G331.5-0.1 central region, we discovered one of the most luminous and massive protostellar objects presently known, which drives a powerful molecular outflow and a thermal jet. The outflow is not resolved at a resolution of $\sim$7\arcsec, so it could be directed along the line of sight. Further high resolution observations with ALMA are underway and will be reported elsewhere.

We found four compact radio sources along the \gtres\ central region. Two of them have a spectral index consistent with ionized stellar winds, which originated from young luminous objects. In particular, one of these radio sources is located at the center of the molecular outflow clump,
which suggests that is associated with its driving source.

\acknowledgements{L.B. and G.G. gratefully acknowledge support from CONICYT through projects FONDAP 15010003 and BASAL PFB-06. M.M. was supported by a Fulbright Fellowship and the NSF grant AST-1109116. We thank our referee, M. Pestalozzi, for very useful comments that improved the clarity of this paper. We would like to dedicate this paper to the memory of our friend and collaborator Jorge May Humeres (1936-2011), whose inspiration and advice in many ways made this work possible. He is missed deeply. }



\clearpage

\begin{deluxetable}{clcccccccc}
\tabletypesize{\scriptsize}
\tablecaption{Parameters of the molecular line observations\label{tbl-obs}}
\tablewidth{0pt}
\tablehead{
\colhead{Telescope} & \colhead{Line} & \colhead{Frequency$$} & 
 \colhead{Beam} &
\colhead{$\eta_{mb}$} & \colhead{Pos. obs.} & \colhead{Spacing} & \colhead{t$_{in}$} & \colhead{$\Delta v$} & \colhead{Noise}\\
 & & \colhead{(MHz)}&\colhead{(FWHM)} & & & &\colhead{(sec)}&\colhead{(\kms)}&\colhead{(K)}
}
\startdata

 NANTEN	&\couc							&	115271.202	& 2.6\arcmin  & 0.89 & 585	& 2.5\arcmin 	& $\le$60	 & 0.15	 &  0.35 \cr
 				&\cDOo  						& 109782.173  & 2.6\arcmin  & 0.89 & 440		& 2.5\arcmin 	& $\le$600	 & 0.15  &  0.1\cr
\tableline
 SEST		&\csdu  						& 97980.968  	& 52\arcsec   & 0.73 & 81  	  & 45\arcsec  	& 180  & 0.130 & 0.06 \cr
 				&\cscc  						& 244935.644 	& 22\arcsec   & 0.48 & 81  	  & 45\arcsec  	& 180  & 0.052 & 0.08 \cr
\tableline
 ASTE		&\tcotd 						& 330587.957 	& 22\arcsec   & 0.61 & 167 	  & 22.5\arcsec	& 240  & 0.113 & 0.1 \cr
 				&\csss  						& 342882.950 	& 22\arcsec   & 0.61 & 167 	  & 22.5\arcsec & 240  & 0.109 & 0.1 \cr
\tableline
 APEX		&\sioss 						& 303926.809  & 20\arcsec   & 0.7  & 1   		& \nodata   				& 670  & 0.48  & 0.04 \cr
 				&\sioos 						& 347330.635 	& 17.6\arcsec & 0.7  & 1   		& \nodata   				& 670  & 0.42  & 0.04 \cr
 				&\soos  						& 304077.844 	& 20\arcsec   & 0.7  & 1   		& \nodata   				& 680  & 0.48  & 0.04 \cr
 				&\sooo  						& 344310.612 	& 17.6\arcsec & 0.7  & 1   		& \nodata   				& 680  & 0.48  & 0.05 \cr
\enddata
\label{tbl:observ_mol}
\end{deluxetable}

\begin{deluxetable}{ccccc}
\tabletypesize{\scriptsize}
\tablecaption{Parameters of continuum observations\label{tbl-obs}}
\tablewidth{0pt}
\tablehead{
\colhead{Telescope} &\colhead{Frequency} & 
 \colhead{Bandwidth} &
\colhead{Beamwidth} & \colhead{Noise}\\
 &\colhead{(GHz)}&\colhead{(GHz)}	&\colhead{(FWHM)} &\colhead{(mJy/beam)}
}
\startdata

 SEST		& 250	& 90		& 24\arcsec										 & 50\cr
 \tableline
 APEX		& 345 & 60		& 18.6\arcsec									 & 50\cr
 \tableline
 ATCA		& 4.8 & 0.128 &	2.7\arcsec$\times$1.8\arcsec & 0.41\cr
 				& 8.6 & 0.128 &	1.5\arcsec$\times$1.0\arcsec & 0.37\cr
\enddata
\label{tbl:observ_cont}
\end{deluxetable}

\begin{deluxetable}{cc|ccccc|ccccc|c}
\tabletypesize{\scriptsize}
\tablecaption{Observed parameters of dust continuum emission\label{tbl:dustpeaks}}
\tablewidth{0pt}
\tablehead{
\colhead{Condensation} & \colhead{$\lambda$}& \colhead{$\alpha$(J2000)} & 
\colhead{$\delta$(J2000)} & \colhead{Peak flux density} &
\colhead{Flux density} & \colhead{Angular size} & \colhead{Diameter} \cr
 		&\colhead{($\mu$m)} & 			    &				&\colhead{(Jy/beam)}&\colhead{(Jy)}&\colhead{(\arcsec)}&\colhead{(pc)} 	}
\startdata

MM1&1200&16 12 07.69&-51 30 07.1&1.2	  &3.1 & 36$\times$32   &1.3 \cr
          & 870&16 12 07.57&-51 30 03.5&5.0	  &27.4&  44$\times$31  & \cr
MM2&1200&16 12 08.43&-51 27 11.1&1.2			&6.4 & 62$\times$40 &1.9 \cr
	& 870&16 12 06.90&-51 27 15.6&  4.5		&38.3& 59$\times$49	& \cr
MM3&1200&16 12 09.34&-51 28 39.0&2.9			&11.6& 50$\times$34 &1.2 \cr
	& 870&16 12 10.13&-51 28 39.5& 13.8	  &66.7&  36$\times$30	& \cr
MM4&1200&16 12 10.08&-51 25 51.0&1.2			&4.9 & 56$\times$38 &1.9 \cr
	& 870&16 12 10.73&-51 25 45.5&  4.6		&42.7&  72$\times$45  & \cr
MM5&1200&16 12 14.45&-51 27 42.6&1.2		  &7.6 & 76$\times$34 &1.9 \cr
	& 870&16 12 15.25&-51 27 39.41&  5.5		&52.5&  69$\times$38	& \cr
MM6&1200&16 12 27.28&-51 27 41.8&1.6			&3.5 & 32$\times$32 &1.2 \cr
	&870&16 12 27.45&-51 27 39.1&  8.0    &27.6&  30$\times$21	& \cr
\enddata
\label{tbl:dustpeaks}
\end{deluxetable}

\clearpage

\begin{deluxetable}{lccccccc}
\tabletypesize{\scriptsize}
\tablecaption{Molecular Lines: Observed Parameters}
\tablewidth{0pt}
\tablehead{
\colhead{Line}	&\multicolumn{3}{c}{Peak}&\colhead{\ \ \ }&\multicolumn{3}{c}{Average}\cr \cline{2-4} \cline {6-8}
&\colhead{$T^*_A$}&\colhead{$V$}&\colhead{$\Delta v$}&&\colhead{$T^*_A$}&\colhead{$V$}&\colhead{$\Delta v$} 
\cr
&\colhead{(K)}&\colhead{(\kms)}&\colhead{(\kms)}&&\colhead{(K)}&\colhead{(\kms)}&\colhead{(\kms)}
}
\startdata
&&&MM1\cr
\hline
\csdu&0.896 $\pm$0.052&-102.39$\pm$0.03&5.7$\pm$0.1&& \nodata & \nodata & \nodata\cr
&0.734 $\pm$0.052&-88.06$\pm$0.05&5.7$\pm$0.1&& \nodata & \nodata & \nodata\cr
\cscc &0.134 $\pm$0.073&-102.31$\pm$0.20&5.2$\pm$0.5&& \nodata & \nodata & \nodata\cr
&0.181 $\pm$0.073&-88.43$\pm$0.13&3.1$\pm$0.4&& \nodata & \nodata & \nodata\cr
\csss  &0.634 $\pm$0.084&-102.42$\pm$0.07&4.6$\pm$0.2&&0.148 $\pm$0.035&-102.21$\pm$0.11&4.3$\pm$0.3\cr
\tcotd &6.466 $\pm$0.110&-101.30$\pm$0.03\tablenotemark{a}&5.2$\pm$0.1\tablenotemark{b}&&3.283 $\pm$0.046&-101.32$\pm$0.03\tablenotemark{a}&5.1$\pm$0.1\tablenotemark{b}\cr
&2.442 $\pm$0.110&-87.78$\pm$0.04&6.0$\pm$0.1&&3.246 $\pm$0.046&-87.94$\pm$0.03\tablenotemark{a}&6.2$\pm$0.1\tablenotemark{b}\cr
\hline
&&&MM2\cr
\hline
\csdu &1.591 $\pm$0.053&-89.67$\pm$0.02&4.3$\pm$0.1&& \nodata & \nodata & \nodata\cr
\cscc &0.738 $\pm$0.083&-89.65$\pm$0.03&3.6$\pm$0.1&& \nodata & \nodata & \nodata\cr
\csss &1.080 $\pm$0.120&-89.72$\pm$0.05&4.5$\pm$0.1&&0.680 $\pm$0.049&-89.76$\pm$0.04&4.4$\pm$0.1\cr
\tcotd &10.753 $\pm$0.106&-88.93$\pm$0.03\tablenotemark{a}&6.1$\pm$0.1\tablenotemark{b}&&9.386 $\pm$0.067&-89.40$\pm$0.01&5.5$\pm$0.1\cr
\hline
&&&MM3\cr
\hline
\csdu &2.650 $\pm$0.067&-89.21$\pm$0.01&5.3$\pm$0.1&& \nodata & \nodata & \nodata\cr
\cscc &1.033 $\pm$0.086&-88.89$\pm$0.03\tablenotemark{a}&5.0$\pm$0.4\tablenotemark{b}&& \nodata & \nodata & \nodata\cr
\csss &1.398 $\pm$0.120&-90.24$\pm$0.05&6.1$\pm$0.1\tablenotemark{c}&&0.422 $\pm$0.048&-89.68$\pm$0.06&6.0$\pm$0.2\tablenotemark{d} \cr
&0.916 $\pm$0.120&-91.79$\pm$2.46&31.1$\pm$2.8\tablenotemark{c}&&0.146 $\pm$0.049&-91.91$\pm$0.40&26.3$\pm$0.9\tablenotemark{d} \cr
\tcotd &0.996 $\pm$0.110&-99.93$\pm$0.07&6.3$\pm$0.2&&0.770 $\pm$0.051&-100.38$\pm$0.01&4.1$\pm$0.1\cr
&15.915 $\pm$0.110&-89.44$\pm$0.01&6.8$\pm$0.1&&11.869 $\pm$0.051&-88.98$\pm$0.03\tablenotemark{a}&6.0$\pm$0.1\tablenotemark{b}\cr
\soos &2.709 $\pm$0.042&-90.20$\pm$0.03\tablenotemark{a}&29.8$\pm$0.5\tablenotemark{b}&& \nodata & \nodata & \nodata\cr
\sooo &2.024 $\pm$0.046&-90.31$\pm$0.03\tablenotemark{a}&32.6$\pm$0.6\tablenotemark{b}&& \nodata & \nodata & \nodata\cr
\sioss &0.859 $\pm$0.035&-89.31$\pm$0.03\tablenotemark{a}&43.3$\pm$1.8\tablenotemark{b}&& \nodata & \nodata & \nodata\cr
\sioos &0.899 $\pm$0.045&-89.64$\pm$0.03\tablenotemark{a}&47.2$\pm$2.4\tablenotemark{b} && \nodata & \nodata & \nodata\cr
\hline
&&&MM4\cr
\hline
\csdu &2.262 $\pm$0.055&-88.34$\pm$0.01&4.7$\pm$0.1&& \nodata & \nodata & \nodata\cr
\cscc &0.776 $\pm$0.064&-89.17$\pm$0.03&4.6$\pm$0.1&& \nodata & \nodata & \nodata\cr
\csss &1.498 $\pm$0.120&-88.21$\pm$0.04&3.9$\pm$0.1&&0.413 $\pm$0.046&-88.83$\pm$0.06&4.6$\pm$0.1\cr
\tcotd &15.320 $\pm$0.090&-87.79$\pm$0.01&4.5$\pm$0.1&&8.616 $\pm$0.047&-88.33$\pm$0.01&5.1$\pm$0.1\cr
\hline
&&&MM5\cr
\hline
\csdu &2.949 $\pm$0.056&-88.44$\pm$0.01&5.1$\pm$0.1&& \nodata & \nodata & \nodata \cr
\cscc &1.057 $\pm$0.085&-88.89$\pm$0.03&4.7$\pm$0.1&& \nodata & \nodata & \nodata\cr
\csss &0.899 $\pm$0.130&-89.21$\pm$0.07&4.5$\pm$0.2&&0.463 $\pm$0.048&-88.76$\pm$0.05&4.9$\pm$0.1\cr
\tcotd &0.701 $\pm$0.100&-100.32$\pm$0.05&2.1$\pm$0.1&&0.714 $\pm$0.066&-100.54$\pm$0.02&2.9$\pm$0.05\cr
&19.143 $\pm$0.100&-88.41$\pm$0.03\tablenotemark{a}&5.4$\pm$0.1\tablenotemark{b}&&14.312 $\pm$0.066&-88.07$\pm$0.01&5.2$\pm$0.1\cr	
\hline
&&&MM6\cr
\hline
\csdu &1.039 $\pm$0.075&-100.48$\pm$0.04&4.5$\pm$0.1&& \nodata & \nodata & \nodata\cr
\cscc &0.205 $\pm$0.073&-100.35$\pm$0.12&4.0$\pm$0.3&& \nodata & \nodata & \nodata\cr
\csss &1.837 $\pm$0.099&-100.05$\pm$0.03&5.7$\pm$0.1&&0.423 $\pm$0.045&-100.00$\pm$0.05&6.3$\pm$0.1\cr
\tcotd &10.806 $\pm$0.110&-99.84$\pm$0.02&6.2$\pm$0.1&&5.684 $\pm$0.042&-100.48$\pm$0.03\tablenotemark{a}&5.6$\pm$0.1\tablenotemark{b}	 \cr
\enddata
\label{tbl:linepeaks}
\tablenotetext{a}{The error in the estimation of the velocity is fixed to $\sigma_V=0.03$   }
\tablenotetext{b}{Non-gaussinan profile. $\Delta V\ =\ I(T_A^*)\, /\, T^*_{A}\, _{Peak}$. }
\tablenotetext{c}{For this spectrum, we considered two gaussian fittings. The first is related with the broad emission related with outflowing gas, and the second gaussian fit is made to the spectrum after the subtraction of the broad emission and is related with the ambient gas.}
\tablenotetext{d}{Same as previous note, but for the composited integrated spectrum.}
\end{deluxetable}

\begin{deluxetable}{ccccc}
\tabletypesize{\scriptsize}
\tablecaption{Sizes (FWHM) obtained with \csss\ line \label{tbl:cs_sizes}}
\tablewidth{0pt}
\tablehead{
\colhead{Clump} & \colhead{$\Delta v$} & 
\colhead{Maj axis} & \colhead{Min axis} & \colhead{Radius} \\
    & \colhead{$\left(\kms \right)$} & \colhead{$\left( \arcsec \right)$}& \colhead{$\left( \arcsec \right)$}& \colhead{$\left( \mathrm{pc} \right)$}
}
\startdata

MM1	& $4.3\pm 0.3$ & $35\pm 3.6$  &$23\pm 2.4$ & $0.30\pm 0.03$    \cr
MM2 & $4.4\pm 0.1$ & $68\pm 7.6$  &$47\pm 5.3$ & $0.94\pm 0.04$    \cr
MM3 & $6.0\pm 0.2$ & $37\pm 1.4$  &$21\pm 0.8$ & $0.29\pm 0.01$    \cr
MM4 & $4.6\pm 0.1$ & $44\pm 3.7$  &$24\pm 1.9$ & $0.42\pm 0.02$    \cr
MM5 & $4.9\pm 0.1$ & $110\pm 17.3$&$44\pm 6.9$ & $1.19\pm 0.07$    \cr
MM6 & $6.3\pm 0.1$ & $29\pm 0.7$  &$25\pm 0.6$ & $0.25\pm 0.01$    \cr
\enddata
\label{tbl:cs_sizes}
\end{deluxetable}


{\setlength{\tabcolsep}{2.4pt} 
\begin{deluxetable}{cccccccccccccccccc}
\tabletypesize{\scriptsize}
\tablecaption{Fluxes from continuum observations\label{tbl:seds}}
\tablewidth{1.0\textwidth}
\tablehead{
&\multicolumn{2}{c}{MIllimeter cont.}&&\multicolumn{4}{c}{IRAS bands}&&\multicolumn{4}{c}{MSX bands}&&\multicolumn{4}{c}{Spitzer bands}\cr
\cline{2-3} \cline{5-8} \cline{10-13} \cline{15-18}
\colhead{Region} & \colhead{1.2mm}& \colhead{0.87mm}&& \colhead{100\um}& \colhead{60\um}& \colhead{25\um}& \colhead{12\um}&& \colhead{21.3\um}& \colhead{14.7\um}& \colhead{12.1\um}& \colhead{8.3\um}&& \colhead{8\um}& \colhead{5.8\um}& \colhead{4.5\um}& \colhead{3.6\um}\cr
&\colhead{(Jy)}&\colhead{(Jy)}&&\colhead{(Jy)}&\colhead{(Jy)}&\colhead{(Jy)}&\colhead{(Jy)}&&\colhead{(Jy)}&\colhead{(Jy)}&\colhead{(Jy)}&\colhead{(Jy)}&&\colhead{(Jy)}&\colhead{(Jy)}&\colhead{(Jy)}&\colhead{(Jy)}
}
\startdata
Complex&37.0&227.9&&36517.0&30998.2&3855.8&649.1&&2680.8&1143.3&1005.1&373.0&&712.2&262.2&36.3&34.1\cr
MM1     &3.1 &27.3  &&2254.0 &629.3  &24.9  &3.7  &&22.9  &13.1  &21.4  &13.1 &&25.0 &7.5  &0.7 &0.9 \cr
MM2     &6.4 &38.3  &&8548.7 &7948.4 &1177.2&232.5&&855.4 &387.9 &305.7 &86.4 &&136.7&48.8 &6.9 &5.7 \cr
MM3     &11.6&66.7  &&7687.6 &6682.3 &496.3 &72.3 &&731.5 &309.9 &269.5 &99.2 &&171.8&73.6 &10.7&9.6 \cr
MM4     &4.9 &42.7  &&5098.2 &5243.1 &707.9 &122.2&&500.8 &222.7 &169.7 &54.3 &&84.3 &29.9 &4.4 &3.9 \cr
MM5     &7.6 &52.5  &&8618.1 &6530.5 &923.5 &147.2&&482.0 &187.7 &173.7 &71.5 &&102.8&38.7 &4.6 &4.2 \cr
MM6     &3.5 &27.6  &&2086.1 &1710.6 &192.1 &33.9 &&115.0 &47.5  &51.4  &24.6 &&40.63&14.1 &2.3 &2.0 \cr

\enddata
\label{tbl:flujos}
\end{deluxetable}

}

\begin{deluxetable}{ccccccccc}
\tabletypesize{\scriptsize}
\tablecaption{Observed parameters of radio sources detected towards \gtres \label{tbl:radio_param}}
\tablewidth{0pt}
\tablehead{
\colhead{Component} & \colhead{Freq.}& \colhead{$\alpha$(J2000)} & 
\colhead{$\delta$(J2000)} & \colhead{Flux Peak} &
\colhead{Flux density} & \colhead{Beam} &\colhead{Deconv. angular size}&\colhead{Spectral index} \\
 		      &\colhead{(GHz)}& 			     &			 			 &\colhead{(Jy/beam)}&\colhead{(Jy)}&\colhead{(\arcsec)} &\colhead{(\arcsec)} 
}
\startdata

   A      &4.80 &16 12 07.510&-51 30 02.23 &0.155 &0.255 &2.7$\times$1.8 &1.97$\times$1.54&\nodata\cr
          &8.64 &16 12 07.471&-51 30 02.39 &0.101 &0.217 &1.5$\times$1.0 &1.49$\times$1.34&\cr
\tableline
   B      &4.80 &16 12 09.036&-51 25 47.83 &0.125 &0.141 &2.7$\times$1.8 &0.81$\times$0.60&0.8$\pm$0.2\cr
          &8.64 &16 12 09.038&-51 25 47.79 &0.166 &0.222 &1.5$\times$1.0 &0.84$\times$0.36&\cr
\tableline
   C      &4.80 &16 12 10.039&-51 28 37.72 &0.085 &0.095 &2.7$\times$1.8 &0.86$\times$0.37&1.2$\pm$0.2\cr
          &8.64 &16 12 10.030&-51 28 37.66 &0.158 &0.194 &1.5$\times$1.0 &0.61$\times$0.40&\cr
\tableline
   D      &4.80 &16 12 27.308&-51 27 32.54 &0.066 &0.135 &2.7$\times$1.8 &2.30$\times$2.06&\nodata\cr
          &8.64 &16 12 27.264&-51 27 32.81 &0.022 &0.092 &1.5$\times$1.0 &2.15$\times$1.87&\cr
\enddata
\label{tbl:radio_param}
\end{deluxetable}

\begin{deluxetable}{ccccc|cccc|cccc}
\tabletypesize{\scriptsize}
\tablecaption{Spectral Energy Distributions\label{tbl:seds}}
\tablewidth{0pt}
\tablehead{
\colhead{Region} & \colhead{T$^c$}& \colhead{$\beta^c$}&
\colhead{$\nu^c_0$}& \colhead{$\theta^c$}& \colhead{T$^w$}&
\colhead{$\beta^w$}& \colhead{$\nu^w_0$}& \colhead{$\theta^w$}&
\colhead{T$^h$}& \colhead{$\beta^h$}& \colhead{$\nu^h_0$}&
\colhead{$\theta^h$} \cr \colhead{} &\colhead{(K)}&\colhead{}&\colhead{(THz)}&\colhead{(\arcsec)}&\colhead{(K)}&  &\colhead{(THz)} &\colhead{(\arcsec)}&\colhead{(K)}&\colhead{}&\colhead{(THz)}&\colhead{(\arcsec)} 
}
\startdata
Complex&30 &2.2&7.3&200&114&1.0&25&4.3&302&1.0&316&0.85   \cr
MM1      &29 &1.7&6.1&33& \nodata & \nodata & \nodata& \nodata & \nodata & \nodata & \nodata & \nodata \cr
MM2      &34 &1.9&5.8&50&136&1.0&26&1.8& \nodata & \nodata & \nodata & \nodata \cr
MM3      &35 &1.6&5.9&40&109&1.0&29&2.0& \nodata & \nodata & \nodata & \nodata \cr
MM4      &30 &1.9&5.8&50&112&1.0&24&2.3& \nodata & \nodata & \nodata & \nodata \cr
MM5      &33 &1.8&5.7&51&114&1.0&28&2.4& \nodata & \nodata & \nodata & \nodata \cr
MM6      &31 &1.7&5.8&30&112&1.0&28&1.2& \nodata & \nodata & \nodata & \nodata \cr
\enddata
\label{tbl:seds_clumps}
\tablenotetext{c}{Cold component}
\tablenotetext{w}{Warm component}
\tablenotetext{h}{Hot component}
\end{deluxetable}

\begin{deluxetable}{ccc}
\tabletypesize{\scriptsize}
\tablecaption{Bolometric luminosities of complex and individual clumps\label{tbl:luminosities}}
\tablewidth{0pt}
\tablehead{
\colhead{Region} &\colhead{Integrated flux}
&\colhead{Bolometric luminosity} \cr \colhead{}&\colhead{$10^6$ (Jy GHz)}&\colhead{$10^6$ \Lsun}
}
\startdata

Complex&252.2                 & 4.4\\
MM1&7.2											 	  & 0.1	\cr
MM2&57.6												& 1.0\cr
MM3&40.9 												& 0.7 \cr
MM4&35.5												& 0.6\cr
MM5&48.2 												& 0.8\cr
MM6&12.3 												& 0.2\cr
\enddata
\label{tbl:lumin}
\end{deluxetable}

\begin{deluxetable}{cccccccc}
\tabletypesize{\scriptsize}
\tablecaption{Masses and densities derived for the clumps\label{tbl:seds}}
\tablewidth{0pt}
\tablehead{
&0.87 mm&&\multicolumn{3}{c}{1.2 mm}&& \csss \cr 
 \cline{2-2} \cline {4-6}  \cline{8-8}
\colhead{Clump} &\colhead{Mass}
&&\colhead{Mass}&\colhead{Density}&\colhead{Surface density}&&\colhead{Virial mass}\cr
\colhead{}&\colhead{$10^3\: \Msun$} &&\colhead{$10^3\: \Msun$}&\colhead{$10^4 \:\left(\mathrm{cm}^{-3}\right)$}&\colhead{$\:\left(\mathrm{g\: cm}^{-2}\right)$}&& \colhead{10$^3\:$\Msun}
}
\startdata 
MM1&5.1&&1.9& 2.9 &  0.30&& $1.2\pm 0.2$\cr
MM2&5.8&&3.2& 1.6 &  0.24&& $3.8\pm 0.3$\cr
MM3&9.7&&5.6& 10.9 &  1.03&& $2.2\pm 0.2$\cr
MM4&7.6&&2.8& 1.4 &  0.21&& $1.9\pm 0.1$\cr
MM5&8.2&&3.9& 1.9 &  0.29&& $6.0\pm 0.4$\cr
MM6&4.7&&1.9& 3.7 &  0.35&& $2.1\pm 0.1$\cr
\enddata
\label{tbl:dust_cs_masses}
\end{deluxetable}

\begin{deluxetable}{lcccc}
\tabletypesize{\scriptsize}
\tablecaption{Outflow parameters obtained from rotational diagrams\label{tbl:seds}}
\tablewidth{0pt}
\tablehead{
\colhead{Molecule} & \colhead{Velocity range}
& \colhead{$T_{rot}$} & \colhead{$Q(T_{rot})$} &
\colhead{$N_T$}\cr \colhead{}&\colhead{}&\colhead{$\left(\mathrm{K}\right)$}&\colhead{} &\colhead{10$^{13}\;\left(\mathrm{cm}^{-2}\right)$}
}
\startdata
SiO                &-144.1 $\leftrightarrow$ -95.6&123.2$\pm$15.2								&118.3									 &3.26  \cr
									 &-79.9 $\leftrightarrow$ -16.3 &138.0$\pm$29.4 								&132.5									 &2.56  \cr
\tableline
SO								 &-144.1 $\leftrightarrow$ -95.6&93.7$\pm$4.8									&254.3									 &45.93 \cr
									 &-79.9 $\leftrightarrow$ -16.3 &77.3$\pm$4.4									&207.1									 &37.95\cr

\enddata
\label{tbl:rotation}		
\end{deluxetable}

\clearpage



\clearpage

\clearpage

\appendix

\section{Online material}

\setcounter{figure}{0}
\renewcommand{\thefigure}{A\arabic{figure}}
\clearpage
\begin{figure}[h] 
	\begin{center}
		\includegraphics[angle=0,scale=0.8]{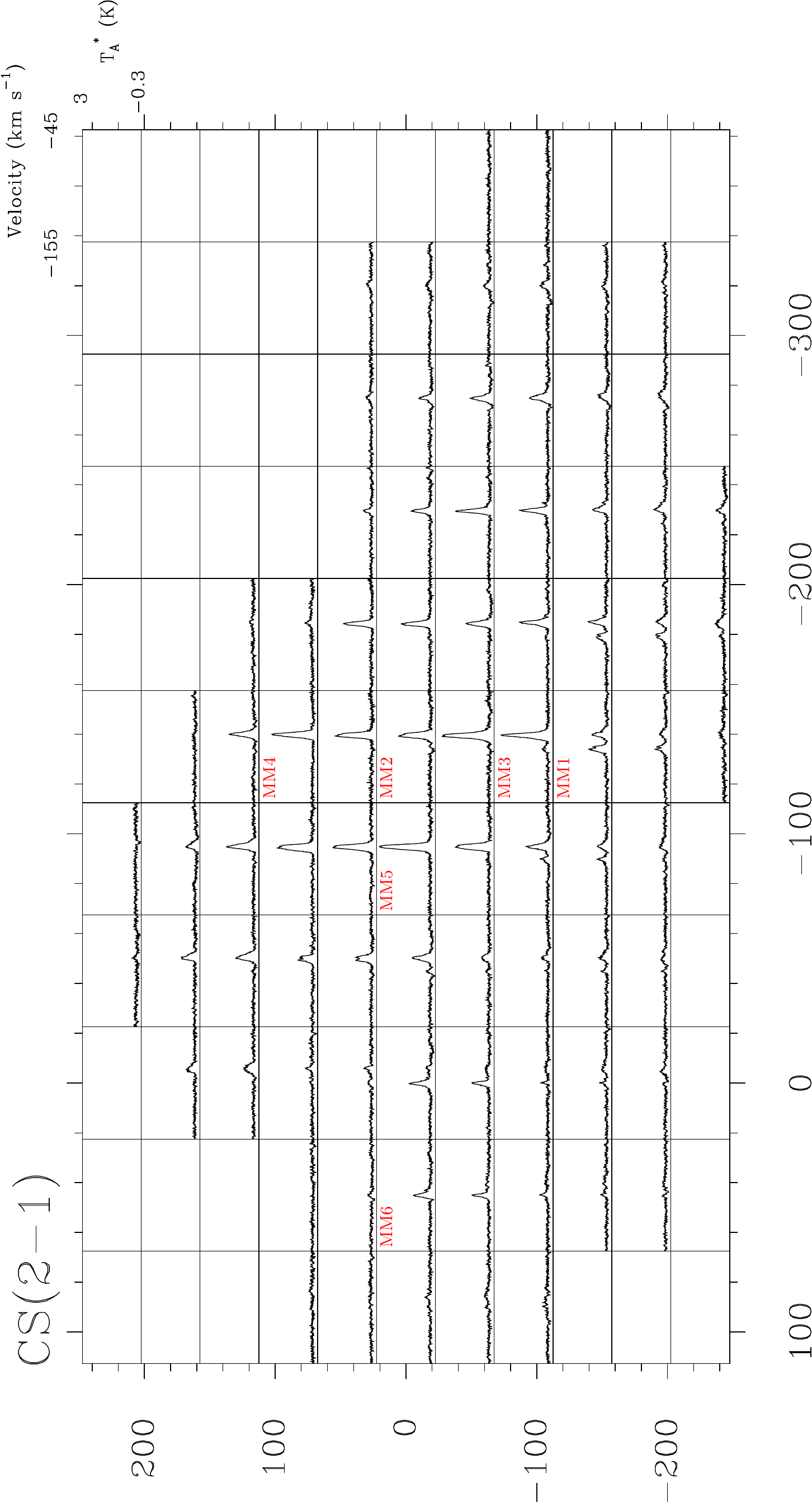}
	\end{center}
	\caption[Observed spectra of the \csdu\ line emissions toward the core \gtres.]{Observed spectra of the \csdu\ line emissions toward the \gtres\ GMC central region. The grid spacing is 45\arcsec. Offsets are from the reference position at $\alpha_{2000} = 16^h12^m24.5^s$ and $\delta_{2000} = -51\arcdeg
27\arcmin 29.98\arcsec$. In each box the velocity scale ranges from -155 to -45 \kms. The antenna temperature scale is from -0.3 to 3 K. The labels MM1 to MM6 show the positions in the grid closer to the peak position of the millimeter clumps, defined in section 3.2.1.} 
	\label{fig:grid_cs21}
\end{figure}
\clearpage
\begin{figure}[h]
	\begin{center}
		\includegraphics[angle=0,scale=0.8]{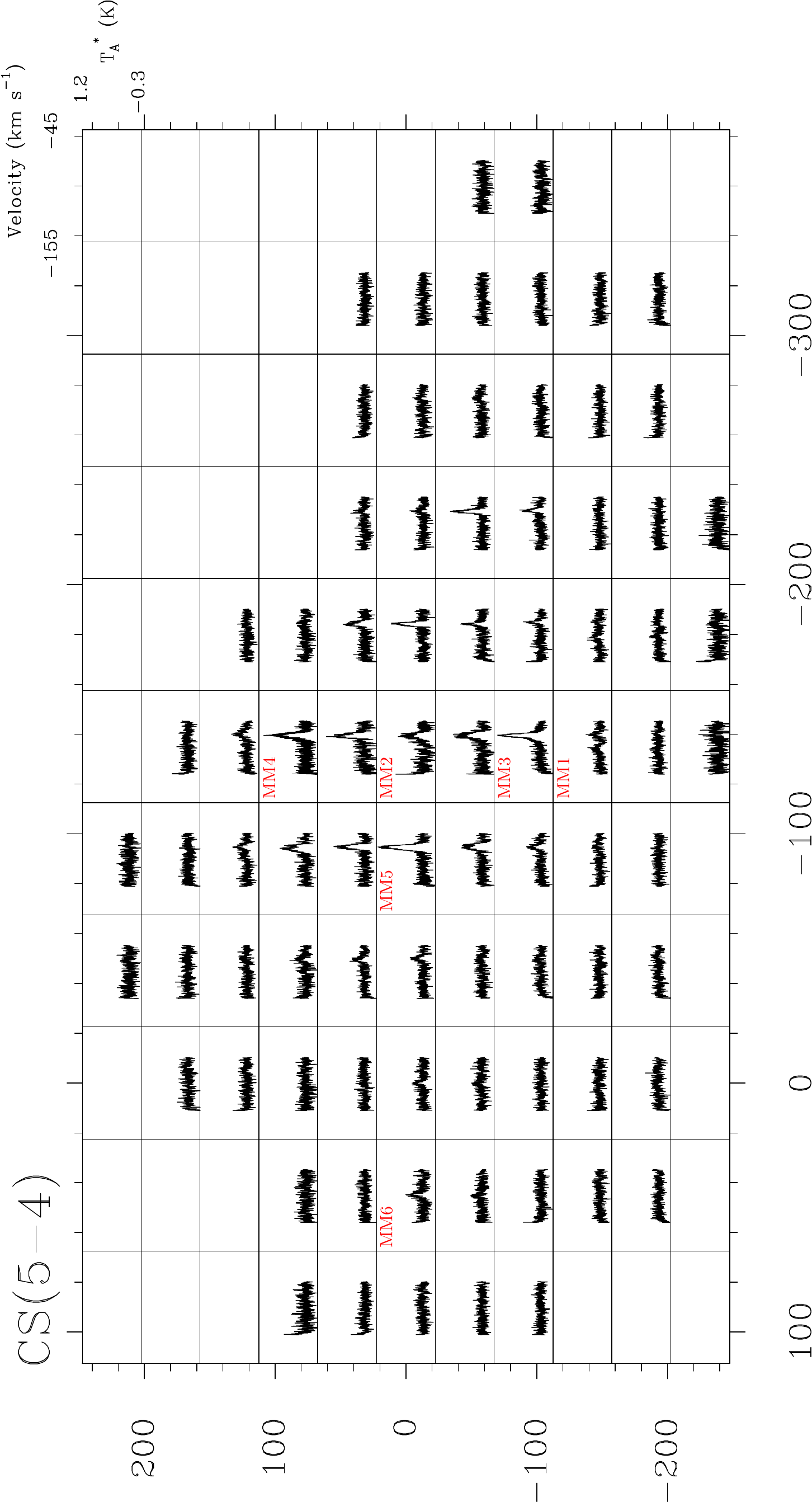}
	\end{center}
	\caption[Observed spectra of the \cscc\ line emissions toward the core \gtres. ]{Observed spectra of the \cscc\ line emissions toward the \gtres\ GMC central region. The grid spacing is 45\arcsec. Offsets are from the reference position at $\alpha_{2000} = 16^h12^m24.5^s$ and $\delta_{2000} = -51\arcdeg 27\arcmin 29.98\arcsec$. In each box the velocity scale ranges from -155 to -45 \kms. The antenna temperature scale is from -0.3 to 1.2 K. The labels MM1 to MM6 show the positions in the grid closer to the peak position of the millimeter clumps, defined in section 3.2.1.} 
	\label{fig:grid_cs54}
\end{figure}
\clearpage
\begin{figure}[h]
\begin{center}
		\includegraphics[angle=0,scale=0.8]{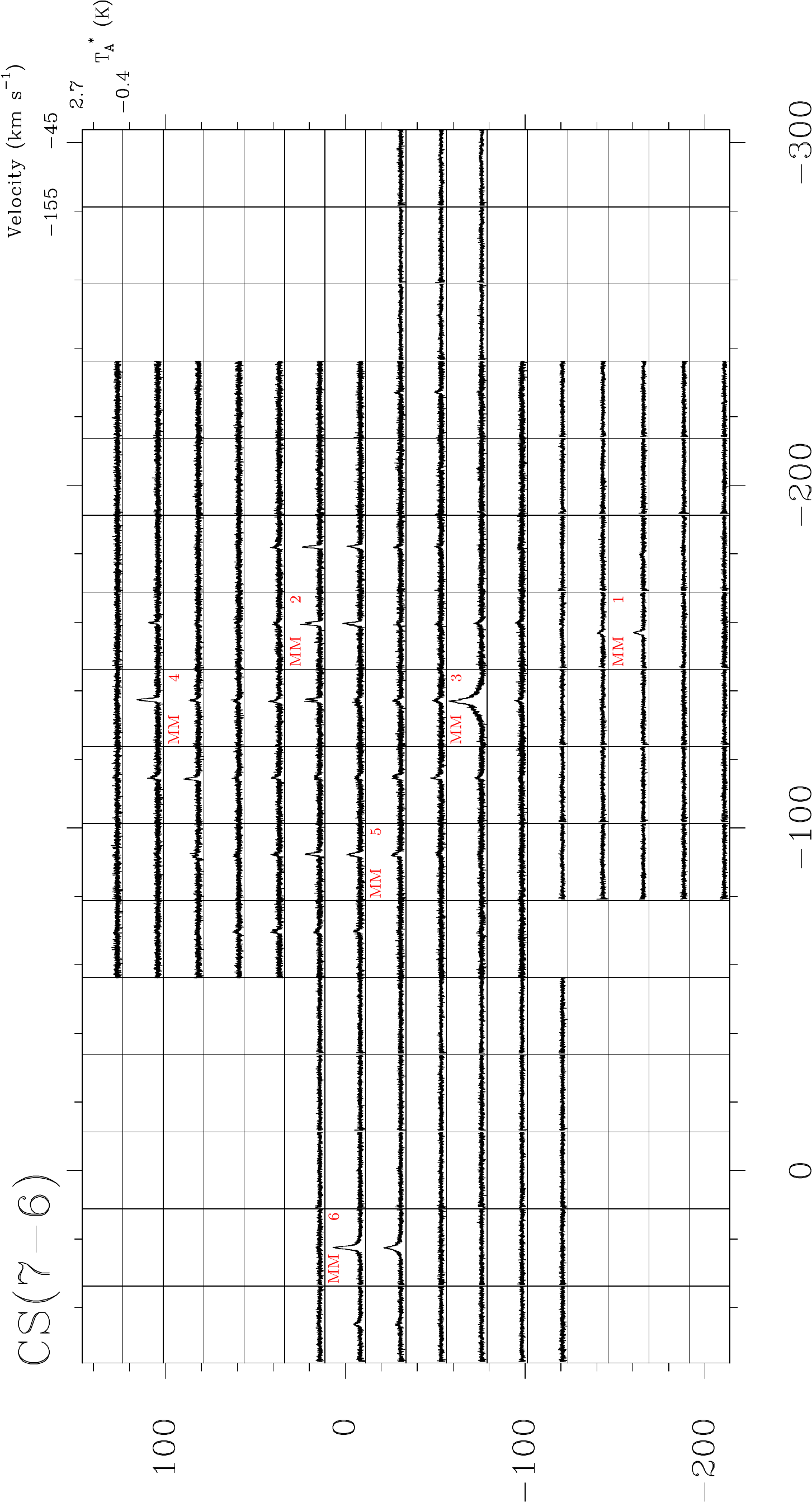}
	\end{center}
	\caption[Observed spectra of the \csss\ line emissions toward the core \gtres.]{Observed spectra of the \csss\ line emissions toward the \gtres\ central region. The grid spacing is 22.5\arcsec. Offsets are from the reference position at $\alpha_{2000} = 16^h12^m24.5^s$ and $\delta_{2000} = -51\arcdeg
27\arcmin 29.9\arcsec$. In each box the velocity scale ranges from -155 to -45 \kms. The antenna temperature scale is from -0.4 to 2.7 K. The labels MM1 to MM6 show the positions in the grid closer to the peak position of the millimeter clumps, defined in section 3.2.1.} 
	\label{fig:grid_cs76}
\end{figure}
\clearpage
\begin{figure}[h]
	\begin{center}
		\includegraphics[angle=0,scale=0.8]{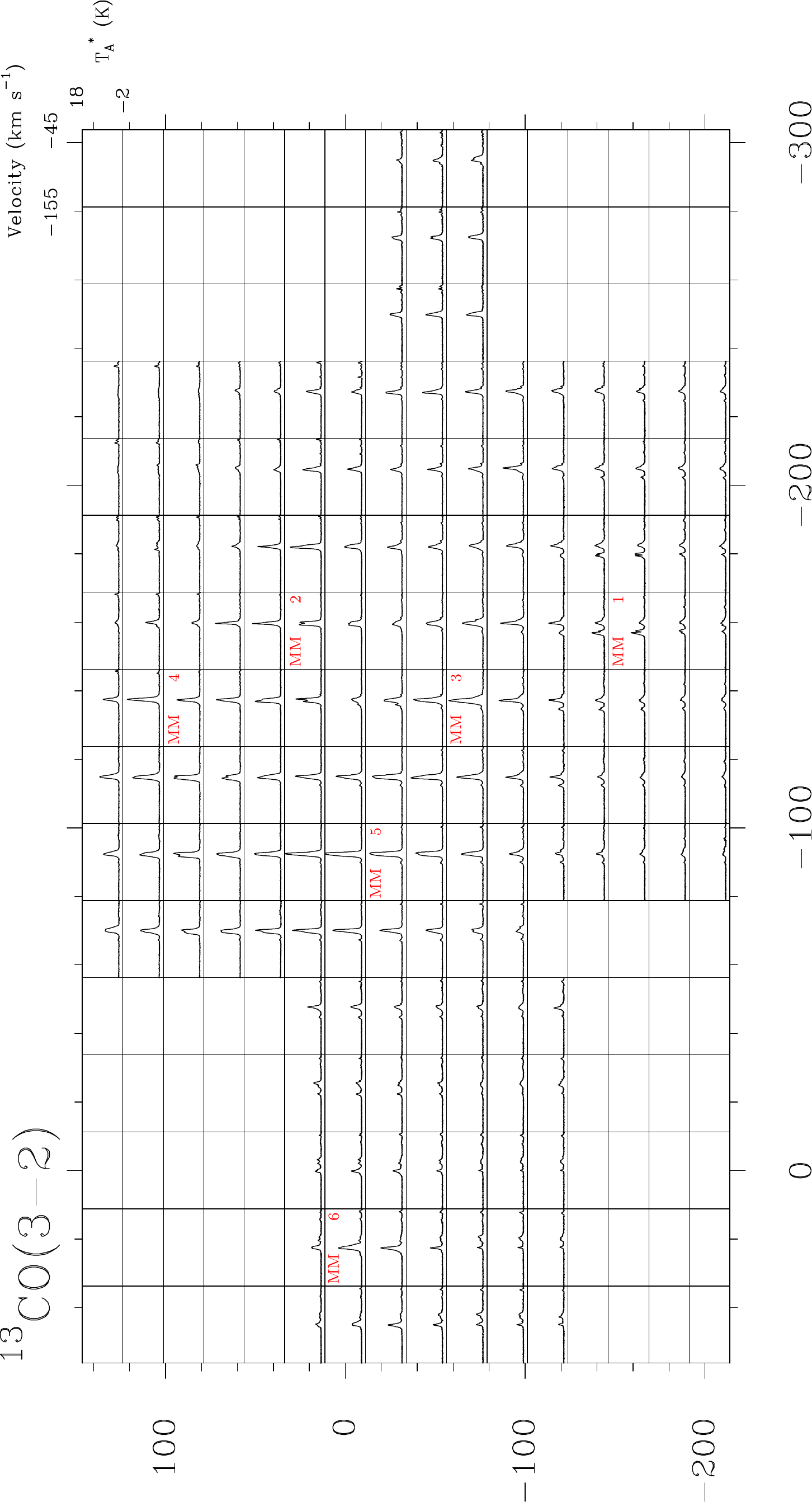}
	\end{center}
	\caption[Observed spectra of the \tcotd\ line emissions toward the core \gtres.]{Observed spectra of the \tcotd\ line emissions toward the \gtres\ central region. The grid spacing is 22.5\arcsec. Offsets are from the reference position at $\alpha_{2000} = 16^h12^m24.5^s$ and $\delta_{2000} = -51\arcdeg 27\arcmin 29.9\arcsec$. In each box the velocity scale ranges from -155 to -45 \kms. The antenna temperature scale is from -2 to 18 K. The labels MM1 to MM6 show the positions in the grid closer to the peak position of the millimeter clumps, defined in section 3.2.1.} 
	\label{fig:grid_13co32}
\end{figure}
\clearpage

\end{document}